\documentclass[useAMS,usenatbib]{mn2e}

\usepackage{times}
\usepackage{amssymb}
\usepackage{graphicx}
\usepackage{rotating}
\usepackage{natbib} %This uses the natbib style for the bibliography.

\title{Radio observations of the Horologium-Reticulum Supercluster -I. A3158: Excess star-forming galaxies in a merging cluster?}
\author[M. Johnston-Hollitt et al.]
       {M. Johnston-Hollitt$^{1, 2}$\thanks{E-mail: Melanie.JohnstonHollitt@utas.edu.au}, M. Sato$^{3,4}$, J. A. Gill$^1$, M. C. Fleenor$^5$ and A.-M. Brick$^1$\\
        1. School of Mathematics \& Physics, University of Tasmania, 7001, Australia\\
        2. Excellence Cluster Universe, Technische Universit\"{a}t M\"{u}nchen, Boltzmannstr. 2, D-85748, Garching, Germany\\
        3. Department of Astronomy, Graduate School of Science, The University of Tokyo, 7-3-1 Hongo, Bunkyo-ku, Tokyo 113-0033, Japan\\
        4. Mizusawa VERA Observatory, National Astronomical Observatory of Japan,
           2-21-1 Osawa, Mitaka, Tokyo 181-8588, Japan\\
        5. Roanoke College, 221 College Lane, Salem, Virginia 24153, USA}
\date{Accepted 200X Month Day.
      Received 200X Month Day;
      in original form 200X Month Day }

\pagerange{\pageref{firstpage}--\pageref{lastpage}}
\pubyear{2008}

\begin{document}

\maketitle

\label{firstpage}

\begin{abstract}
We present 1.4 and 2.5 GHz Australia Telescope Compact Array (ATCA) observations of the galaxy cluster A3158 (z=0.0597) which is 
located within the central part of the Horologium-Reticulum Supercluster (HRS). Spectroscopic data for the central
part of the HRS suggest that A3158 is in a dynamically important position within the supercluster and that it is moving
toward the double cluster system A3125/A3128 which marks the centre of the HRS. A total of 110 radio galaxies
are detected in a 35 arcminute radius about the cluster at 1.4 GHz, of which 30 are also detected
at 2.5GHz. We examine the source counts and compute the Radio Luminosity Function (RLF) at 1.4 GHz from the 
subset of 88 sources found within the full-width half-power area of the ATCA beam. Comparison of the source counts in the
area over the background, as computed by Prandoni et al.~(2001), shows some evidence of an excess of galaxies with 
L$_{1.4 GHz} \leq 2 \times 10^{22} W Hz^{-1}$. This result seems to indicate a star forming population and is a result similar to 
that found recently by Owen et al.~(2005) for the merging cluster A2125. In addition we find that the radio luminosity function
for early-type galaxies (E and S0) below log P$_{1.4} \sim$ 22.5 is lower than that found for a composite cluster 
environment (Ledlow \& Owen, 1996) but is similar to the early-type RLF for clusters in the centre of the Shapley Super 
cluster (Venturi et al.~ 2000) which are believed to be in the latter stages of merging. This result implies that the cores 
of superclusters are environments where radio emission, particularly resultant from AGN, is suppressed in the later stages
of merging. Thus, radio observations of clusters might be sensitive indicators of the precise merger stage 
of the cluster but more observational evidence is still required to establish this trend.

\end{abstract}

\begin{keywords}
galaxies:clusters: general - galaxies:clusters: individual: A3158 - radio continuum:galaxies.
\end{keywords}

\section{Introduction}
Observational and numerical studies suggest that dynamical interactions are frequent occurrences in clusters of galaxies 
and that these events are the most energetic in the recent history of the Universe. Matter is believed to stream along
preferred axes (so-called filaments and sheets) often accelerating in-falling galaxies to velocities of the order of
10$^3$ km s$^{-1}$. Rich clusters form at the intersection of the filament and sheets, and larger mass congregations, such as
superclusters, are often detected as a collection of several clusters and filaments. These dynamical processes cause shocks,
turbulence and bulk flows along the filamentary axes resulting in disturbance of the intracluster medium (ICM). Such 
perturbations are frequently detected with X-ray observations which show a variety of features including sharp temperature
jumps, distorted X-ray isophotes and regions of enhanced pressure and entropy. At radio wavelengths, cluster mergers are
thought to generate large areas of diffuse synchrotron emission either as a central, unpolarised halo or one or occasionally
two peripheral ``radio relics''. Cluster ``weather'' and bulk flows have also been cited as causes of the bending of the jets
in head-tail galaxies (Burns et al.~1998, Johnston-Hollitt et al.~2004), although recent results suggest that more local
conditions might play a more important role in their formation (Mao et al.~2008). 

  \begin{figure*}
   \begin{center}
    \includegraphics {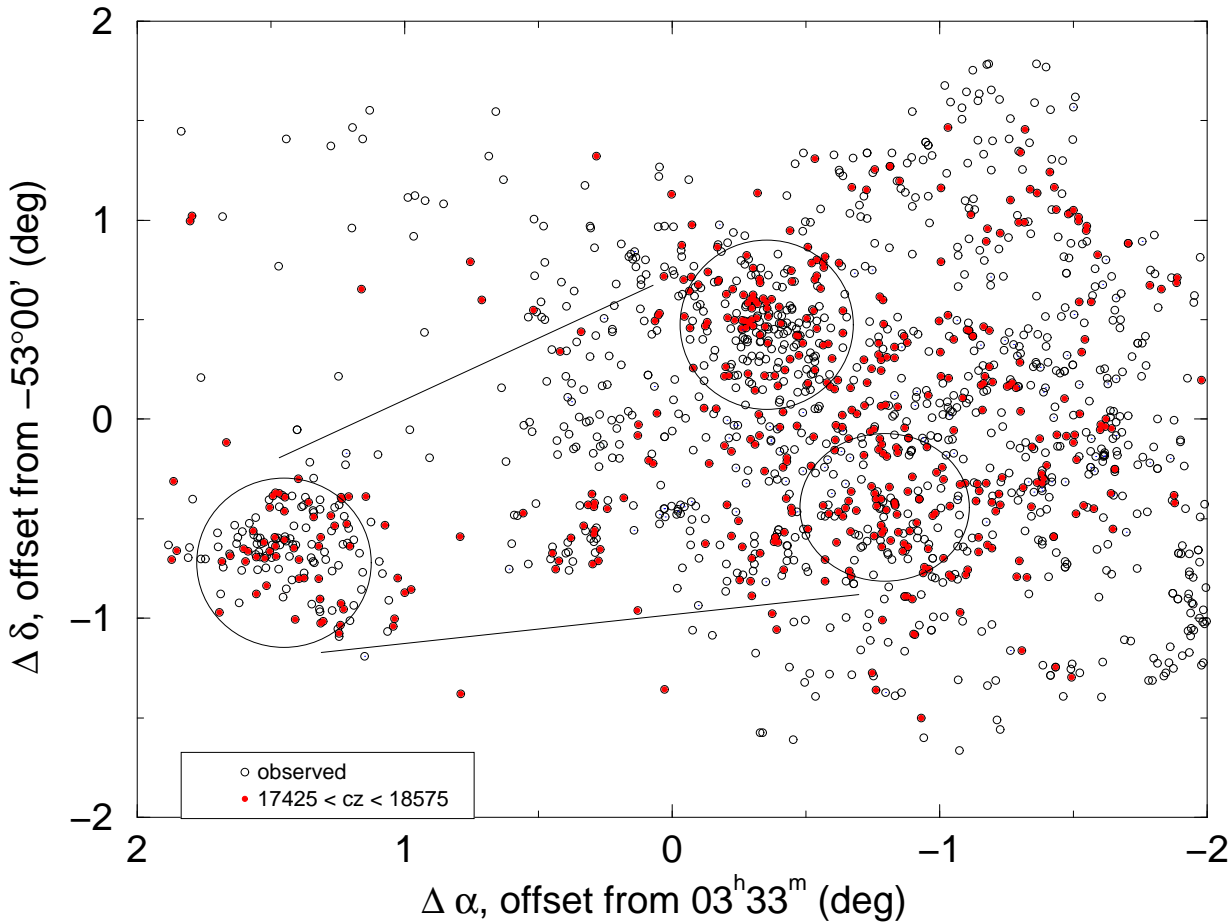}
%[width=8.5cm, height=7cm]{decra_obs_17425_18575.eps}
   \caption{Plot of all spectroscopic redshifts available in the region between A3158 (shown on the left, marked with a circle) and the A3125/A3128 complex on the
right, again with the clusters denoted by large circles. Filled circles represent objects which have velocities in the range of 
17425 km s$^{-1}$ and 18575 km s$^{-1}$ which encompasses those in the bridge investigated by Fleenor (2006).}
   \label{bridge}
 \end{center}
 \end{figure*}

In addition to the large-area phenomena discussed above, the dynamical history of galaxy clusters should also be imprinted
on the constituent galaxies. In particular, increased secondary star formation either observed in the optical 
(Burns et al.~1994; Caldwell \&
Rose 1997) or through increased low-power radio emission (Owen et al.~1999) is one possible signature of the history of a dynamical cluster.
However, the balance between mergers driving gas towards the centre of the galaxies and hence feeding the central AGN to induce
a starburst (Bekki 1999) and an increase in the ram pressure which would strip gas from galaxies and hence curtail the feeding
of the AGN (e.g. Balogh et al.~1998; Fujita et al.~1999) still needs to be addressed observationally. To date, few studies have 
been conducted 
with both sufficient sensitivity in the radio to probe the low-powered radio sources, and good optical observations to 
confirm cluster membership. The most notable progress to date has been made on sections of the Shapley Supercluster with
several areas analysed in both the radio and optical, including radio observations of A3556 (Venturi et al.~1997), 
A3571 (Venturi et al. 2002), the A3528 complex (Venturi et al.~2001) and the A3558 complex (Venturi et al.~2001; Miller 2005) and its 
outskirts (Giacintucci et al.~2004, Miller 2005). In addition to clusters in superclusters, the host radio population for
several individual clusters have also been examined including A2255 (Miller \& Owen, 2005), A2256 (Miller et al.~2003),
A2125 (Owen et al.~2005) and A2111 (Miller et al.~2006). However, at present, those observational studies that have been 
undertaken give incomplete results with both instances in which effects are seen on the radio population, and cases where they
are not observed. Sometimes neighbouring clusters in the same supercluster show markedly different results in terms of radio population. 
This may well be due to the radio galaxy population
only being affected in certain phases of dynamical activity (Venturi et al.~2002) however, the exact point at which a merger might 
affect the radio population is, as yet, unclear. Nevertheless, large-area observations of superclusters, such as Shapley, present 
ideal datasets with which to investigate the way in which environment affects the galaxy population.

In this paper, which will be the first in a series, we present a study of the radio population of A3158 which lies within the
kinematic core of the massive Horologium-Reticulum Supercluster (HRS). The HRS spans around 180 square degrees on the sky 
(Fleenor et al.~2006), contains upwards of 20 galaxy clusters (Zucca et al.~1993), is comprised of at least two major filaments 
(Einasto et al.~2003), and is the second largest mass concentration in the local 
300 Mpc (Hudson et al.~1999). A3158 lies adjacent to the so-called center of the HRS, defined by the double
cluster complex A3125/A3128 (Rose et al.~2002). The paper will be laid out as follows: Section 2 will discuss A3158 in detail;
Section 3 will give details of the radio observations of the cluster; Section 4 will present the radio source population and in
Sections 5 and 6 we will discuss the resultant radio source counts and radio luminosity functions; Sections 7 \& 8 will present
the discussion and conclusions. Throughout this paper we will assume H$_0$=71 km s$^{-1}$ Mpc$^{-1}$ which at the average redshift
of A3158 (z=0.0597) means that 1 arcsecond $\sim$ 1.1 kpc. 

\begin{table}
\begin{center}
\label{optp}
\caption{Known cluster parameters for A3158. Column 1 and 2 give the parameter and the most recent value from the literature. The reference
is given in column 3 where 1 means Havlen \& Quintana~(1978), 2 is Ebeling et al.~(1996), 3 is  Lokas et al.~(2006) and 4 is 
Fleenor et al.~(2006).}
\begin{tabular}{lll}
\hline
Parameter & Value & Reference\\
\hline
redshift & 0.0597 & 3\\
ellipticity & 0.3 & 1\\
X-ray Luminosity & 5.31 $\times 10^{44}$ erg s$^{-1}$ & 2\\
velocity & 17,910 kms$^{-1}$ & 4\\
Redshift n=145 & 0.0597 & 3\\
velocity disp. (n=145)& 970$\pm$57 kms$^{-1}$ & 3 \\
Viral Mass & $15.4^{+7.6}_{-5.4} 10^{14}$M$_{\sun}$ & 3\\
\hline
\end{tabular}
\end{center}
\end{table}

\section{A3158}
The ACO cluster A3158 (Abell, Corwin \& Olowin 1989) is a compact cluster of richness class two (Quintana \& Halven 1979)
and estimated viral mass of 15.4$^{+7.6}_{-5.4} \times 10^{14}$ solar masses (Lokas et al.~2006). It has been listed in many works under 
several different names, the most common of which is Calan 0340-538 (Melnick \& Quintana 1975). The cluster is
an elliptically dominated, centrally condensed cluster (Quintana \& Havlen 1979), with three central dominant (cD) 
galaxies (Sersic 1974). Two cD galaxies are found at the very centre of the cluster and have a 45 arcseconds 
separation (Lucey et al.~1983), the third lies approximately 6 arcminutes away to the southeast of the cluster core 
(See Fig \ref{fig:zoom}). The gravitational
centre of the cluster, as determined by the X-ray emission, is found to have a right ascension and declination (J2000) 
of  $\alpha= 03h 42m 39.6s$ and $\delta= -53^\circ 37' 50''$ (Lokas et al.~2006).

The HRS is comprised of two major groupings of galaxy clusters (Einasto et al.~2003), the northern clusters 
($-43 < \delta < -48$) which are situated around $z \sim 0.072$ and the southern 
clusters ($-51 < \delta < -57$) at $z \sim 0.06$ (Fleenor 2006). Abell 3158 is situated in the 
heart of the kinematic core (16,000 -- 23,000 km s$^{-1}$) of the HRS (Fleenor et al.~2004; Fleenor et al.~2005) 
and has close physical proximity to the southern of the two major mass concentrations in the supercluster.
A3158 is located approximately $2.5\deg$ ($\sim 10 h^{-1}$ Mpc) east of the 
A3125/A3128 double-cluster system which is thought to mark the southern centre of the 
supercluster (Rose et al.~2002; Johnston-Hollitt et al.~2004).  These three clusters have long been suggested
to be connected via a ``bridge'' of galaxies
(Lucey et al.~1983) which is part of the larger HRS structure. It is interesting to note that the two central 
cD galaxies in A3158 and the additional one to the south-east appear to line up in a NW-SE direction. 
This direction is significant, as it is the direction of the proposed bridge that connects A3158 with the southern 
kinematic core of the HRS.The galaxy density centre is shown to be 
concentrated on the two more central cD galaxies and is elongated in the direction of the third (Lucey et al.~1983). In addition,
on a larger scale, the optical isodensity plots for A3158 show that the cluster appears to be elongated along the axis
toward the centre of the HRS (Havlen \& Quintana 1978). 

Lucey et al.~(1983) measured the velocities of 129 galaxies in a 6$^{\circ} \times 6^{\circ}$ region about the central part 
of the HRS. Based on these results they established that A3158 was connected via a bridge of galaxies to the double cluster 
A3125/A3128 and that based on a simple two particle model that A3158 and the A3125/A3128 system are moving toward each 
other and will collide in 0.5 of a Hubble time. More recently Fleenor (2006) studied the dynamics of the A3125/A3128/A3158 system
finding that 52 galaxies (aside from those in Automatic Plate Measuring, APM, clusters (Dalton et al.~1992 \& 1994)) were 
likely to be members of the bridge connecting A3158 and the A3125/A3128 system.
Fleenor (2006) gives the bridge velocity range as between 17250 km s$^{-1}$ and 18575 km s$^{-1}$. This $10 h^{-1}$ Mpc intercluster filament 
has a galaxy overdensity of greater than 7$\times$ the average overdensity of the HRS ($\Delta \rho / \rho \sim 2.4$). Figure \ref{bridge} shows
A3158 and the A3125/A3128 complex along with the bridge of galaxies identified by Fleenor (2006). In the plot open circles represent galaxies with measured
spectroscopic redshift and filled, red circles are those corresponding to the bridge velocity, albeit with a slightly higher lower limit 
(17425 km s$^{-1}$ and 18575 km s$^{-1}$) . In addition, 
multi-wavelength data have suggested that A3125/A3128 are 
undergoing hierarchical merging along the axis connecting the two (Rose et al.~2002). Despite the fact that recent large-scale 
surveys have clearly establish that all of the member clusters of the HRS are connected via a filamentary network of galaxies 
(Fleenor et al.~2005) the bridge between A3158 and A3125/A3128 seems particularly significant. The proximity of A3158 to the 
southern core of the HRS, thus suggests the cluster is in a dynamically important position within the supercluster.

\subsection{Optical Analysis of A3158}

Not surprisingly, the two central cD galaxies, appear to dominate the dynamics of the cluster core. Initial analysis of this 
region by Havlen \& Quintana (1978) and Quintana \& Havlen (1979) shows that the core of the cluster is centrally condensed 
with predominantly elliptical galaxies, whereas spiral galaxies in the region appear to show no significant concentrations. 
Lucey et al.~(1983) noted that the two close cD galaxies 
have similar recessional velocities of the order of cz=17,500 km s$^{-1}$, which are similar to values found by Fleenor (2006)
for the bridge galaxies but notably less than the cluster average
(17,910 km s$^{-1}$), whereas the third cD has a recessional velocity of around cz$\approx$19,000 km s$^{-1}$, which is well above the 
cluster average. This suggests the third cD might be the centre of a subcluster. Subclustering containing the third cD galaxy 
is further substantiated by the flattened velocity histogram at 19,000 km s$^{-1}$ found in Smith et al.~(2004), Figure 7.
This form of mass segregation is suggested 
repeatedly throughout the literature, however, as yet, it has not been confirmed due to the insufficient 
number of galaxies with known redshifts to profile in the cluster (see Lokas et al.~(2006)). Mass segregation, however, is an indication 
of dynamical evolution, and as such is a strong indicator that A3158 is in a state of advanced evolution. The velocity 
dispersion profile is seen to be flat out to about 2.1 h$^{-1}_{o}$Mpc (assuming h$_{o}$=0.71) (den Hartog \& Katgert 1996; 
Dupke \& Bregman 2005). Details of the cluster properties are given in Table 1.
%\ref{optp}.

\subsection{X-ray Properties of A3158}
A3158 has a relatively high X-ray luminosity with L$_x$= 5.31 $\times 10^{44}$ ergs$^{-1}$ in the 0.1 - 2.4 keV band (Ebeling et al.~1996).
The X-ray emission emanating from the region has a smooth distribution which is centrally concentrated, and is similar in 
distribution to the optical component of the cluster giving little indication of substructure. While it is clear that the 
two large elliptical cD galaxies are the dynamical centre of the cluster, Ku et al.~(1983) found that the mass of the 
intracluster gas exceeds 10\% of the virial mass, which indicates that it may play a significant role in the dynamics of 
the outer regions of the system. Evidence for cluster mergers previously occurring in the A3158 region is given by 
Ohata et al.~(2001), who found that the X-ray temperature distribution of A3158 has large axis-symmetry along the major axis 
of its optical counterpart. This indicates a possible shock passage and lends support to the suggestion that mergers have occurred 
previously in the region.

\section{Observations}
\subsection{Radio Observations}

Observations of A3158 were carried out at 1.384 and 2.496 GHz on the Australia Telescope 
Compact Array (ATCA) on five different occasions throughout 1999. The data were taken in 
four array configurations (6A, 6B, 6C and 210m) and co-added to improve coverage 
in the uv-plane. The total observing time was 20.5 hours over all four
configurations. The first observation (taken by MJH) had the field centre at 
03 43 54.29, -53 41 36.6 
%$03\h43\m54\fs29,$-$53\arcdeg41\arcmin36\farcs6$ 
whereas the remainder of observations which were 
retrieved from the Australia Telescope Online Archive were centred at 
03 42 53.00, -53 37 43.00. Details of the observations are given in Table 2.
%\ref{tab:obs}.

\begin{table}
\label{tab:obs}
   \begin{center}

 \begin{tabular}{l|c|c|c|c}
        \hline

  Field Centre  &              &  Date & Array         & Integration\\
  RA$_{\rm{J2000}}$  & DEC$_{\rm{J2000}}$& 1999  & Config. & (min)\\
         \hline 
  03 43 54.29 & -53 41 36.6 & 26 Feb. & 6C & 136\\
  03 42 53.00 & -53 37 43.0 & 6  Sep. & 6D & 331\\
  03 42 53.00 & -53 37 43.0 & 9  Sep. & 6A & 215\\
  03 42 53.00 & -53 37 43.0 & 10 Sep. & 6D & 57\\
  03 42 53.00 & -53 37 43.0 & 12 Nov. & 210m & 490\\
         \hline

 \end{tabular}
 \caption{Details of ATCA radio observations. Columns 1 and 2 are the J2000 centres of each observation, column 3 gives the date 
of observation while columns 4 and 5 are the array configuration and integration time respectively.}
 \end{center}
\end{table}

  \begin{figure*}
   \vspace{-0.3cm}
   \begin{center}
    \includegraphics[width=15cm, angle=-90]{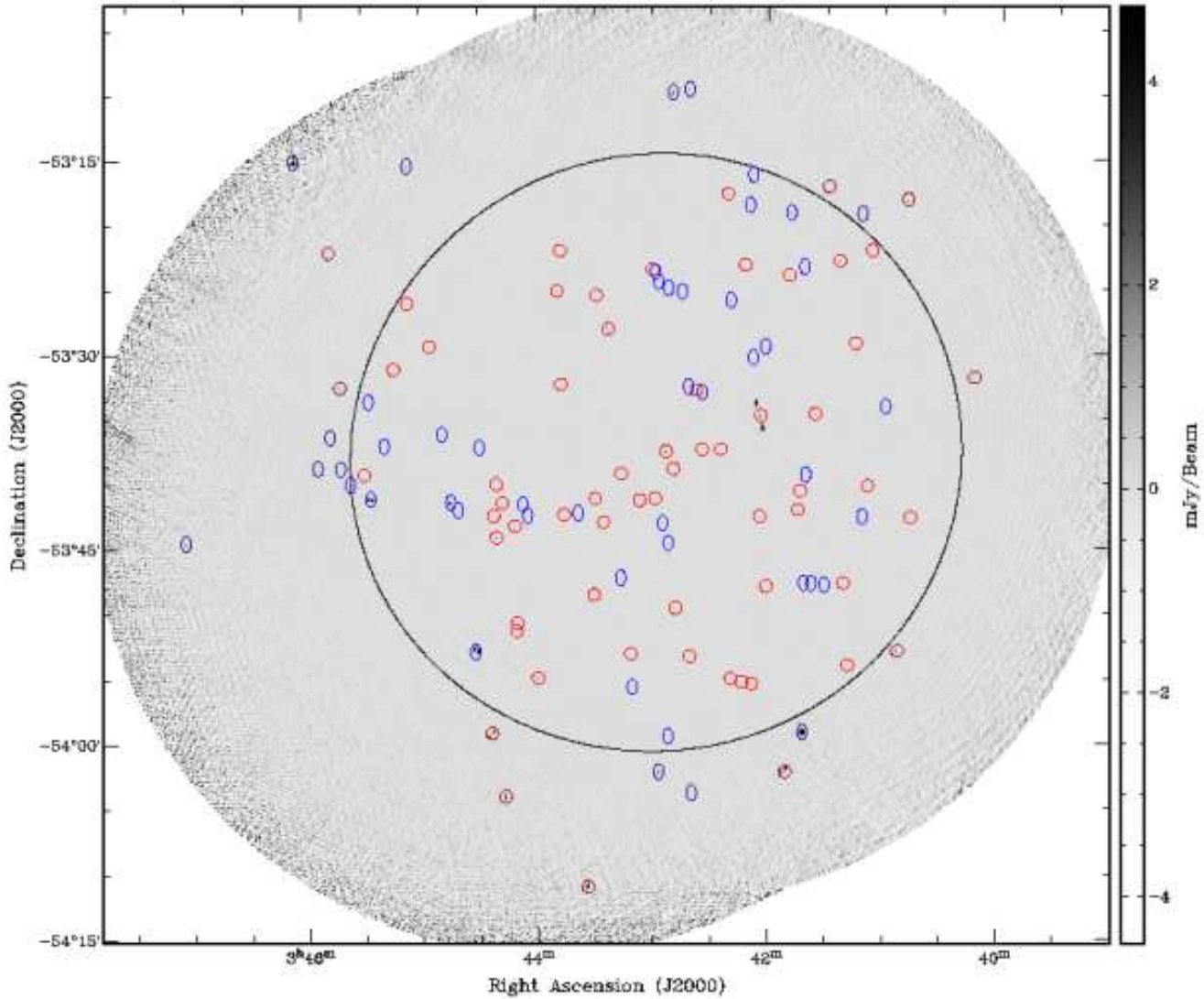}
   \caption{1.4 GHz total intensity image of A3158, the large circle denotes the area over which
the the primary beam attenuation is negligible {\bf ($\leq 1\%$)} and so an RLF may be accurately calculated. Red circles show the 
positions of radio galaxies which also have optical counterparts, while blue ellipses denote radio sources found in 
the image without optical counterparts.}
   \label{fig:1}
 \end{center}
 \end{figure*}

In all cases the ability of the ATCA to simultaneously observe two sections
of the available frequency band was utilised to observe 2 $\times$ 128 MHz bands 
each divided into 13 independent 8 MHz channels. As the channels at the edge of the 
band are removed this gives an effective bandwidth of 2 $\times$ 96 MHz for each 
frequency. The flux density scale was set using observations of PKS B1934-638 
which was assumed to have total fluxes of S$_{1.4}$= 14.9 and S$_{2.5}$= 11.6 Jy 
at 1.4 and 2.5 GHz respectively. The gain and bandpass calibration were 
performed relative to the unresolved calibrators PKS B0355-669 and PKS B0355-483 
which have flux densities of 1.01 and 0.20 Jy at 1.4 GHz and 0.77 and 0.41 Jy at 
2.5 GHz respectively. PKS B0355-669 was used in the initial observation and 
PKS B0355-483 was used in the archival observations by Haida Liang.  
 
Data were reduced in the MIRIAD software suite (Sault et al.~1995) 
using standard calibration routines. At both frequencies both total intensity (Stokes I) 
mosaic images were made using the MIRIAD task {\it LINMOS}. As part of the 
mosaicing process the inner part of the image is corrected for the primary beam 
response of the antennas. In addition, sensitivity and effective gain maps, which give an
estimate of the residual errors due to primary beam attenuation remaining on the edge 
of the field, are produced. 
 
The final 1.4 GHz image has a synthesised beam of 
%8.802$''$ $\times$ 4.638$''$ at position angle 2.396$^\circ$ 
8.8$''$ $\times$ 4.6$''$ at position angle 2.4$^\circ$ 
and has an RMS sensitivity of 0.03 mJy beam$^{-1}$
at the centre of the image. The final 2.5 GHz image has a synthesised beam of 
%4.984$''$ $\times$ 2.616$''$ at position angle 2.466$^\circ$ 
5.0$''$ $\times$ 2.6$''$ at position angle 2.5$^\circ$ 
and has an RMS sensitivity of 0.03 mJy beam$^{-1}$
at the centre of the image. However, because of the slight difference in pointing centres
between the first observation and the rest of the observations the sensitivity does not decrease
uniformly with radius across the resultant images. The 1.4 GHz image is shown in Figure 2. The
 area over which the primary beam attenuation, remaining in the centre of the 
mosaiced image, is negligible ($\leq 1\%$) is marked on the
image as a circle of radius $23\farcm05$ centred at 03 42 53.00 -53 37 43.0. This represents
the area for which the effective gain map is now 1 and is circular due to the unequal integration
times in the pointing centres with this region arising from 89\% of the integration time. Sources found
beyond this region correspond to areas of the effective gain map which are less than one, that is,
they will still have fluxes which are both lower than in reality and have greater uncertainty than
the middle of the field. These sources
had their flux densities corrected for the remaining primary beam attenuation using the 
standard analytical formula for the ATCA (Wieringa \& Kesteven 1992).

The uncertainty in the flux density measurement is estimated after Venturi et al.~(2000) as
a function of the noise in the map, $a$, and the residual calibration error, $b$, such that

\begin{equation}
\Delta S = \sqrt{a^{2} + (bS)^{2}}
\end{equation}

where $S$ is the source flux density and the residual calibration error is assumed to be of order
1$\%$.\\

\begin{table*}

   \begin{center}
\scriptsize{
 \begin{tabular}{l|l|l|l|c|c|c|c|c|c|l}
        \hline
  Radio            &Coordinates   &  S$_{1.4}$ & S$_{2.5}$ & $\alpha^{1.4}_{2.5}$ & Optical ID & &z& Ref. & Mag. &Name\\
  RA$_{\rm{J}2000}$     & DEC$_{\rm{J}2000}$&  mJy           & mJy   &        & RA$_{\rm{J}2000}$  & DEC$_{\rm{J}2000}$&&&&\\
         \hline 
03 40 45.93&	-53 42 52.4&	1.85 $\pm$ 0.04&&		       &03 40 45.86	&-53 42 52.7&	        &&22.63 &\\	
03 41 06.40&	-53 22 16.9&	0.44 $\pm$ 0.03&&		       &03 41 06.33	&-53 22 18.2&	        &&19.06 &\\	
03 41 08.34&	-53 40 28.6&	0.75 $\pm$ 0.03&&		       &03 41 08.44	&-53 40 28.3&	        &&20.08 & APMUKS(BJ) B033949.49-535002.2\\	
03 41 14.84&	-53 29 28.7&	0.64 $\pm$ 0.03&&		       &03 41 14.89	&-53 29 29.4&	        &&19.09 & MSACSS J034114.87-532929.5\\	
03 41 18.40&	-53 54 17.7&	0.83 $\pm$ 0.03&&		       &03 41 18.45	&-53 54 17.1&	        &&19.30 &\\	
03 41 20.89&	-53 47 56.9&	0.89 $\pm$ 0.03&&		       &03 41 20.98	&-53 47 57.0&	        &&21.17 &\\
03 41 23.13&	-53 23 06.4&	4.72 $\pm$ 0.06&&		       &03 41 23.28	&-53 23 06.3&	        &&22.24 &\\
03 41 35.52&	-53 34 56.5&	0.51 $\pm$ 0.03&0.74$\pm$ 0.03&	0.69   &03 41 35.50	&-53 34 55.7&	        &&21.33 &\\	
03 41 43.44&	-53 40 53.0&	0.88 $\pm$ 0.03&&		       &03 41 43.46	&-53 40 53.1&	        &&17.26 &\\	
03 41 44.52&	-53 42 21.0&	3.42 $\pm$ 0.05&1.76$\pm$ 0.03&	-1.23  &03 41 44.46	&-53 42 21.0&	0.066523 $\pm$ 0.000147 & 1 & 16.78 &2MASX J03414449-5342216\\
03 41 49.04&    -53 24 14.2&	1.94 $\pm$ 0.04&&		       &03 41 49.20	&-53 24 13.2&	0.055492 $\pm$ 0.000133 & 1 & 17.59 &APMUKS(BJ) B034029.62-533343.5\\	
03 42 01.07&	-53 48 15.2&	1.09 $\pm$ 0.03&0.67$\pm$ 0.03&	-0.90  &03 42 01.25	&-53 48 16.0&	        && 20.22 & APMUKS(BJ) B034042.78-535747.3\\	
03 42 03.81&	-53 35 06.0&	167.73$\pm$1.68&87.90$\pm$ 0.88& -1.20 &03 42 03.81	&-53 35 05.8&	        &&22.47 &\\
03 42 04.34&	-53 42 51.9&	1.81 $\pm$ 0.04&&		       &03 42 04.36	&-53 42 50.5&	0.066513 $\pm$ 0.000133 & 1 & 17.20 &2MASX J03420435-5342503\\	
03 42 08.39&	-53 55 44.1&	0.81 $\pm$ 0.03&&		       &03 42 08.38	&-53 55 44.7&	0.073641 $\pm$ 0.000290 & 1 & 18.04 &2MASX J03420850-5355443\\	
03 42 11.86&	-53 23 27.0&	0.73 $\pm$ 0.03&&		       &03 42 11.84	&-53 23 27.0&	0.062536 $\pm$ 0.000100 & 2 & 15.77 &2MASX J03421179-5323273\\		
03 42 13.60&	-53 55 36.3&	0.36 $\pm$ 0.03&&		       &03 42 13.44	&-53 55 37.1&	        &&22.69 &\\		
03 42 19.26&	-53 55 22.6&	0.96 $\pm$ 0.03&&		       &03 42 19.45	&-53 55 21.8&	        &&20.37 &\\		
03 42 20.87&	-53 17 58.1&	3.02 $\pm$ 0.04&&		       &03 42 20.99	&-53 17 58.2&	0.062003 $\pm$ 0.000190 & 1 &16.52  &2MASX J03422083-531757\\		
03 42 24.67&	-53 37 42.7&	0.92 $\pm$ 0.03&0.83$\pm$ 0.03&	-0.19  &03 42 24.65	&-53 37 42.3&	        &&20.16 & APMUKS(BJ) B034105.75-534709.6\\		
03 42 34.34&	-53 37 44.9&	0.67 $\pm$ 0.03&0.38$\pm$ 0.03&	-1.05  &03 42 34.39	&-53 37 46.2&	0.056809 $\pm$ 0.000157 & 3 & 16.86 &2MASX J03423440-533745\\		

03 42 37.29&	-53 33 08.8&	0.61 $\pm$ 0.03&0.50$\pm$ 0.03&	-0.37  &03 42 37.24	&-53 33 09.0&	        &&21.94 &\\	%changed from 3.08mJy, split components.

03 42 40.78&	-53 53 39.3&	0.84 $\pm$ 0.03&&		       &03 42 40.75	&-53 53 39.3&	        &&21.40 &\\		
03 42 48.03&	-53 49 56.0&	0.48 $\pm$ 0.03&0.65$\pm$ 0.03&	0.56   &03 42 48.08	&-53 49 55.7&	        &&15.68 &\\		
03 42 49.18&	-53 39 13.0&	0.24 $\pm$ 0.03&0.13$\pm$ 0.03&	-1.14  &03 42 49.29	&-53 39 12.2&	        &&22.16 &\\		
03 42 53.00&	-53 37 53.0&	7.88 $\pm$ 0.08&5.12$\pm$ 0.06 &-0.80  &03 42 52.98	&-53 37 52.7&	0.057653 $\pm$ 0.000180 & 1 & 15.31 &ESO 156- G 008 NED01\\
03 42 58.63&	-53 41 30.9&	1.19 $\pm$ 0.03&1.23$\pm$ 0.03&	0.06   &03 42 58.56	&-53 41 30.9&	0.121077 $\pm$ 0.000197 & 3 & 18.07 &2MASX J03425857-5341306\\		
03 42 59.93&	-53 23 49.7&	0.75 $\pm$ 0.03&&		       &03 42 59.95	&-53 23 51.3&	        &&18.45 &\\		
03 43 06.74&	-53 41 38.8&	1.49 $\pm$ 0.03&1.02$\pm$ 0.03&	-0.70  &03 43 06.72	&-53 41 39.2&	0.120003 $\pm$ 0.000210 & 1 & 17.61 &2MASX J03430668-5341396\\		
03 43 11.33&	-53 53 29.3&	0.36 $\pm$ 0.03&&		       &03 43 11.39	&-53 53 28.9&	        &&17.44 &\\		
03 43 16.18&	-53 39 36.8&	1.27 $\pm$ 0.03&0.87$\pm$ 0.03&	-0.70  &03 43 16.12	&-53 39 35.9&	0.056530 $\pm$ 0.000160 & 4 & 16.79 &2MASX J03431618-5339365\\		
03 43 22.80&	-53 28 26.1&	11.84 $\pm$ 0.12&5.90$\pm$ 0.07& -1.29 &03 43 22.72	&-53 28 28.6&	        &&20.98 &\\			
03 43 25.45&	-53 43 20.4&	0.41 $\pm$ 0.03&0.18$\pm$ 0.03&	-1.53  &03 43 25.47	&-53 43 19.1&	        &&17.93 & APMUKS(BJ) B034207.10-535245.7\\	
03 43 29.03&	-53 25 49.6&	0.18 $\pm$ 0.03&&		       &03 43 29.01	&-53 25 50.1&	        &&21.13 &\\						
03 43 29.70&	-53 41 30.4&	1.61 $\pm$ 0.03&0.77$\pm$ 0.03&	-1.37  &03 43 29.70	&-53 41 31.4&	0.062430 $\pm$ 0.000160 & 4 & 15.15 &2MASX J03432968-5341316\\	
03 43 30.49&    -53 48 55.6&    0.35 $\pm$ 0.03&&                      &03 43 30.48     &-53 48 54.9&          &&19.30 & APMUKS(BJ) B034212.50-535820.4\\% %extra%
03 43 45.94&	-53 42 45.7&	1.03 $\pm$ 0.03&&		       &03 43 45.94	&-53 42 46.2&	0.059691 $\pm$ 0.000310 & 1 & 17.16 &2MASX J03434600-5342466\\	
03 43 47.53&	-53 32 41.6&	0.70 $\pm$ 0.03&0.61$\pm$ 0.03&	-0.26  &03 43 47.46	&-53 32 40.0&	        &&20.86 &\\						
03 43 47.75&	-53 22 22.3&	0.24 $\pm$ 0.03&&		       &03 43 47.53	&-53 22 23.4&	        &&19.46 &\\						
03 43 49.39&	-53 25 28.8&	3.45 $\pm$ 0.05&2.00$\pm$ 0.04&	-1.01  &03 43 49.53	&-53 25 28.3&	        &&15.56 &\\		
03 43 59.57&	-53 55 19.3&	4.12 $\pm$ 0.05&&		       &03 43 59.58	&-53 55 18.2&	0.079431 $\pm$ 0.000290 & 1 & 15.93 &2MASX J03435962-5355181\\
03 44 10.30&	-53 51 05.5&	0.36 $\pm$ 0.03&&		       &03 44 10.31	&-53 51 06.0&	        &&20.24 & APMUKS(BJ) B034252.48-540029.1\\		
03 44 10.78&	-53 51 41.3&	0.65 $\pm$ 0.03&&		       &03 44 10.77	&-53 51 41.3&	        &&21.20 &\\						
03 44 11.65&	-53 43 36.3&	1.51 $\pm$ 0.03&1.25$\pm$ 0.03&	-0.35  &03 44 11.72	&-53 43 34.9&	        &&21.89 &\\						
03 44 18.13&	-53 41 52.0&	1.43 $\pm$ 0.03&1.70$\pm$ 0.03&	0.32   &03 44 18.06	&-53 41 51.7&	        &&20.31 &\\						
03 44 21.15&	-53 44 29.6&	2.82 $\pm$ 0.04&1.21$\pm$ 0.03&	-1.57  &03 44 21.10	&-53 44 30.6&	0.160084 $\pm$ 0.000240 & 3 & 17.96 &2MASX J03442114-5344299\\		
03 44 21.23&	-53 40 23.8&	0.15 $\pm$ 0.03&&		       &03 44 21.40	&-53 40 23.8&	        &&21.31 &\\
03 44 22.67&    -53 42 49.6&    0.66 $\pm$ 0.03&&                      &03 44 22.55     &-53 42 47.2&  0.059551 $\pm$ 0.000133 & 1 & 17.83 &APMUKS(BJ) B034304.39-535208.5\\%extra2%
03 44 55.60&    -53 29 44.5&    0.71 $\pm$ 0.03&&                      &03 44 55.45     &-53 29 43.9&          &&21.78 &\\%extra2
03 45 07.08&	-53 26 22.9&	0.85 $\pm$ 0.03&&		       &03 45 07.19	&-53 26 22.0&	        &&21.47 &\\						
03 45 14.30&	-53 31 28.7&	1.27 $\pm$ 0.03&&		       &03 45 14.29	&-53 31 26.7&	        &&22.29 &\\				
03 45 29.61&	-53 39 37.1&	1.04 $\pm$ 0.03&&		       &03 45 29.45	&-53 39 37.4&	0.062323 $\pm$ 0.000170 & 1 & 16.18 &2MASX J03452956-5339370\\			
        \hline	
 \end{tabular}
}
 \caption{Details of detected radio sources with optical counterparts detectable in the DSS.Columns 1 and 2 give the J2000 coordinates 
of the radio source; column 3 is the total flux at 1.4 GHz; column 4 gives the total flux at 2.5 GHz; columns 5 and 6 give the position of
the corresponding optical source from the DSS; column 7 gives the redshift of the optical source; column 8 is the
reference for the redshift where 1 is Katgert et al.~(1998), 2 is Loveday et al.~(1996), 3 is Smith et al.~(2004), 4 is De Vaucouleurs
et al.~(1991) and 5 is Lucey et al.~(1983); column 9 gives the blue magnitude; finally column 9 gives the optical source name as returned 
by the NASA Extragalactic Database.}
 \end{center}
 \label{tab:a}
\end{table*}

\begin{figure*}
%\vspace{-0.2cm}
\centering
\resizebox{\hsize}{!}{\includegraphics[angle=-90]{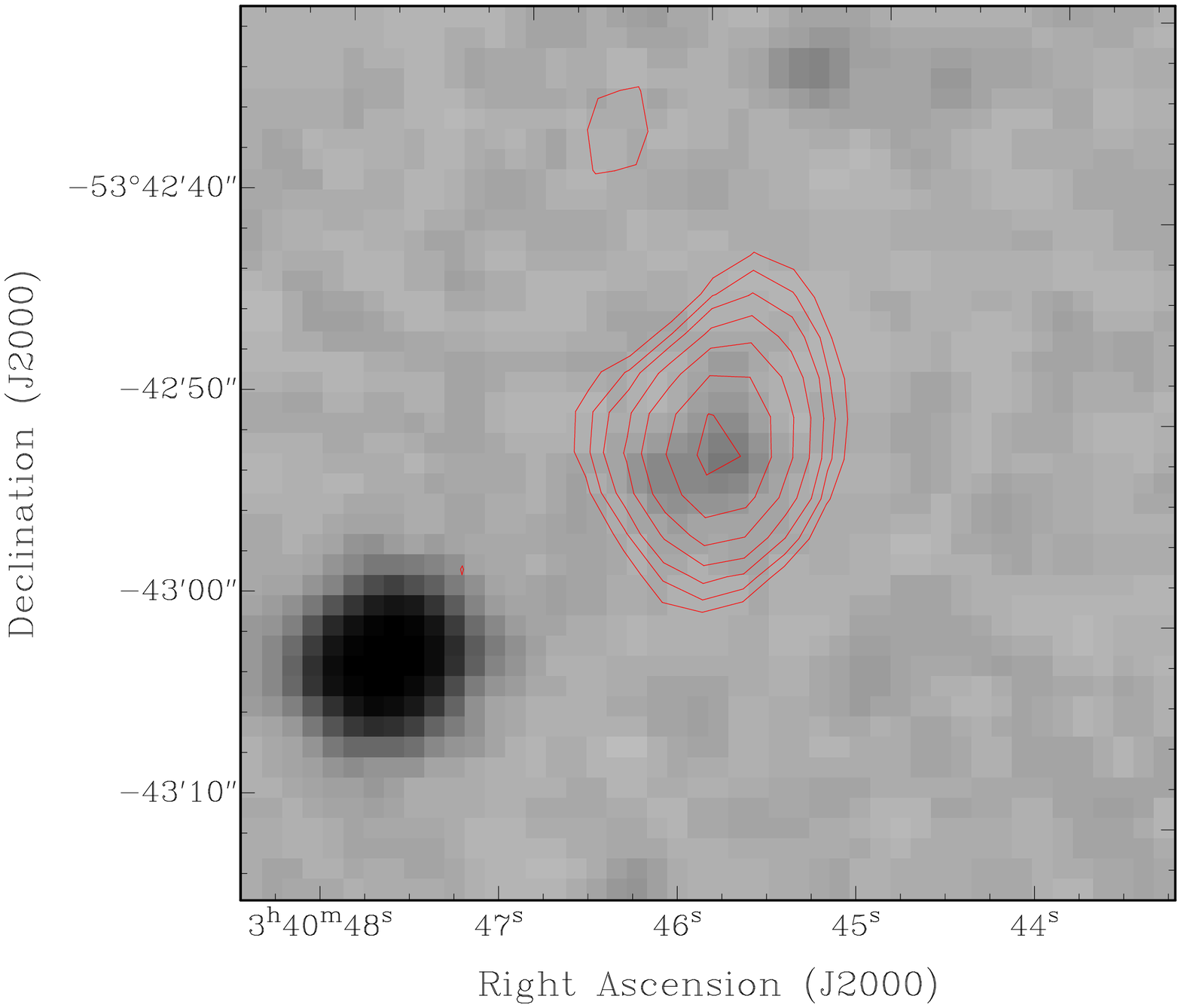}\includegraphics[angle=-90]{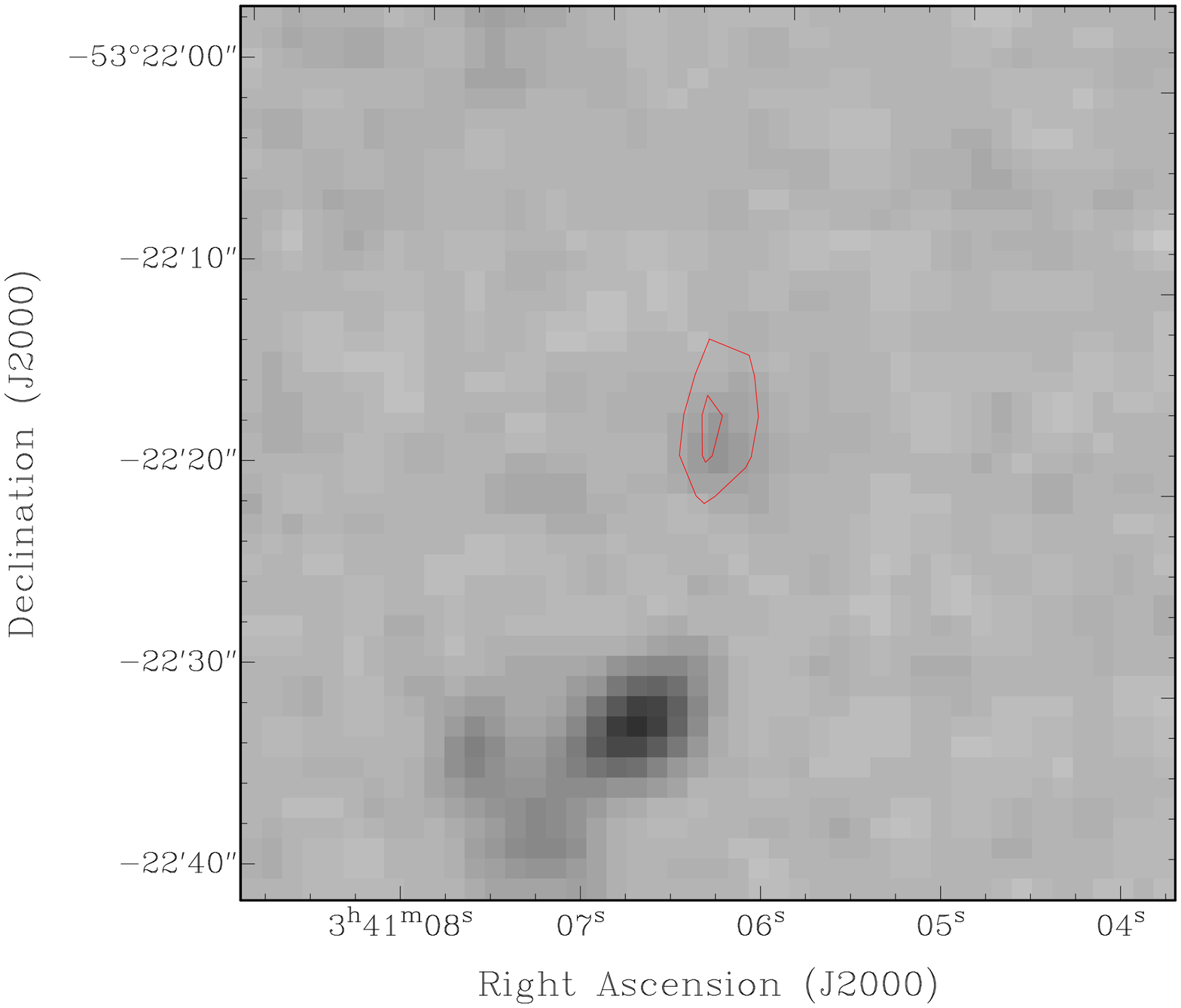}\includegraphics[angle=-90]{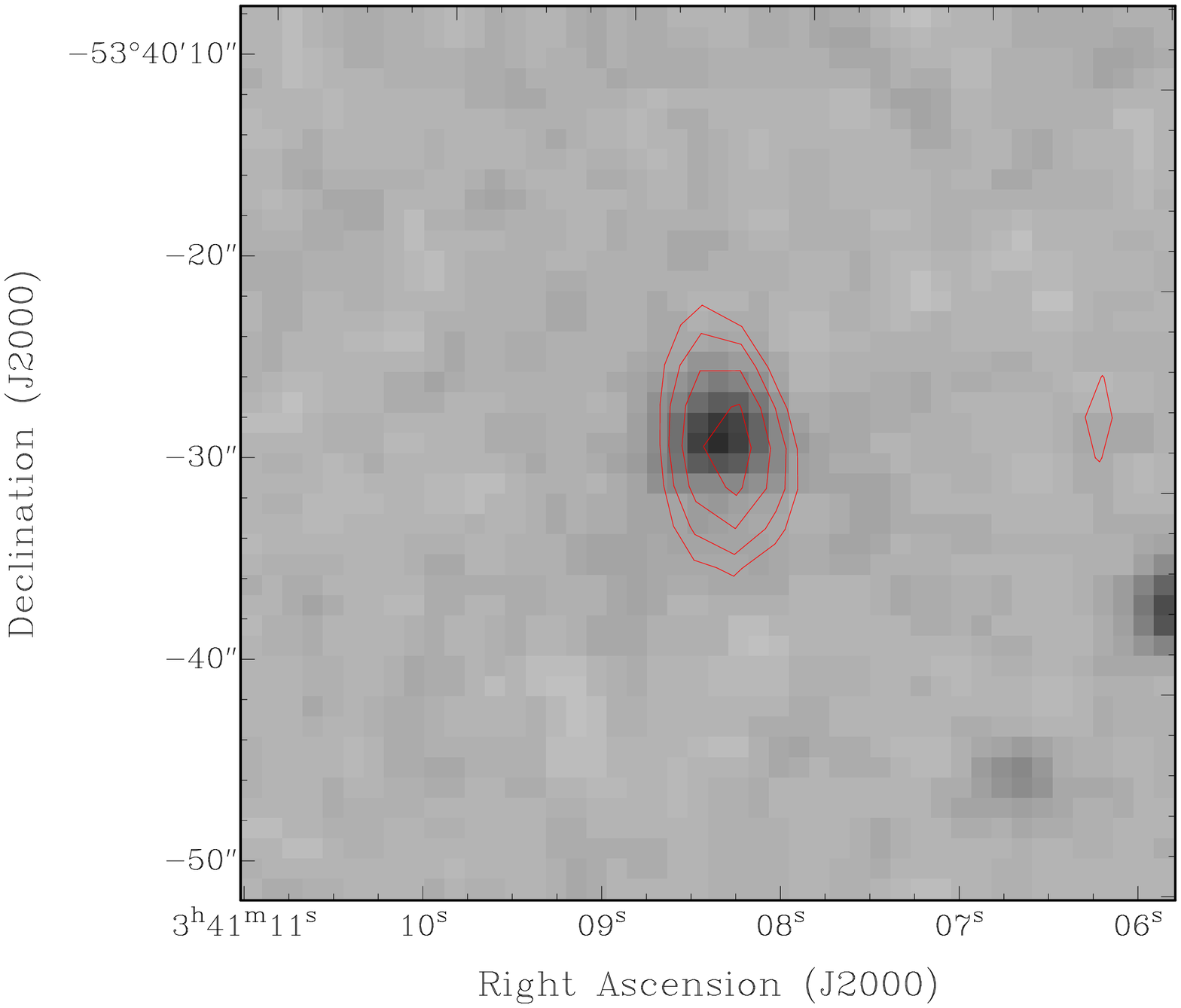}\includegraphics[angle=-90]{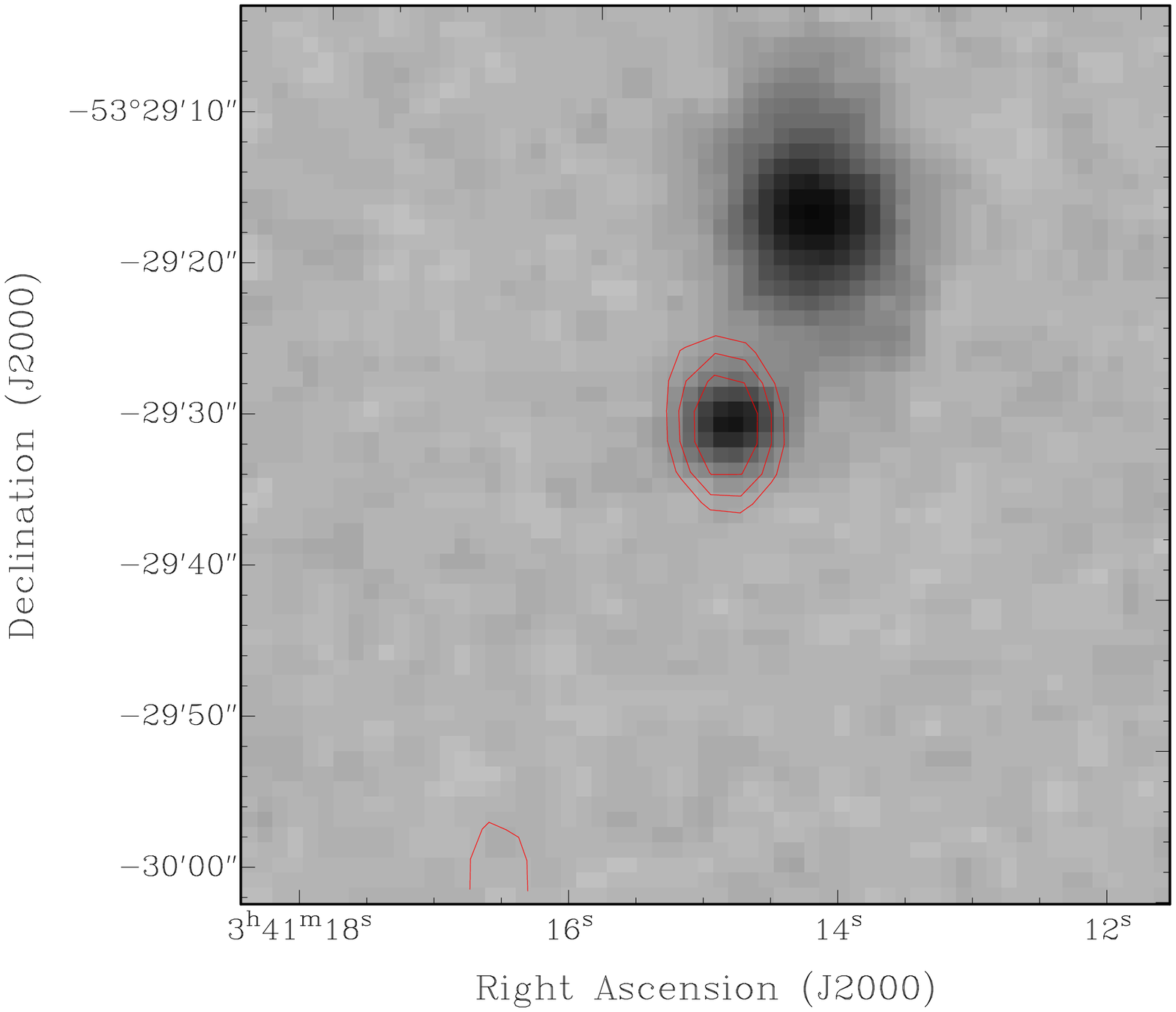}\includegraphics[angle=-90]{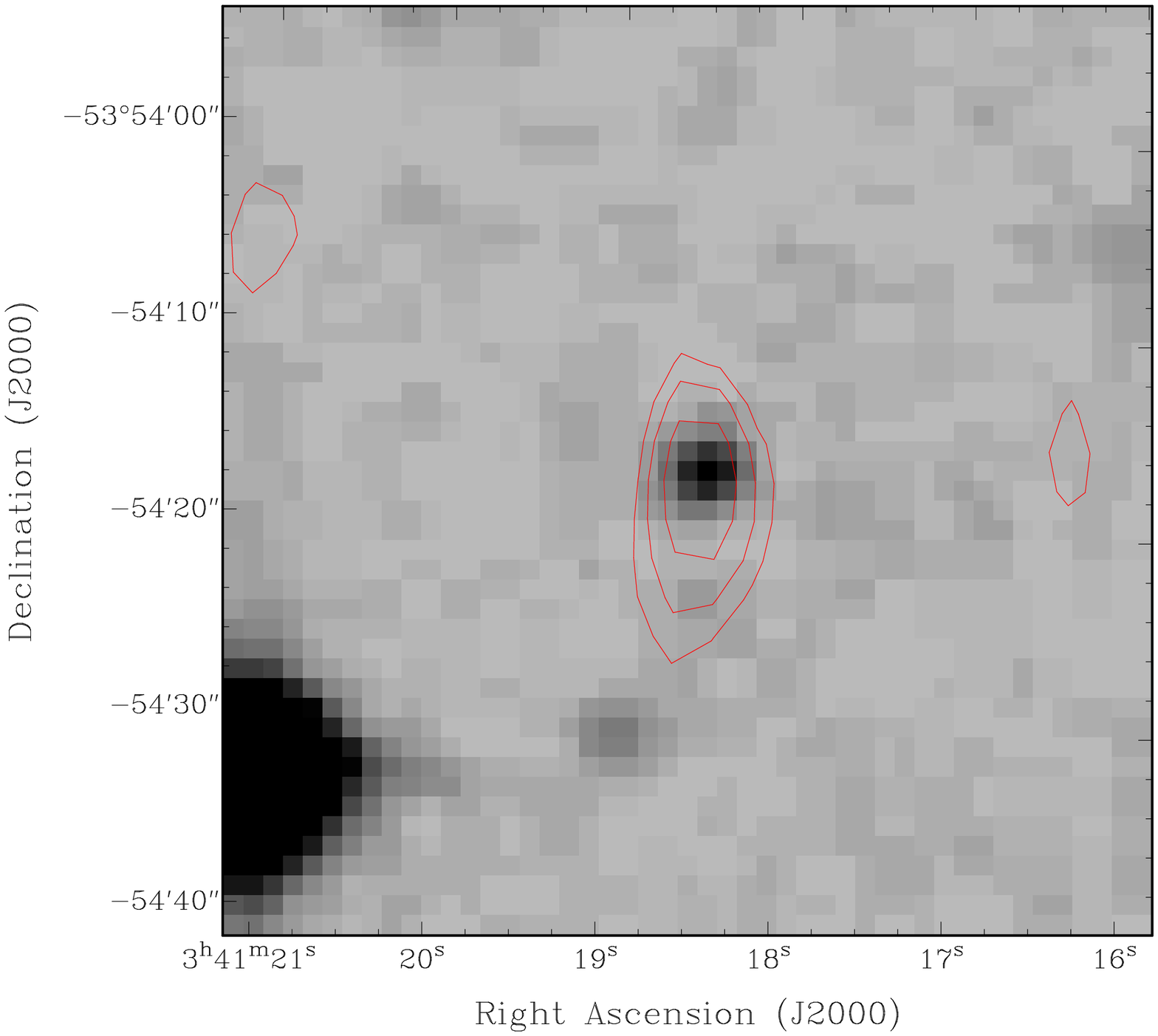}\includegraphics[angle=-90]{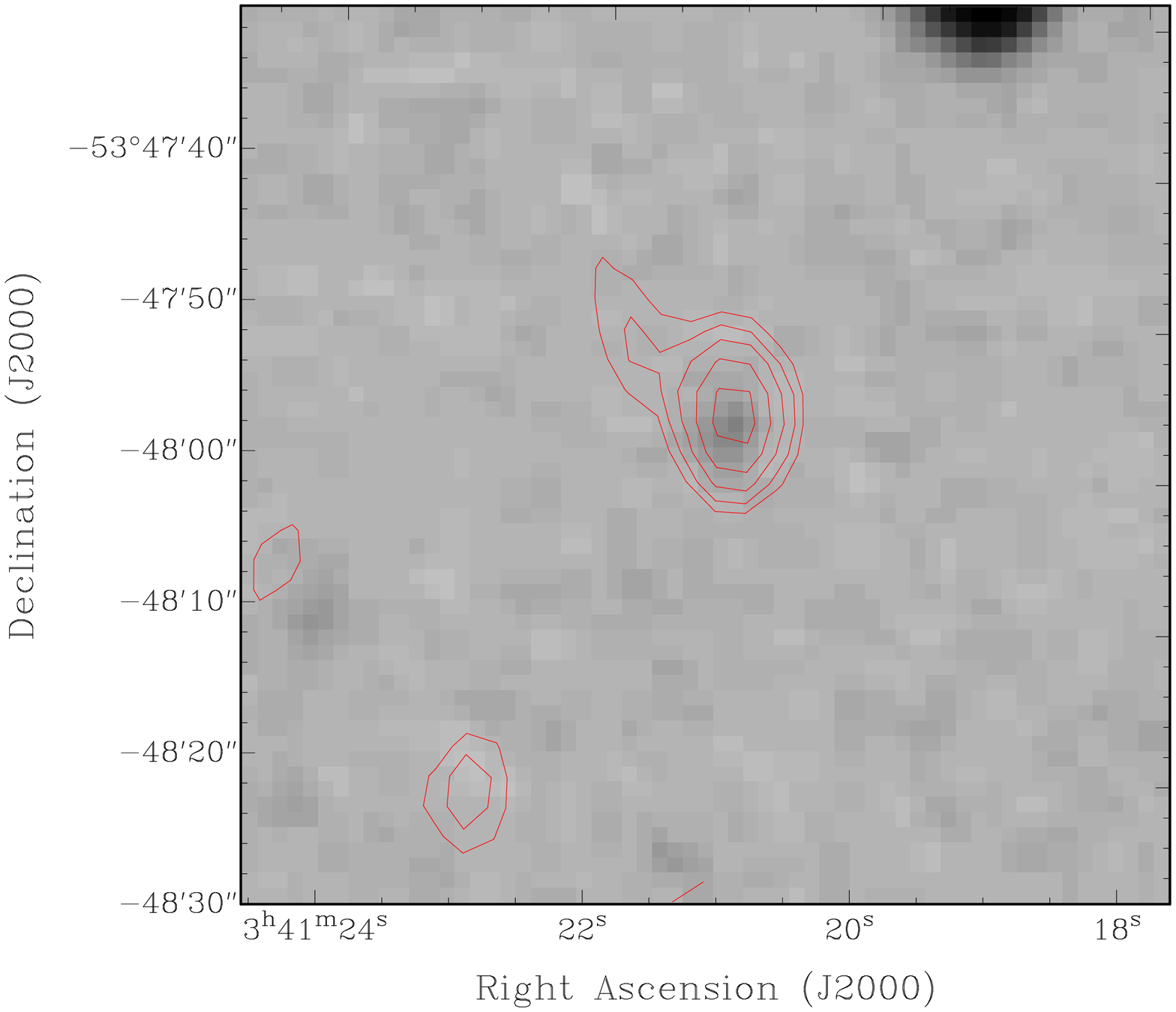}}
\centering
\resizebox{\hsize}{!}{\includegraphics[angle=-90]{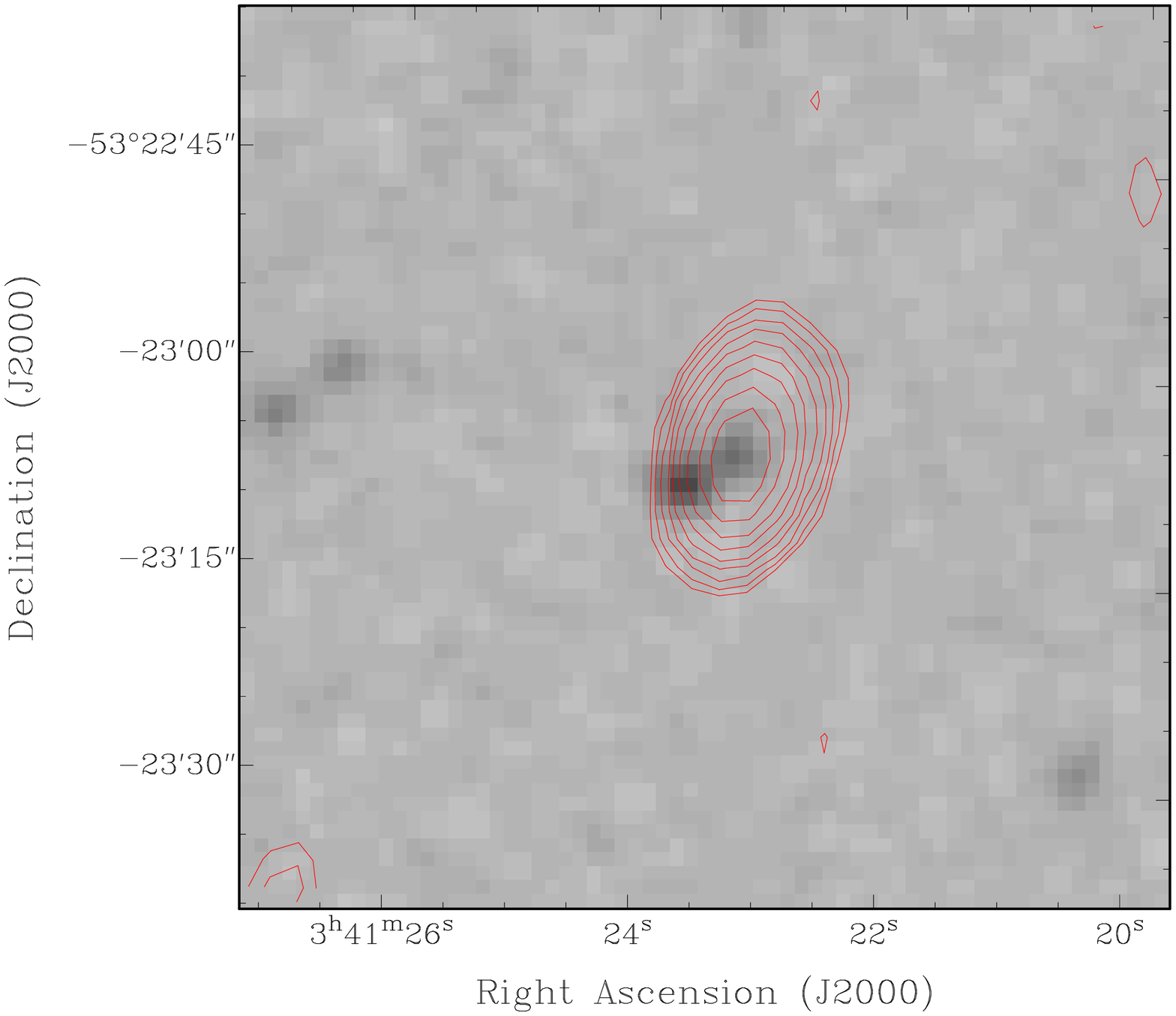}\includegraphics[angle=-90]{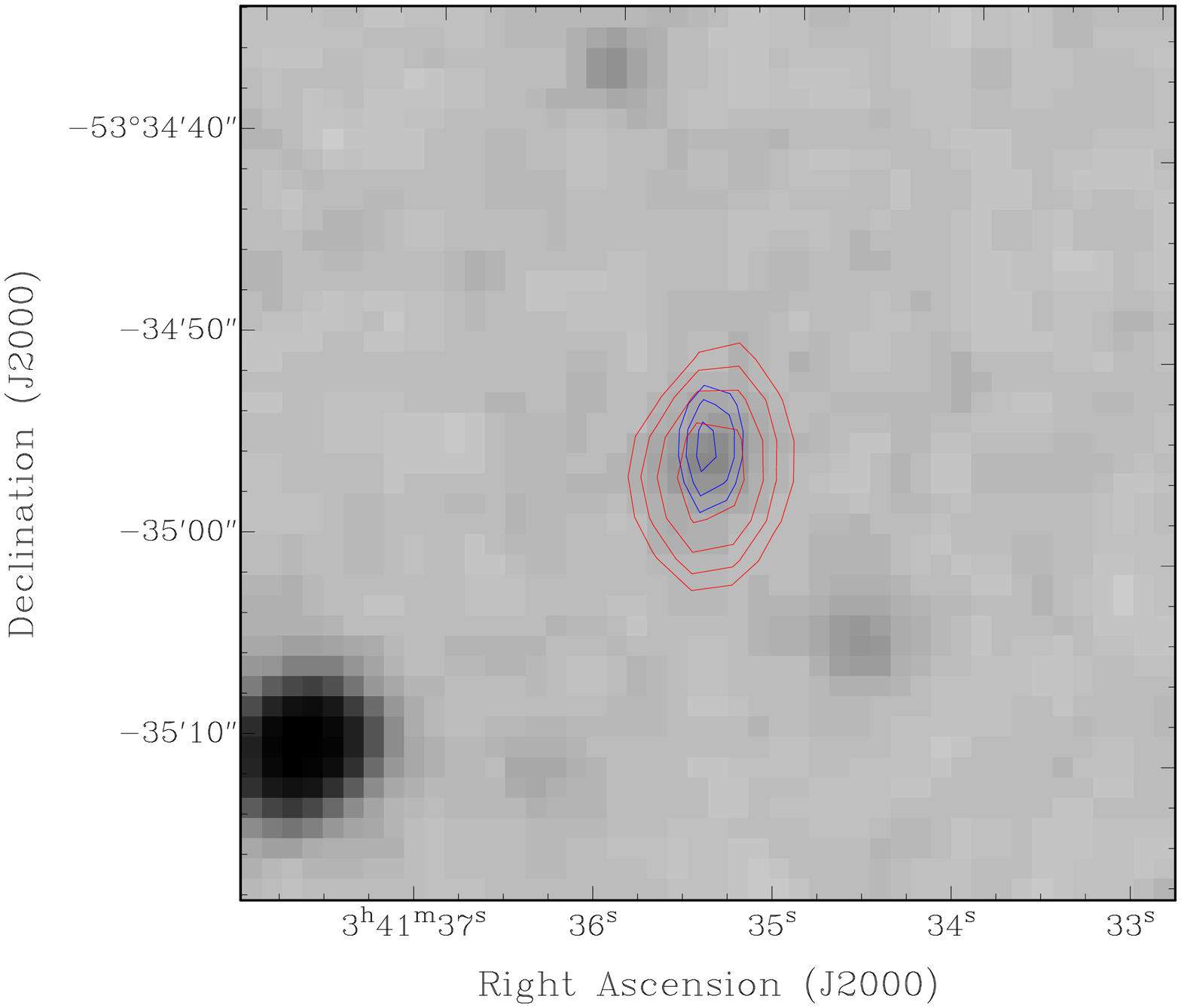}\includegraphics[angle=-90]{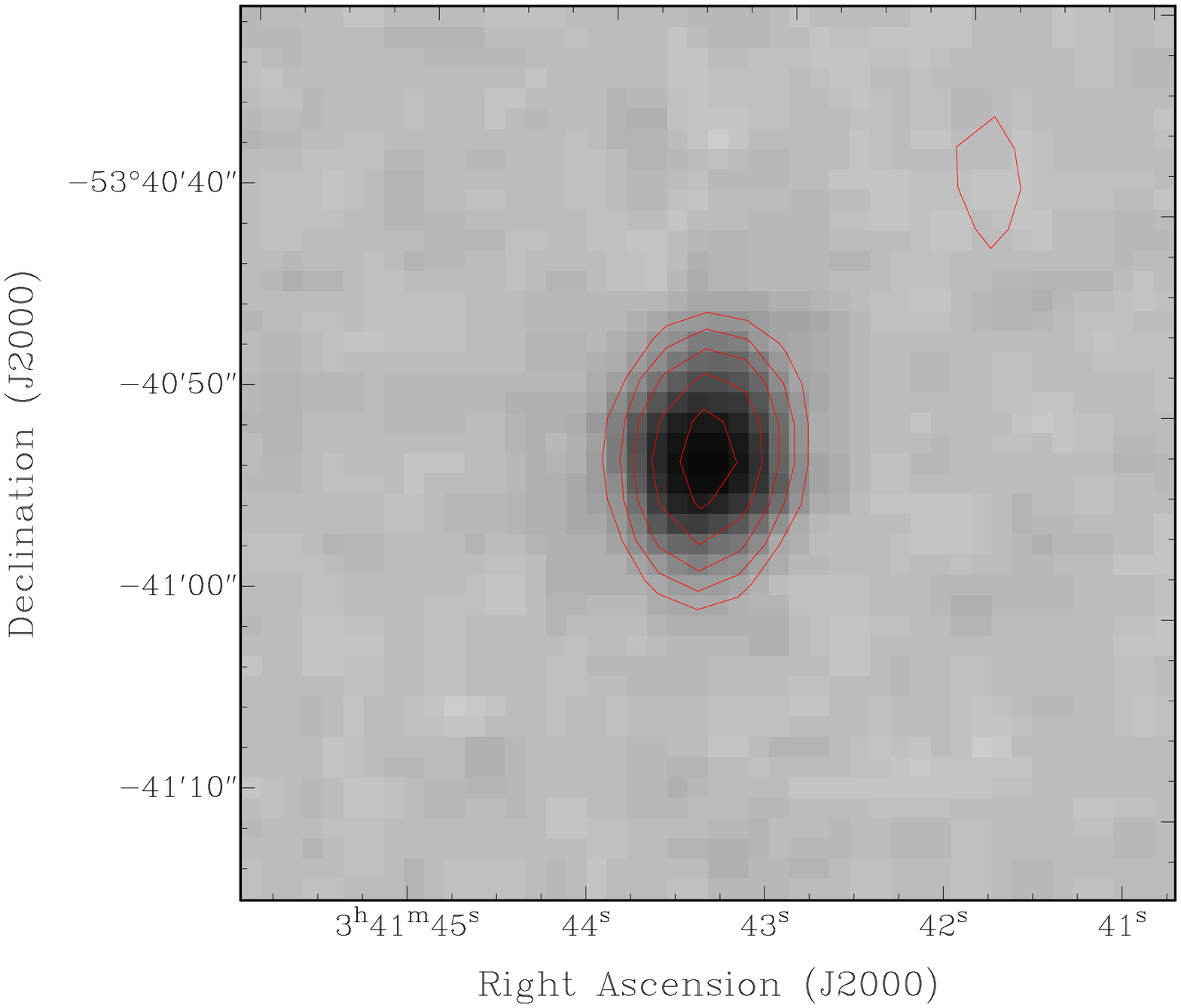}\includegraphics[angle=-90]{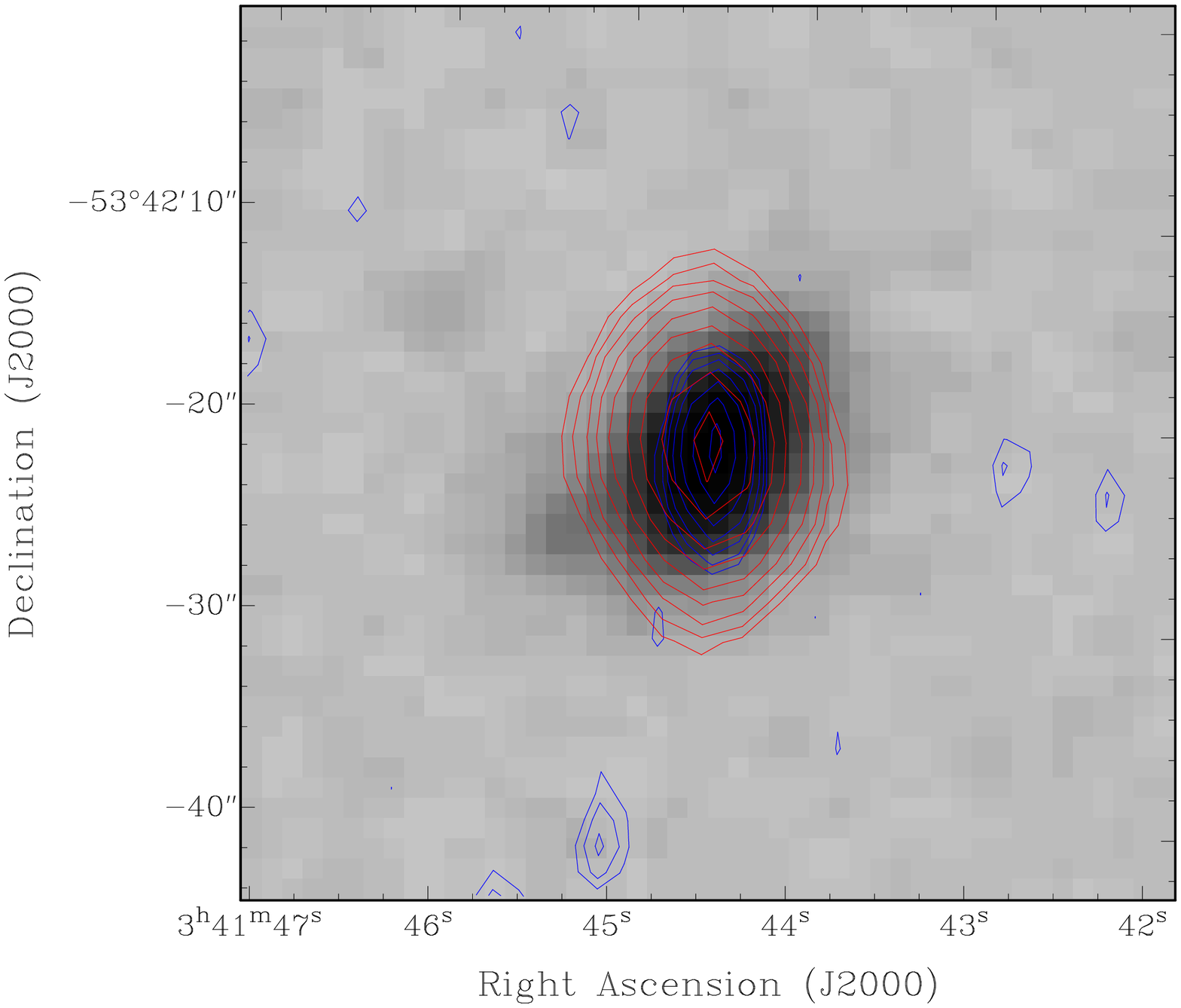}\includegraphics[angle=-90]{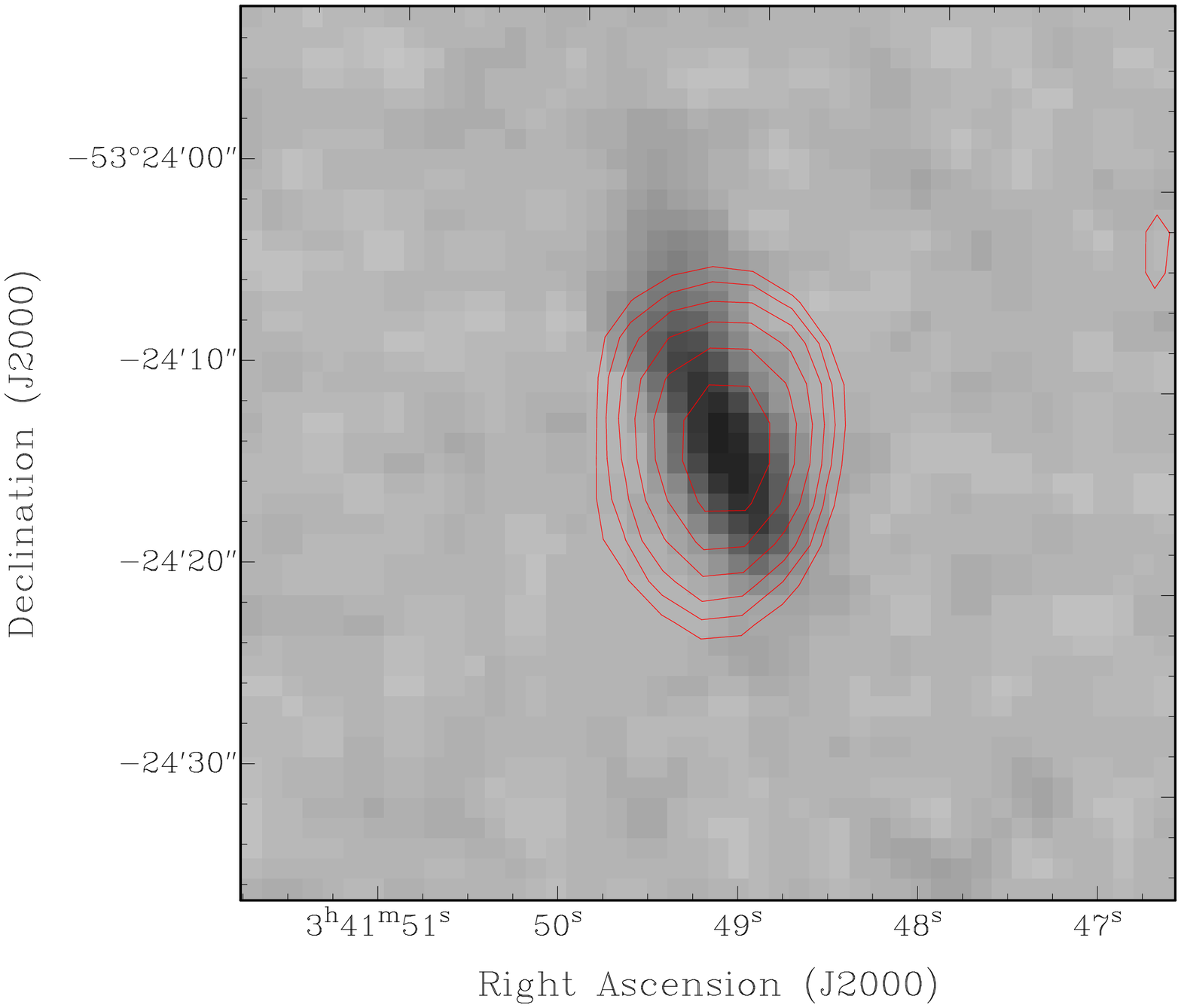}\includegraphics[angle=-90]{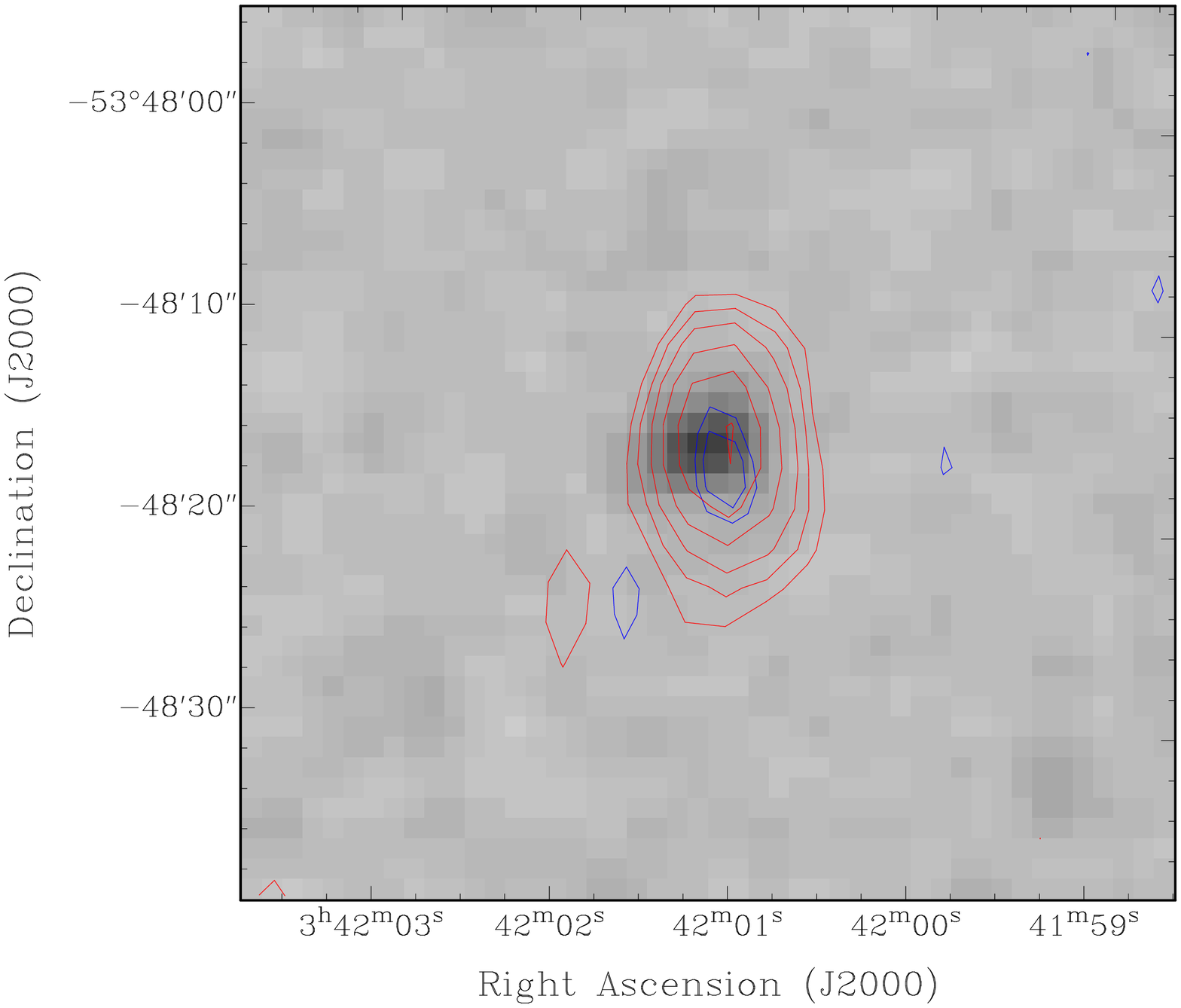}}
\resizebox{\hsize}{!}{\includegraphics[angle=-90]{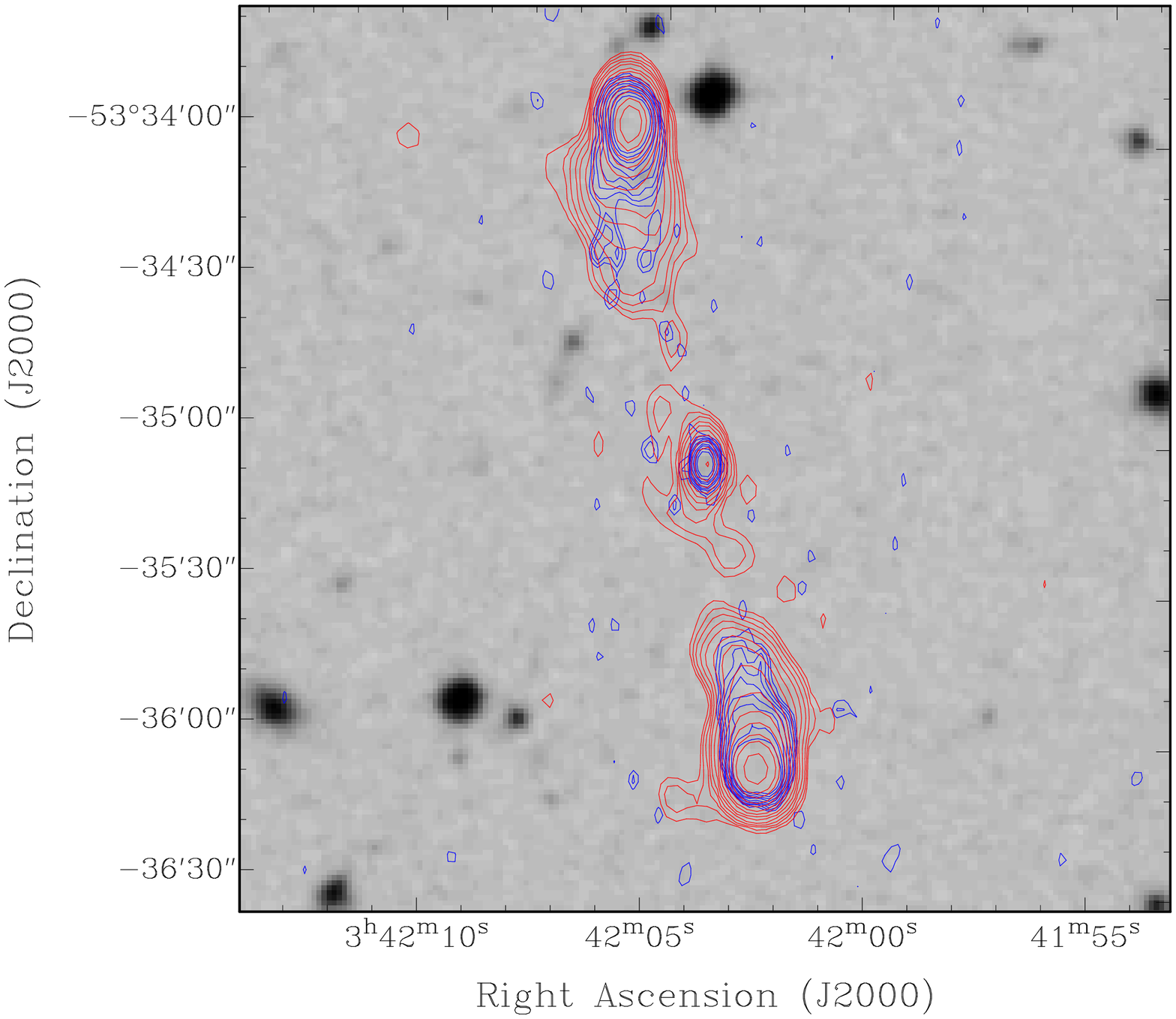}\includegraphics[angle=-90]{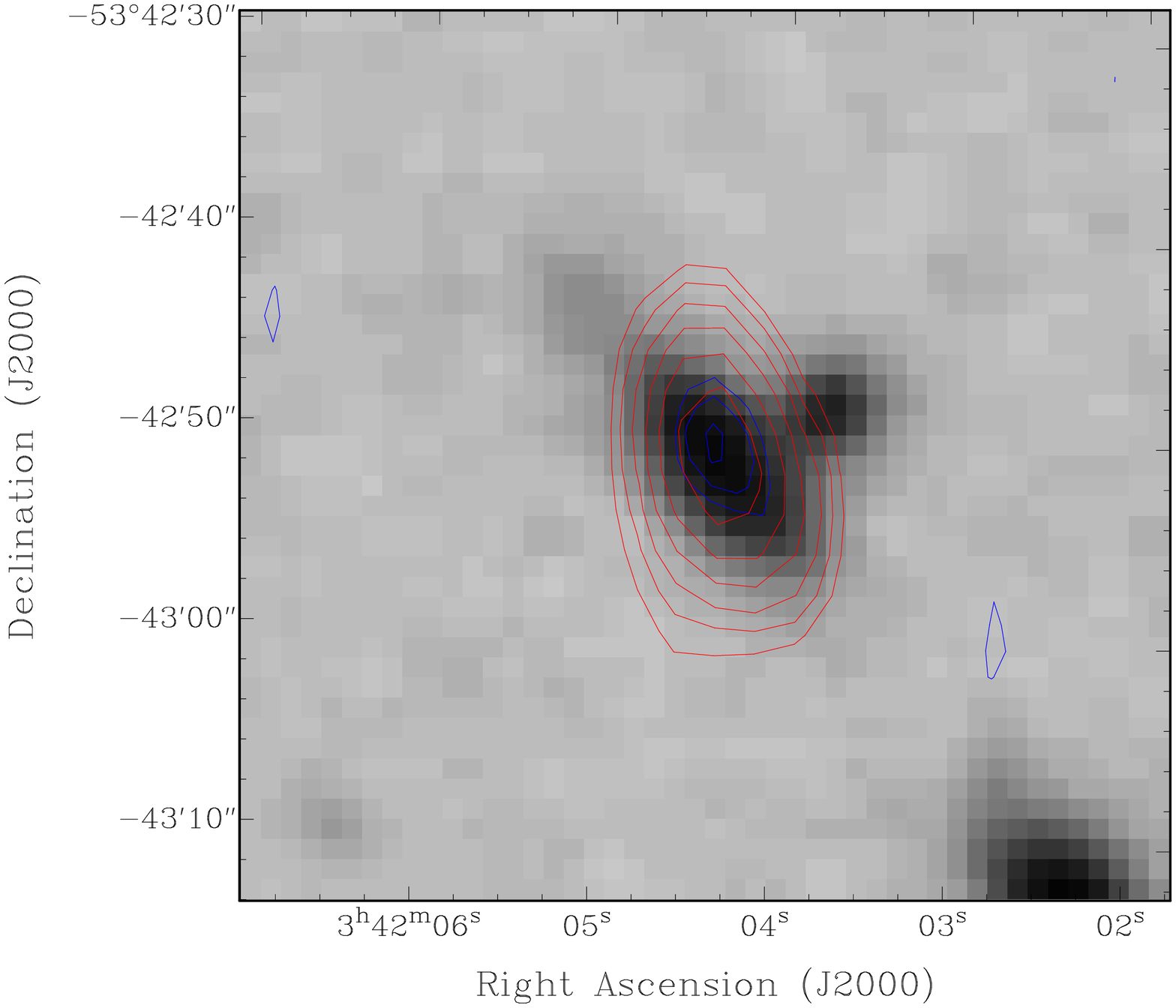}\includegraphics[angle=-90]{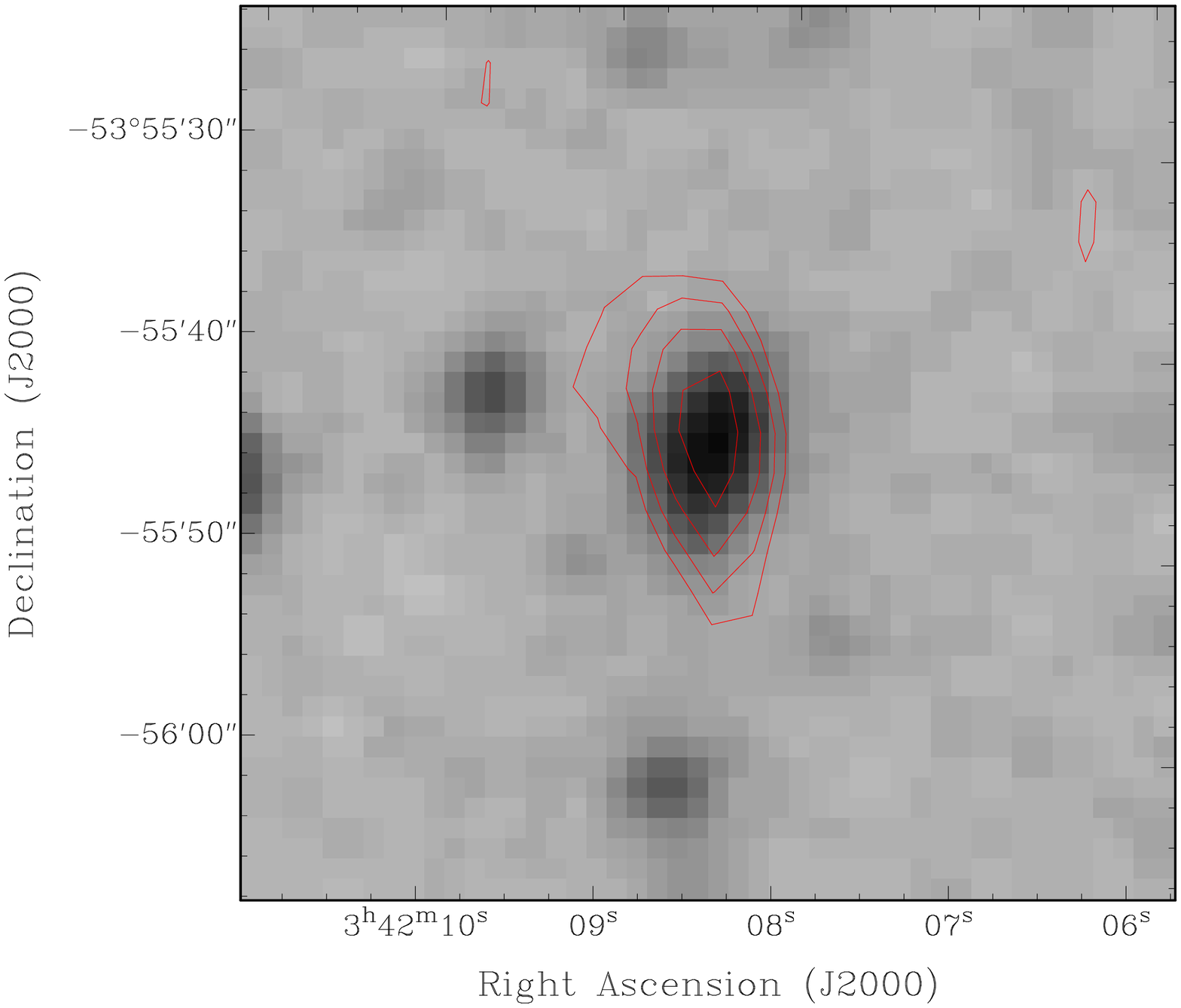}\includegraphics[angle=-90]{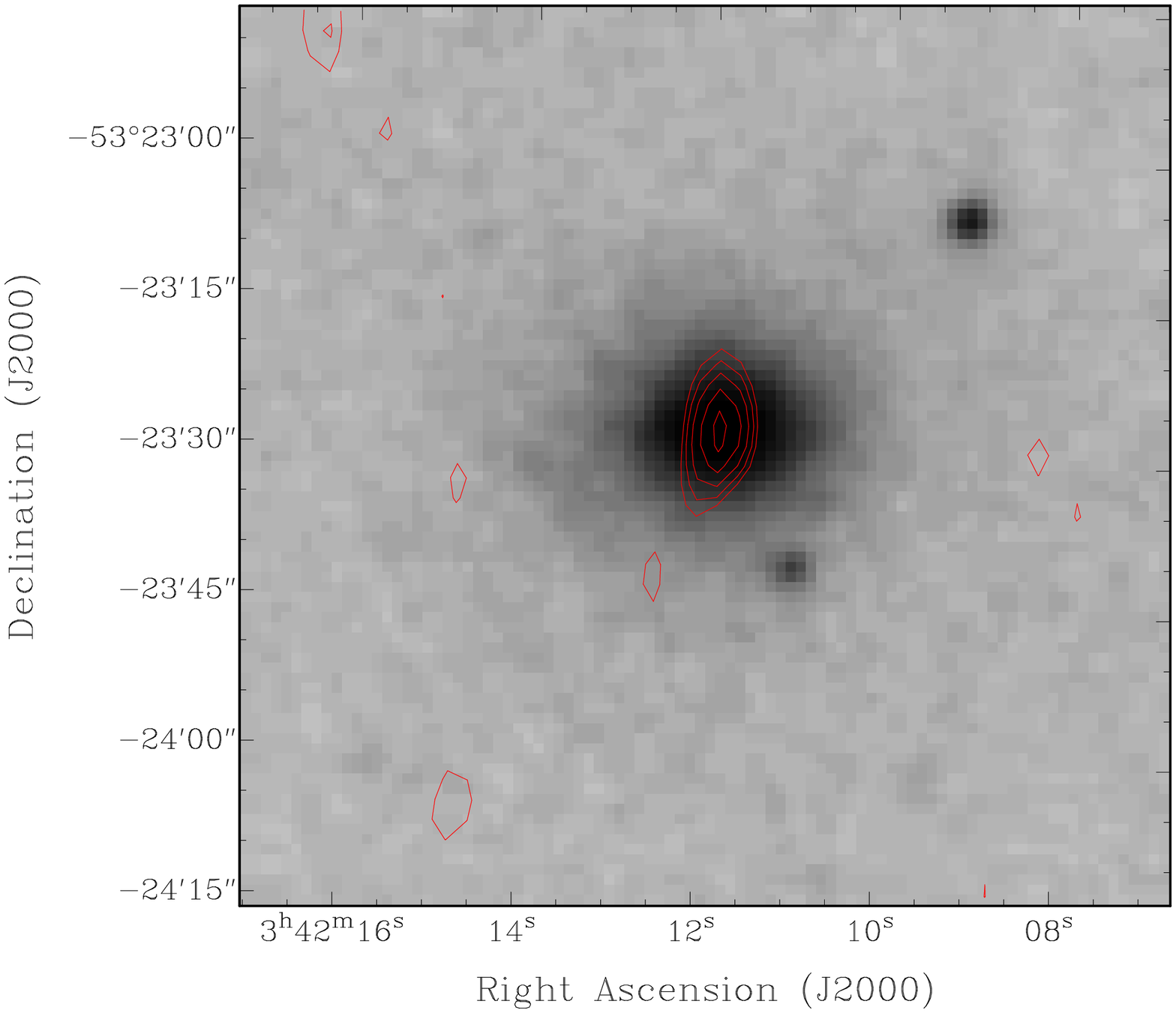}\includegraphics[angle=-90]{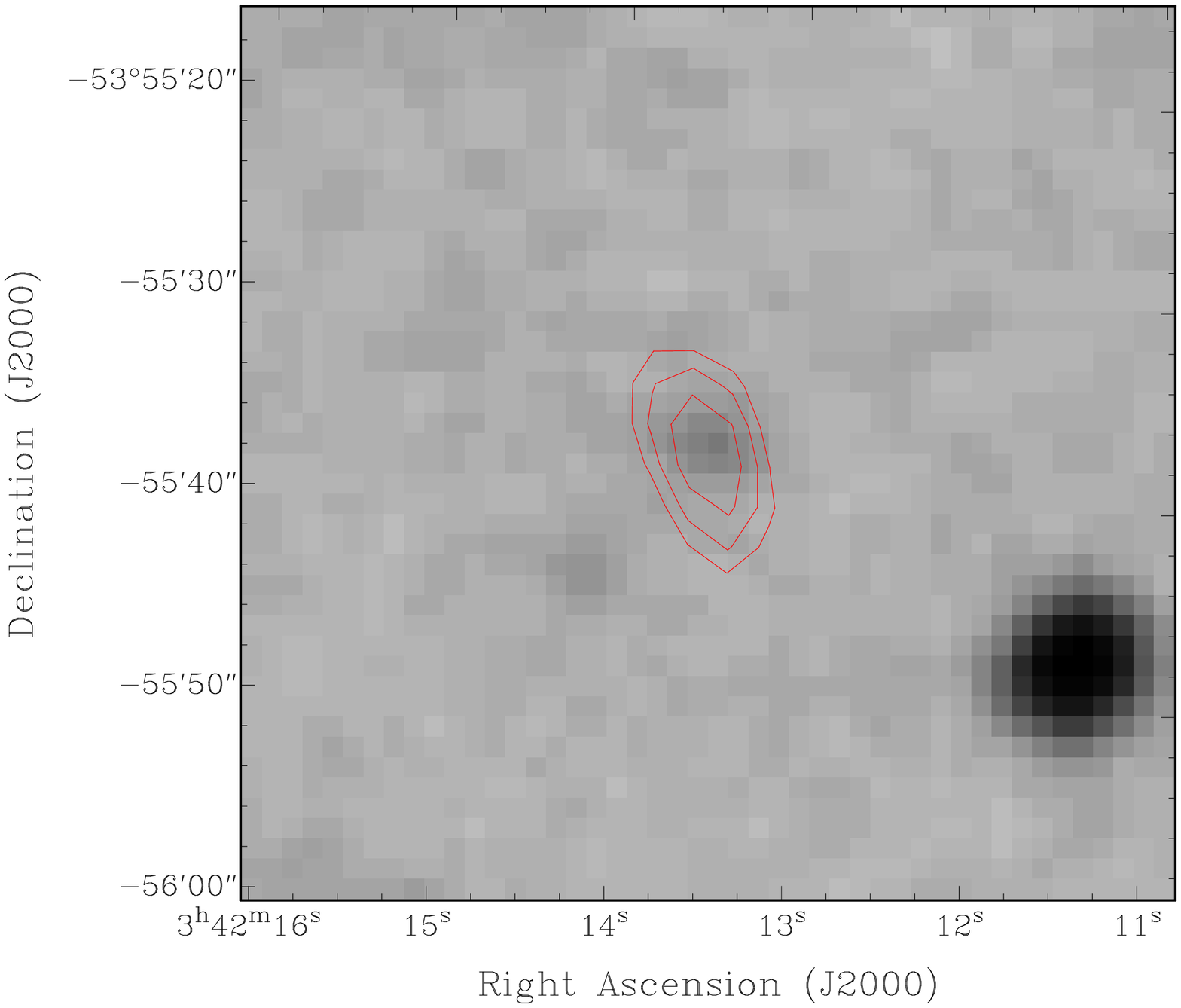}\includegraphics[angle=-90]{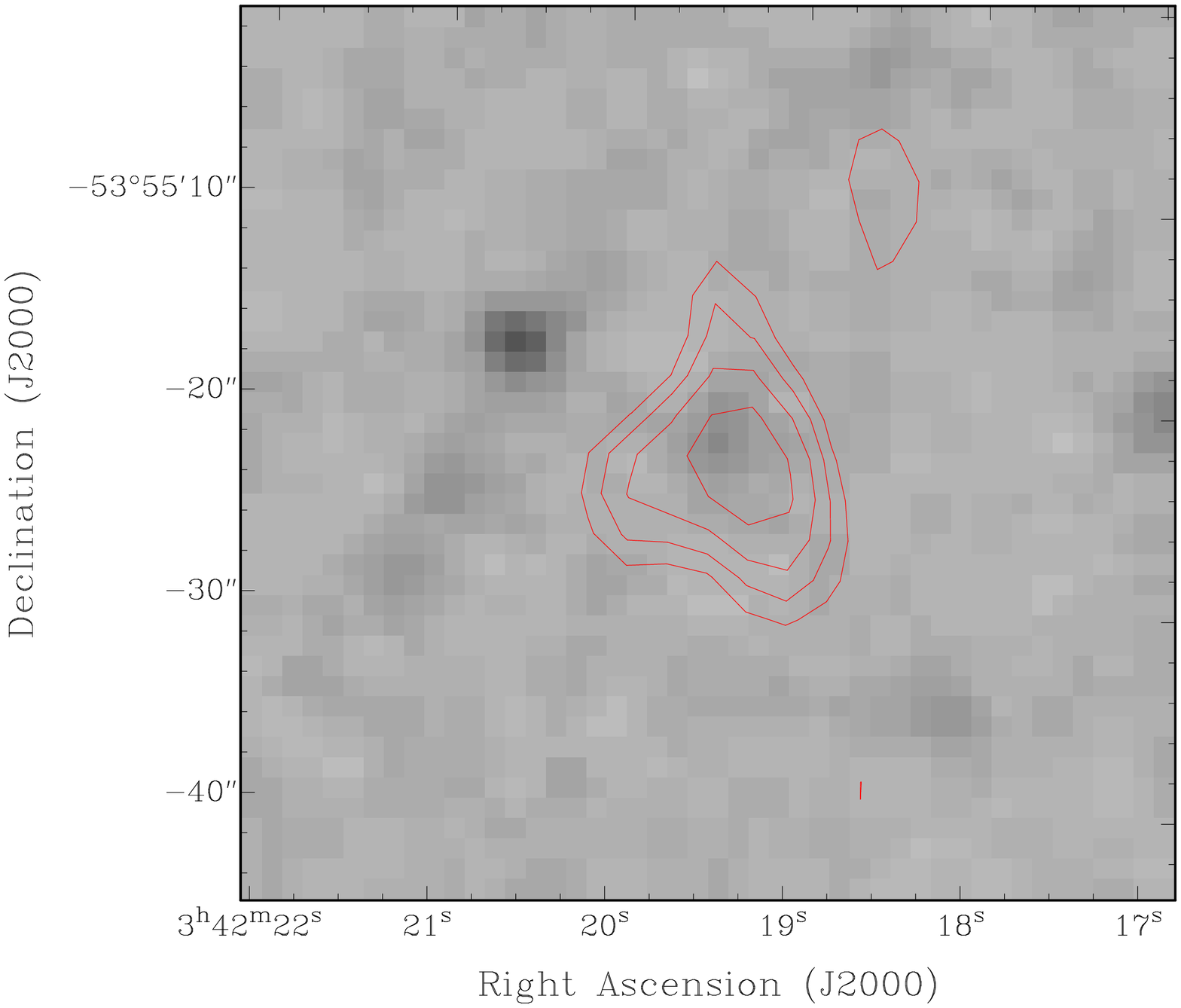}}
\resizebox{\hsize}{!}{\includegraphics[angle=-90]{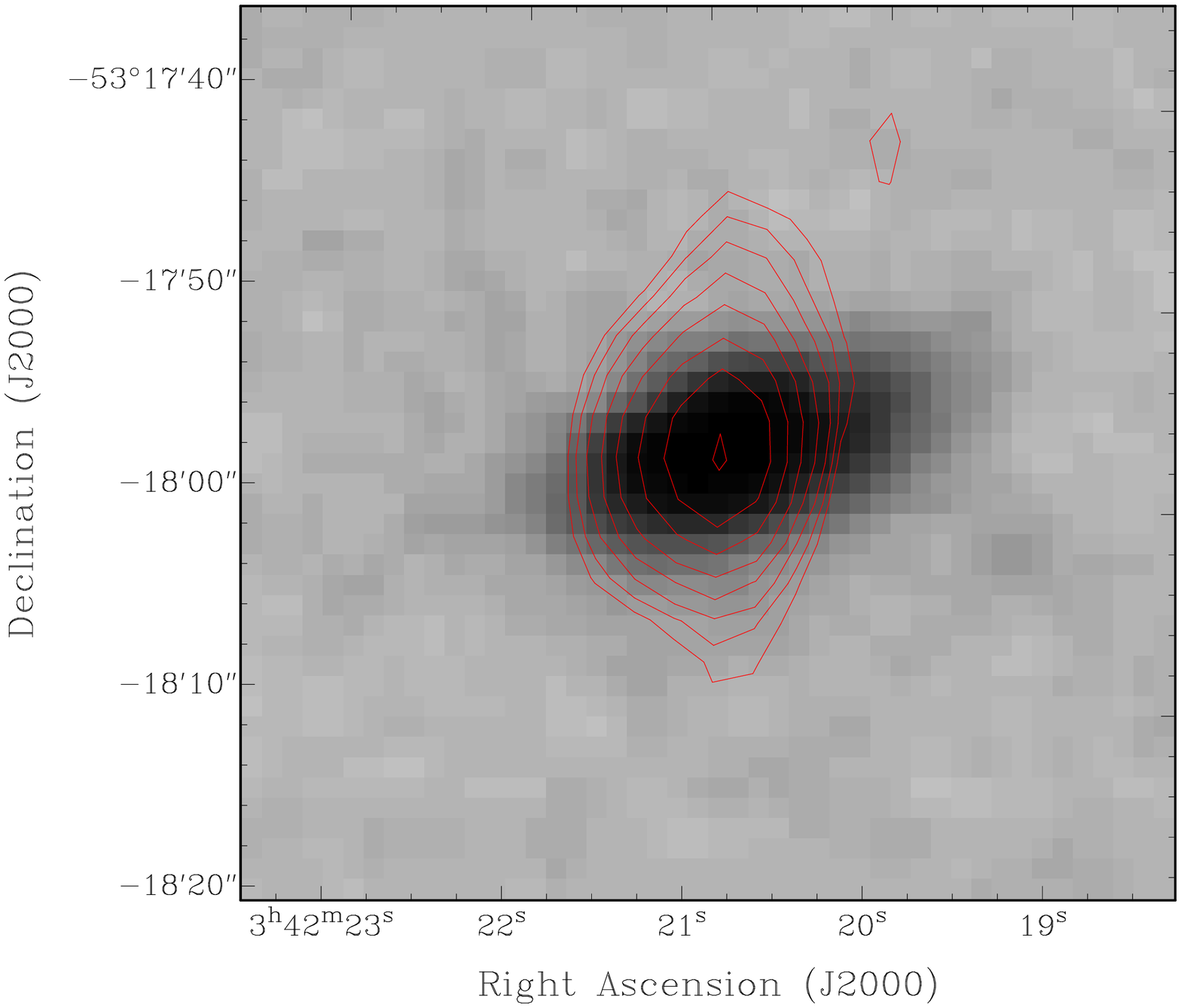}\includegraphics[angle=-90]{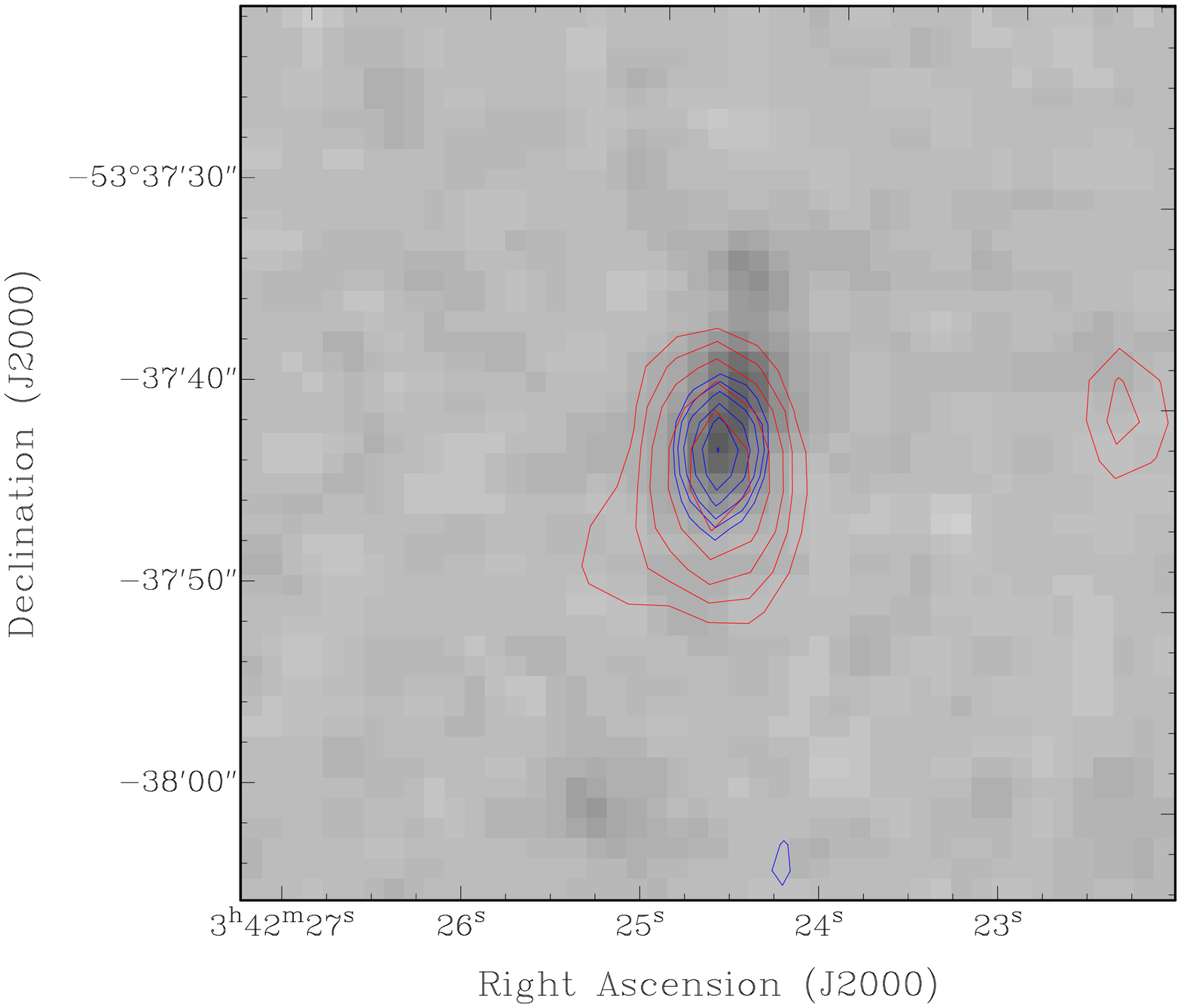}\includegraphics[angle=-90]{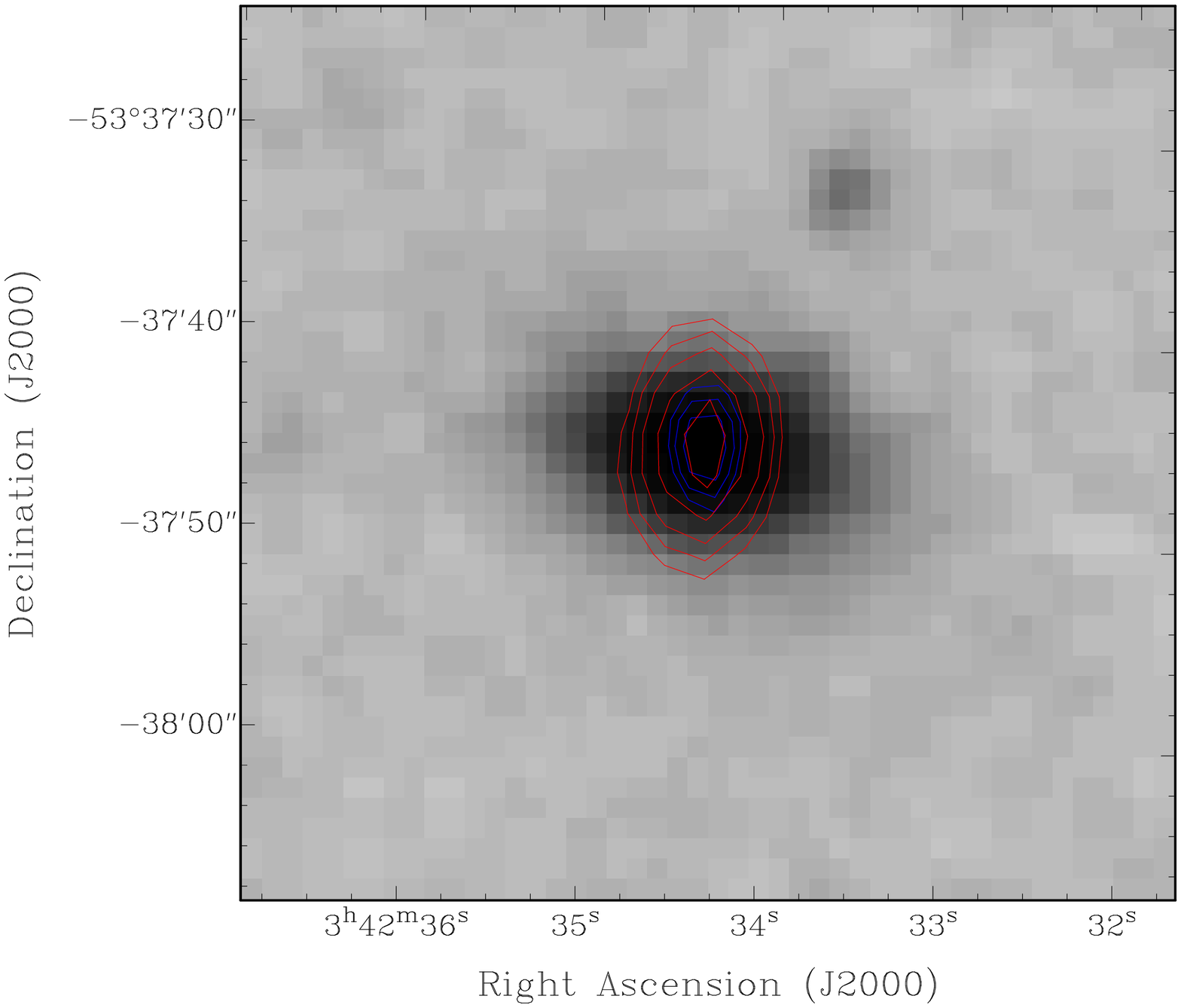}\includegraphics[angle=-90]{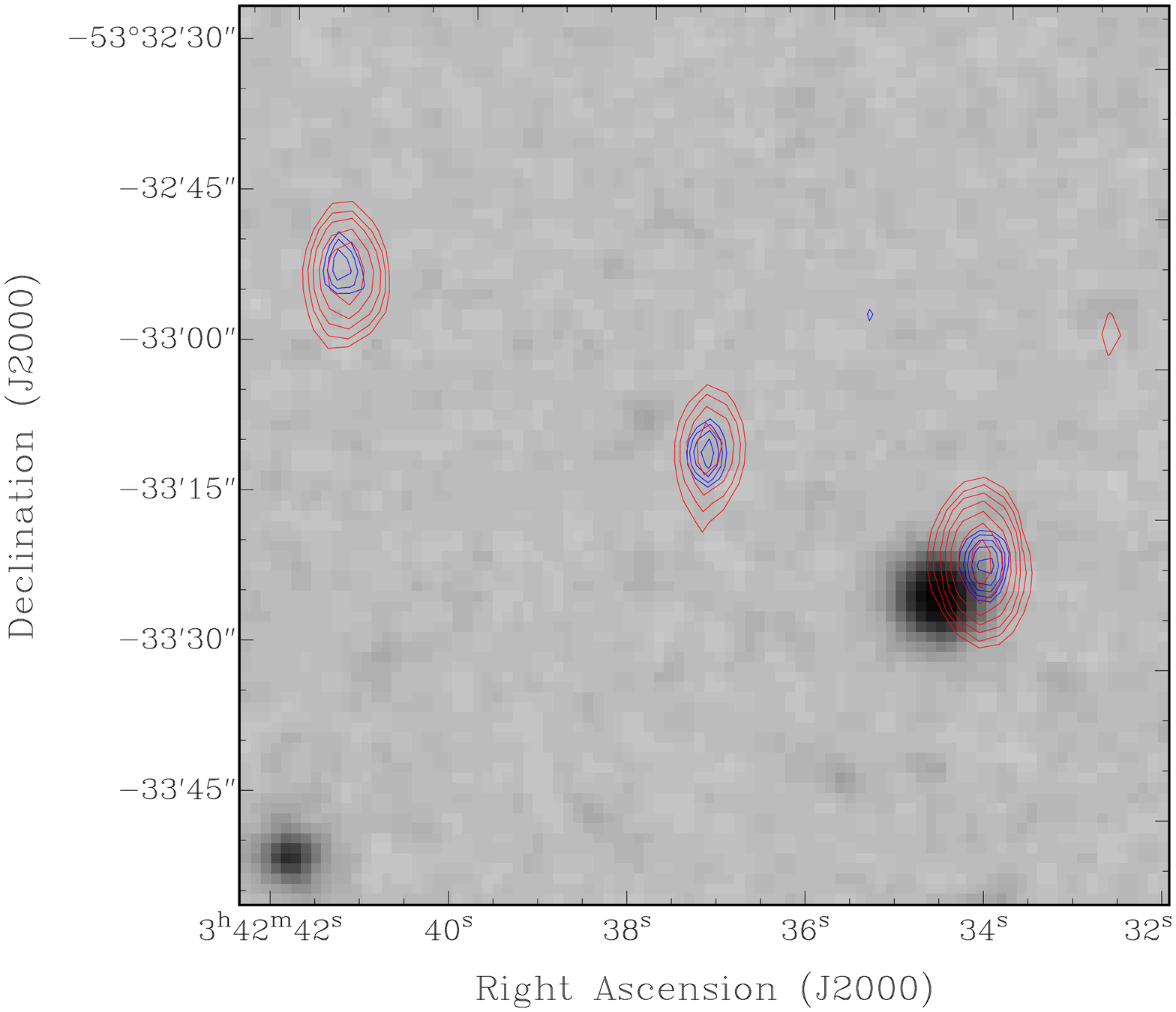}\includegraphics[angle=-90]{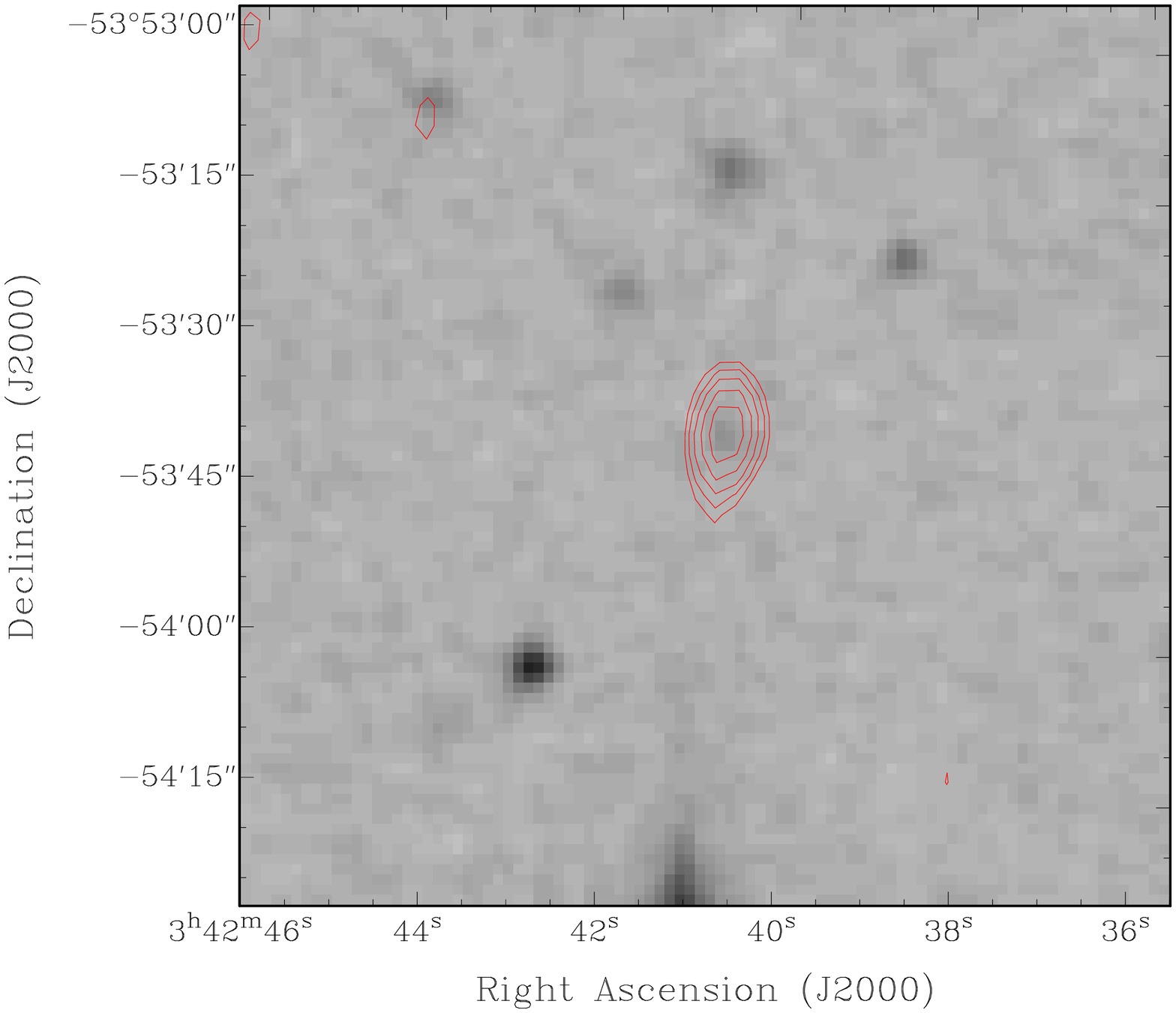}\includegraphics[angle=-90]{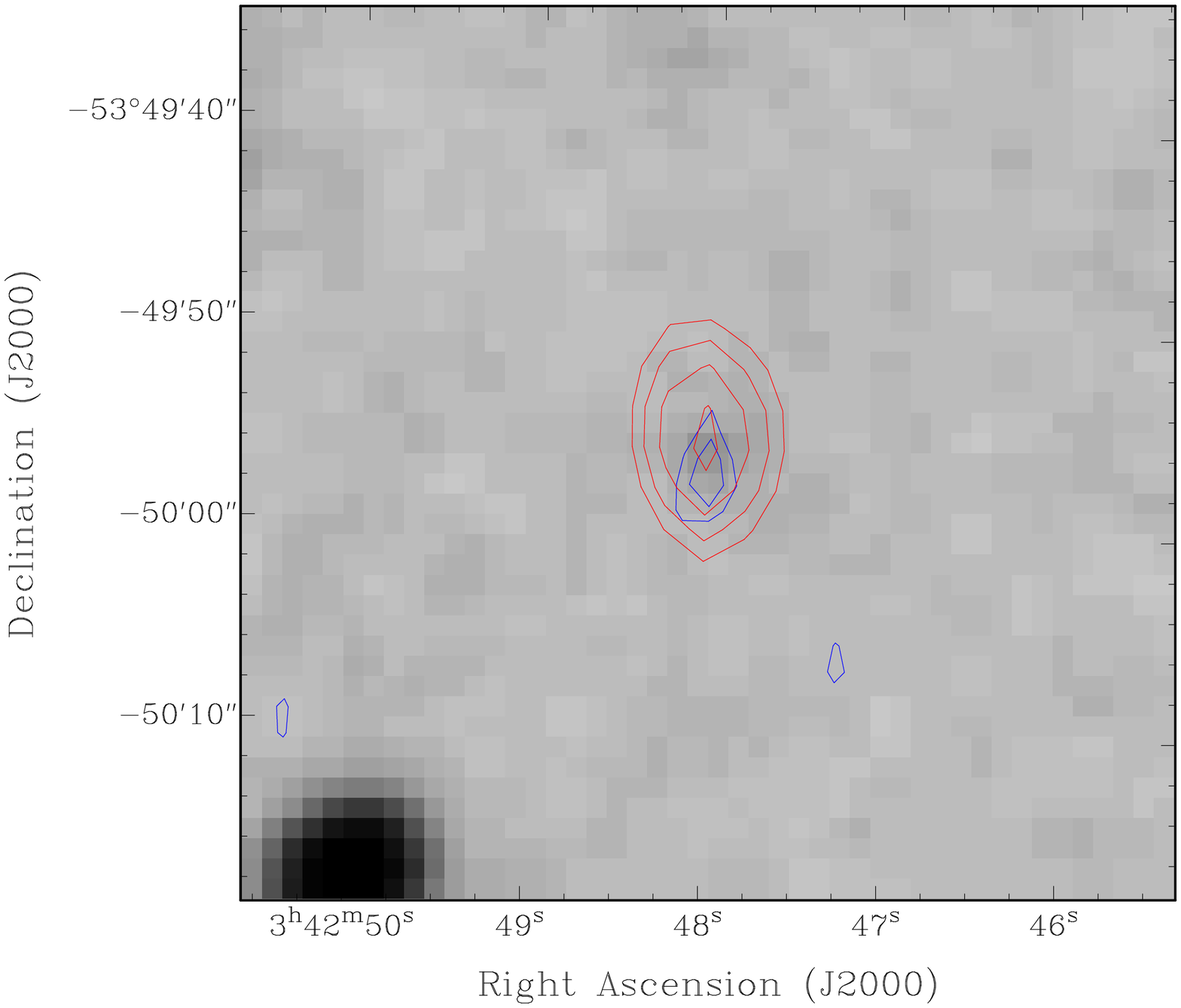}}
\resizebox{\hsize}{!}{\includegraphics[angle=-90]{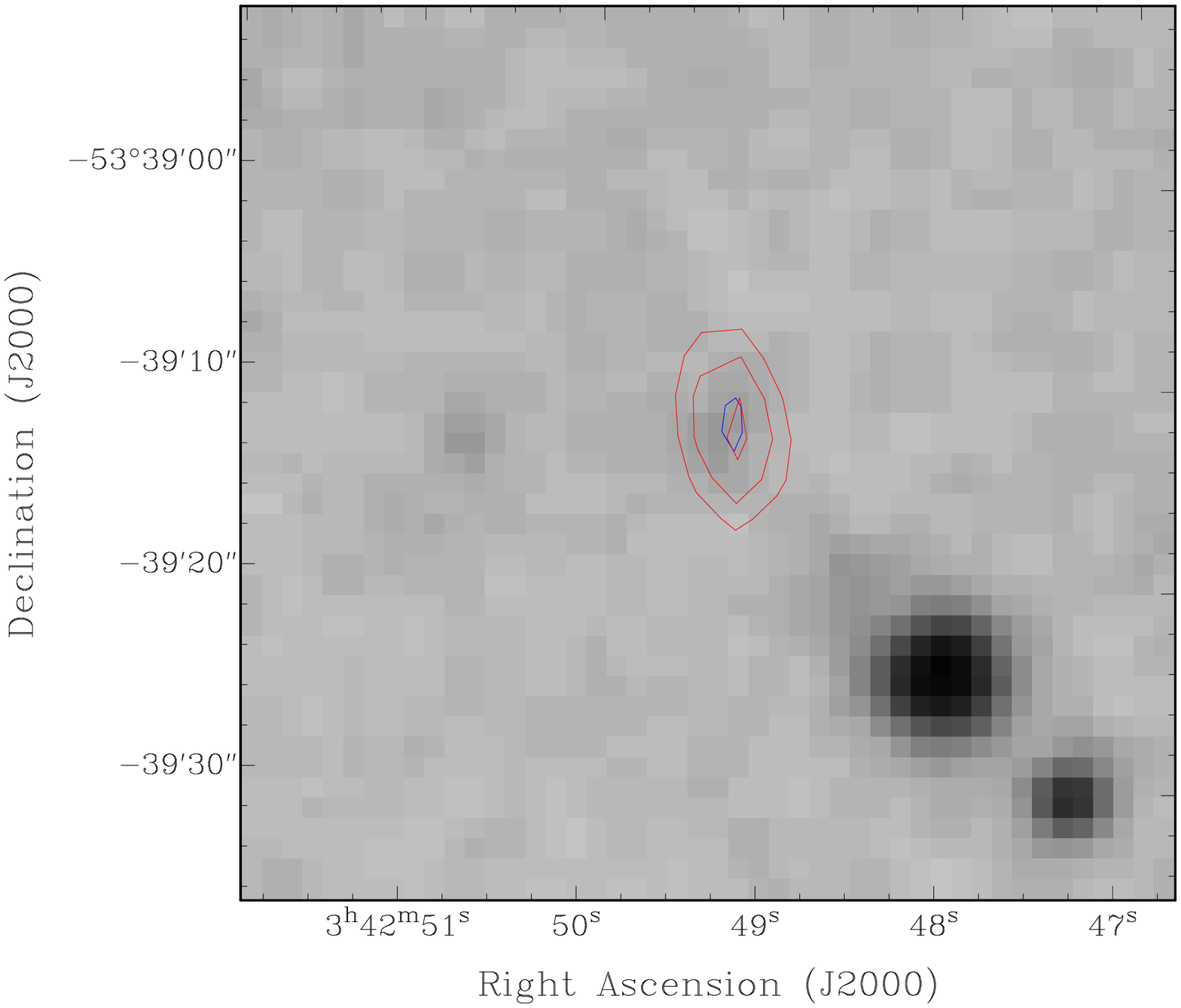}\includegraphics[angle=-90]{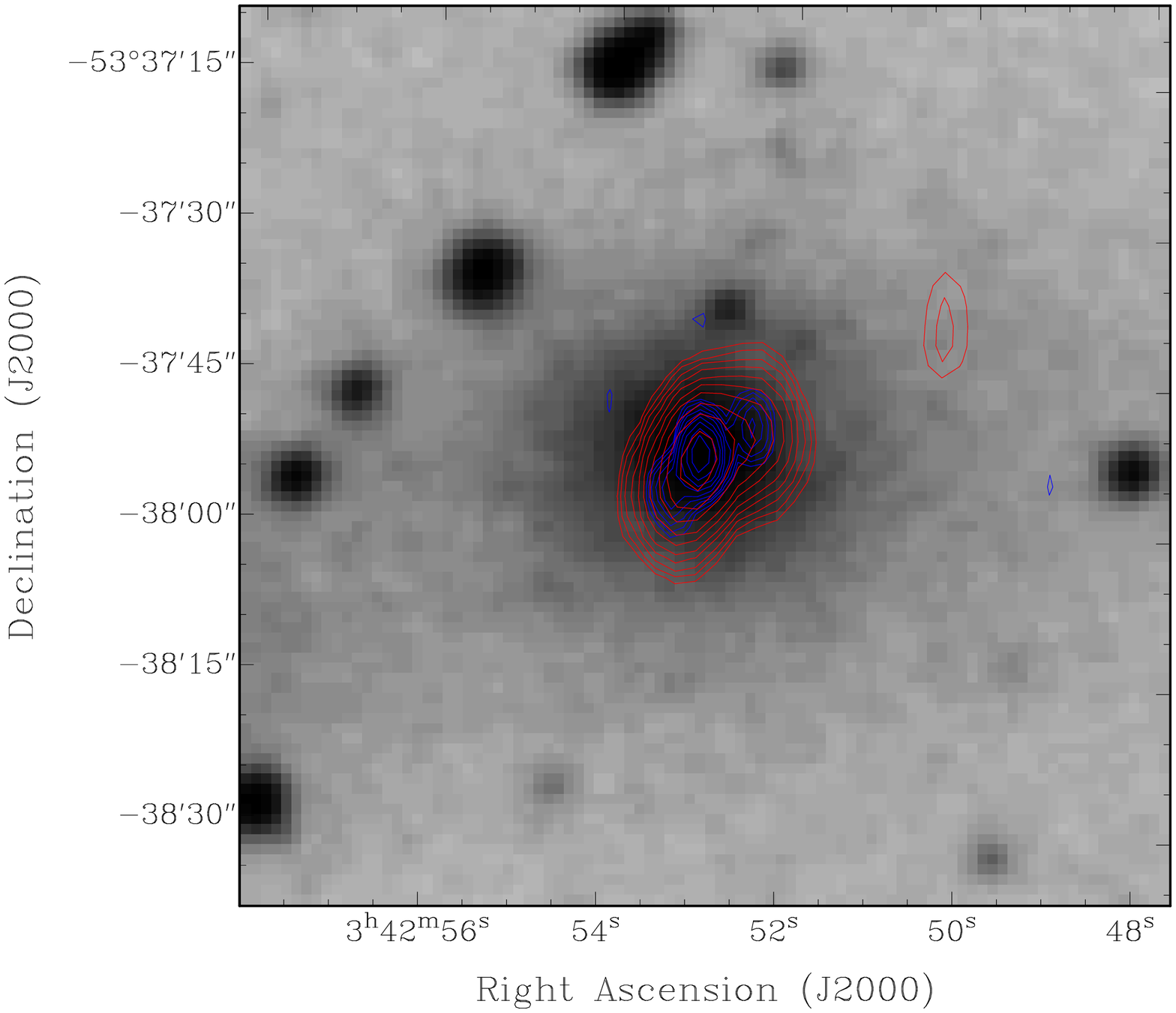}\includegraphics[angle=-90]{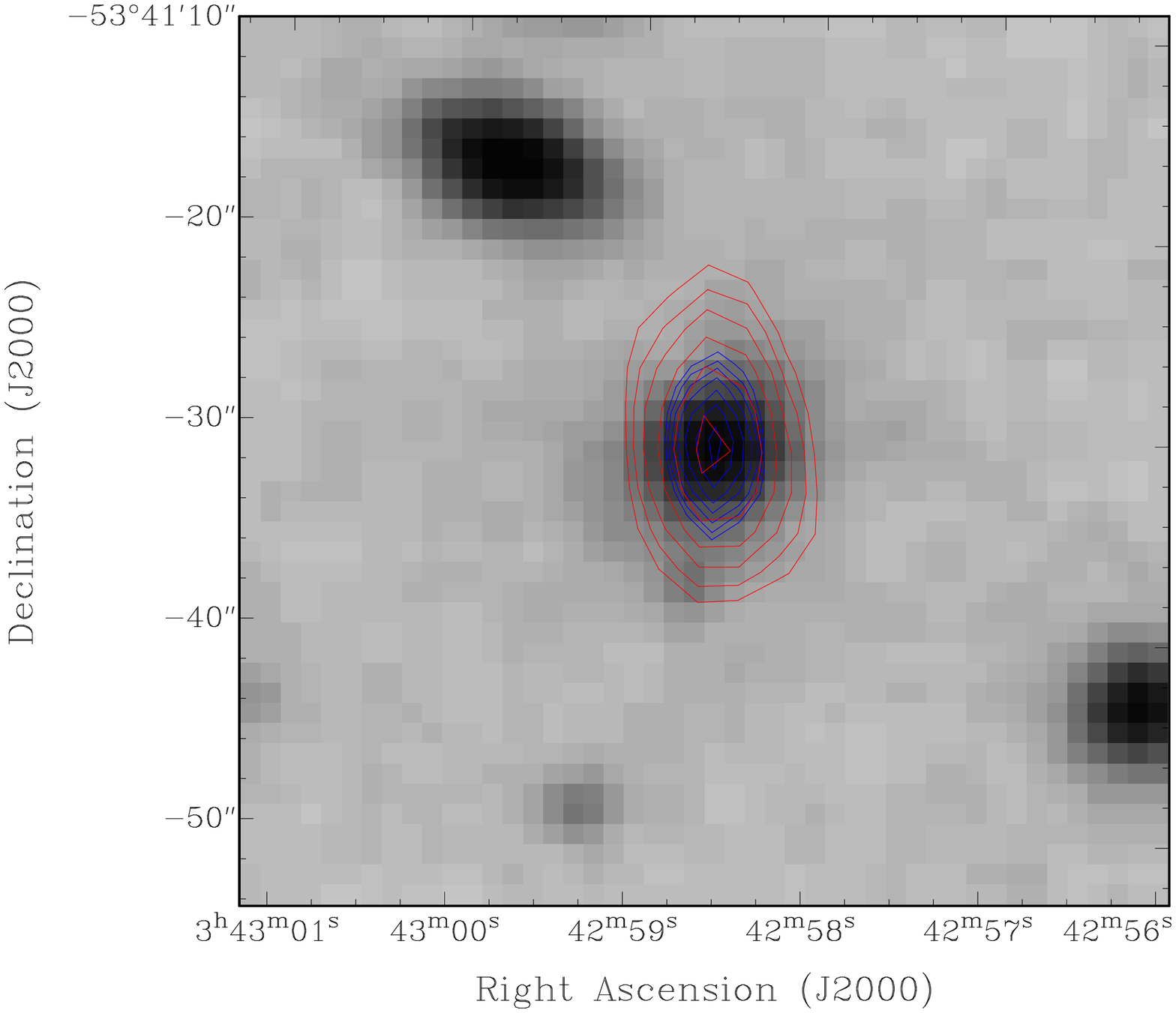}\includegraphics[angle=-90]{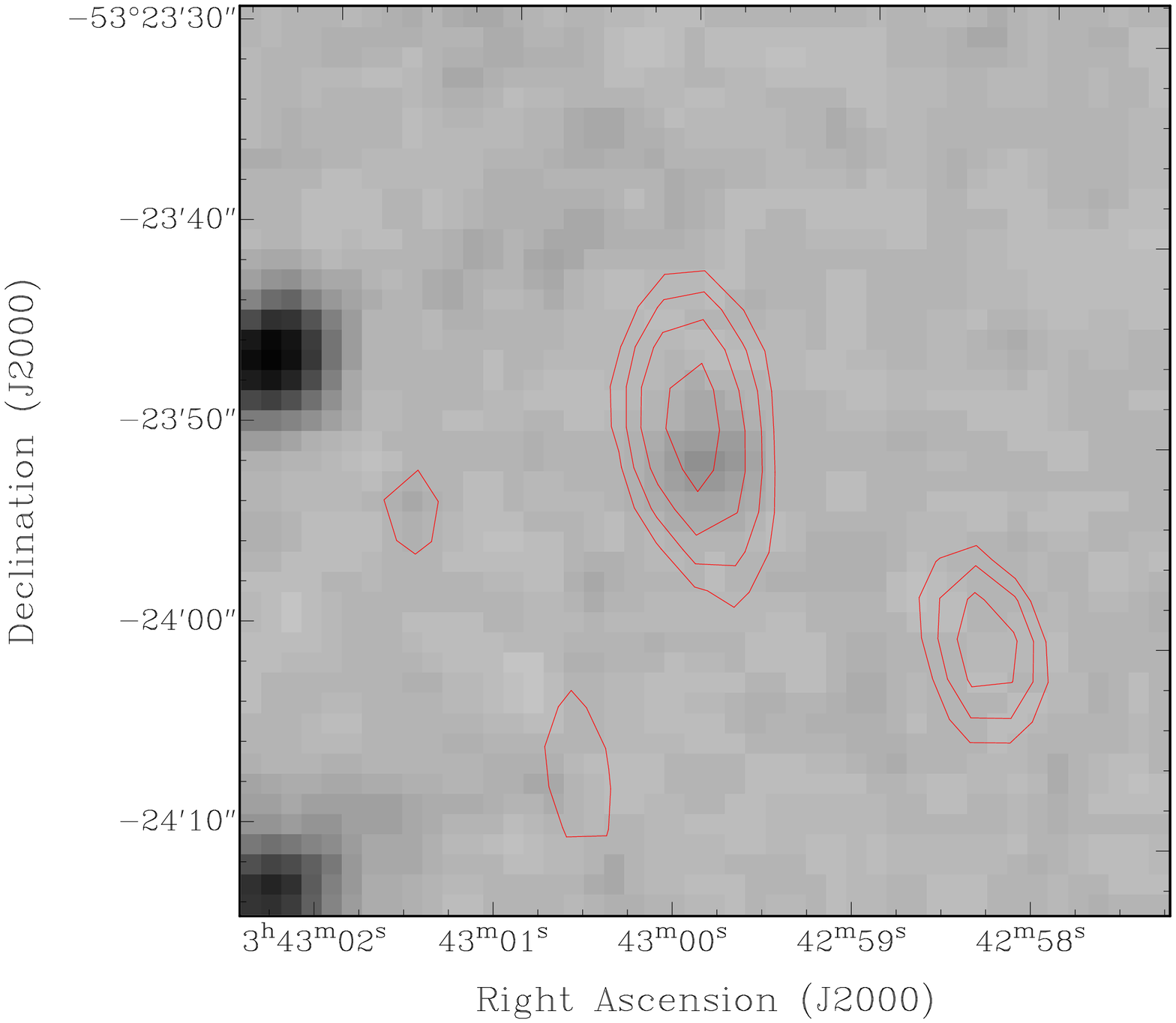}\includegraphics[angle=-90]{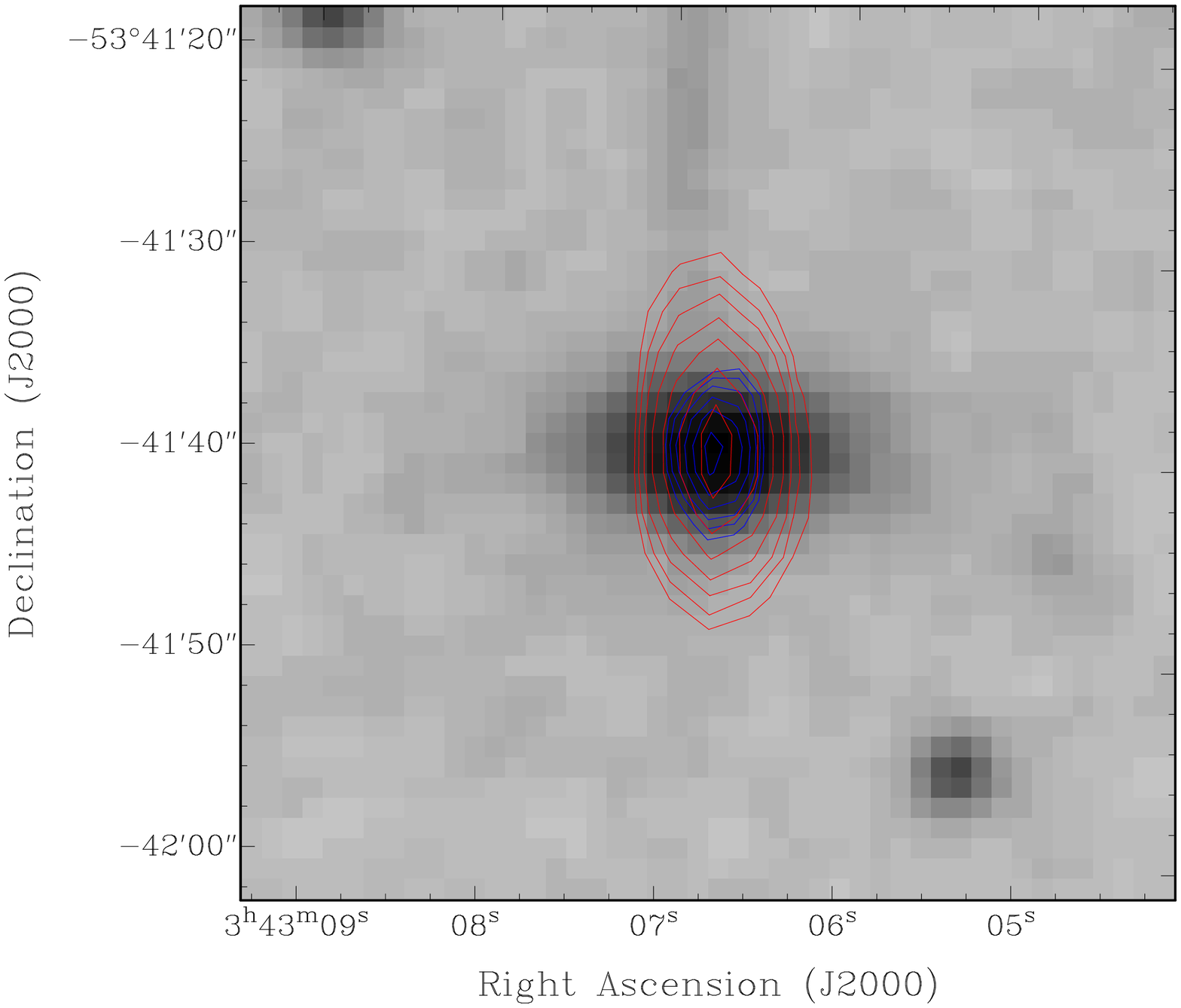}\includegraphics[angle=-90]{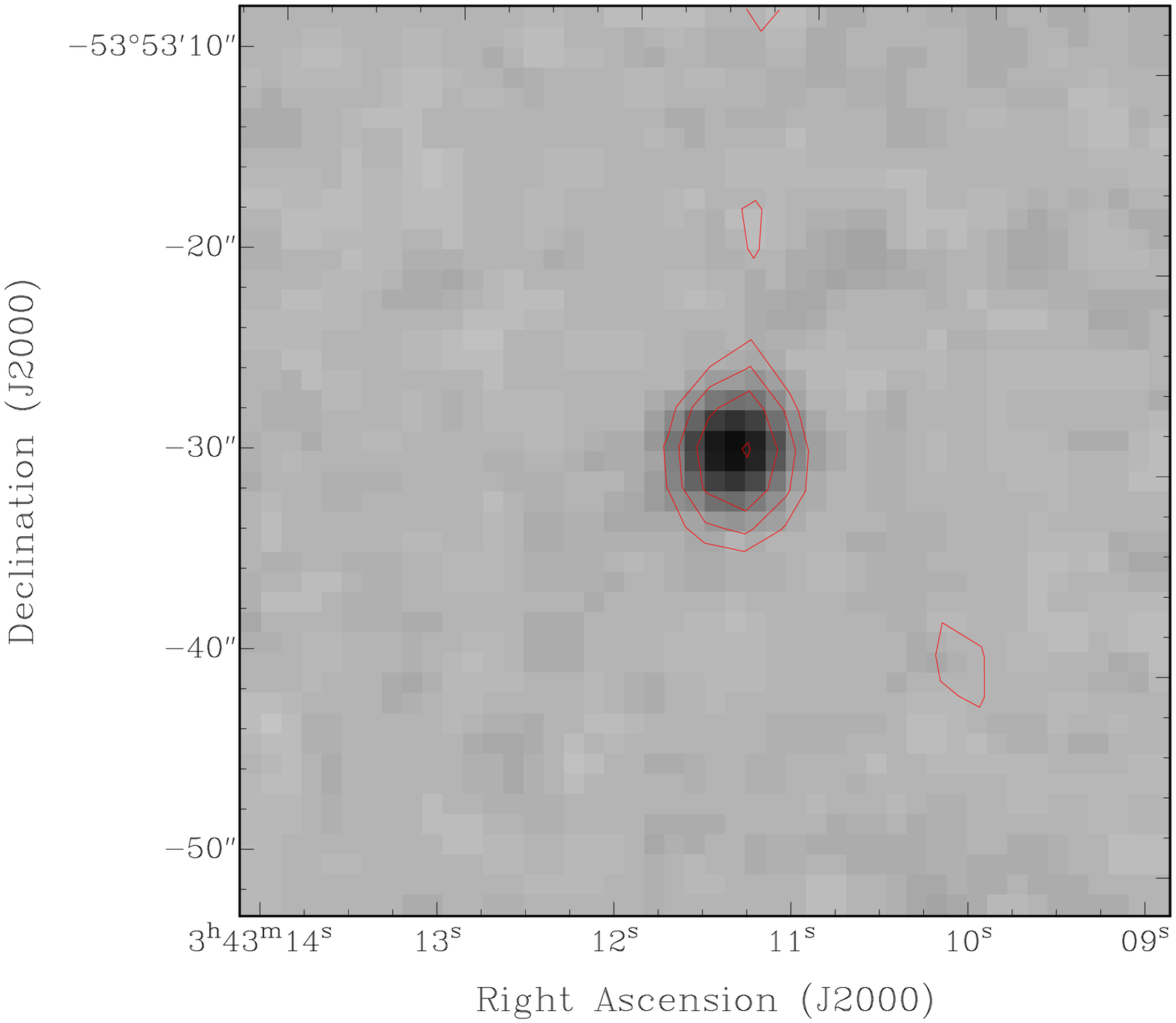}}
\resizebox{\hsize}{!}{\includegraphics[angle=-90]{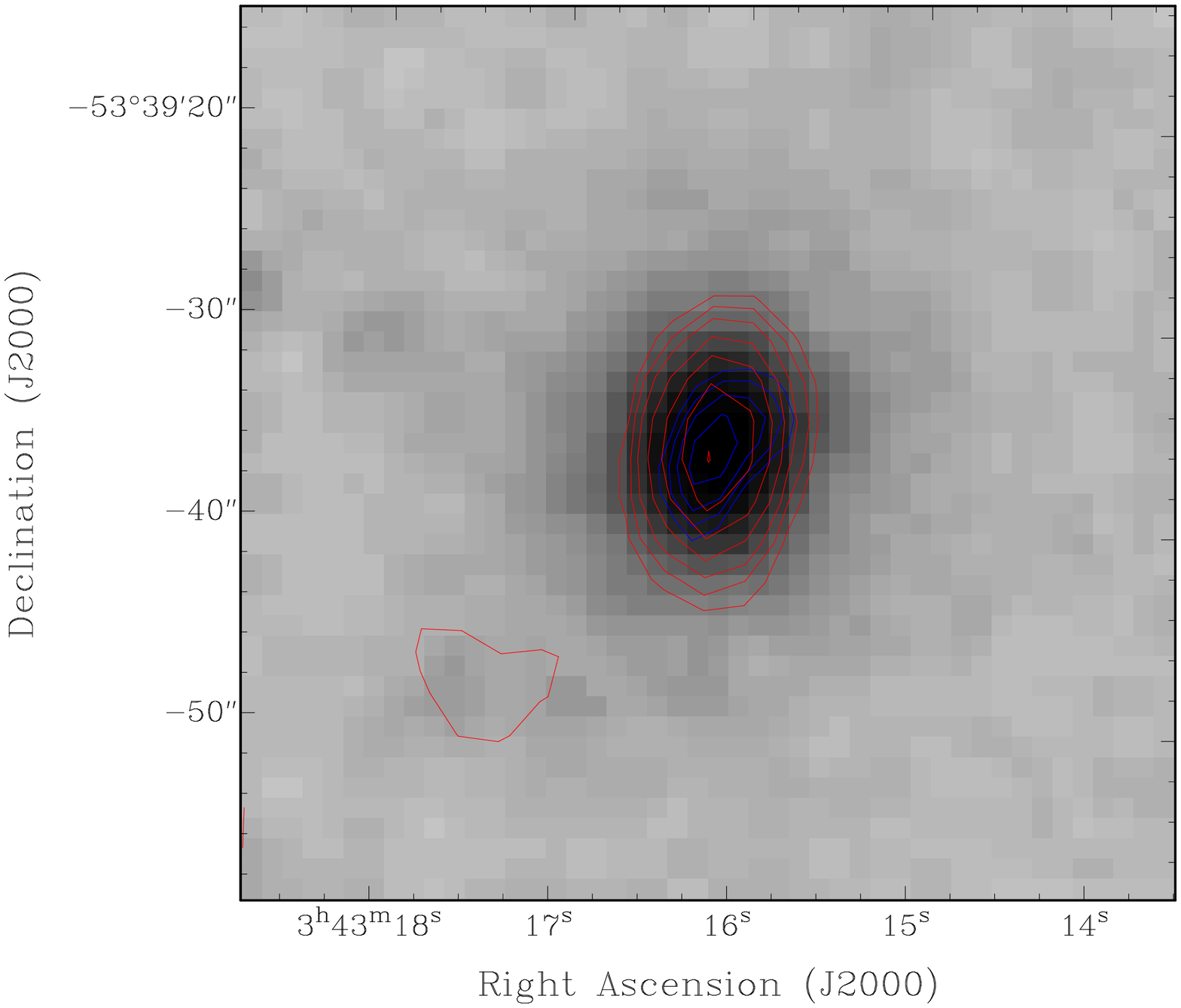}\includegraphics[angle=-90]{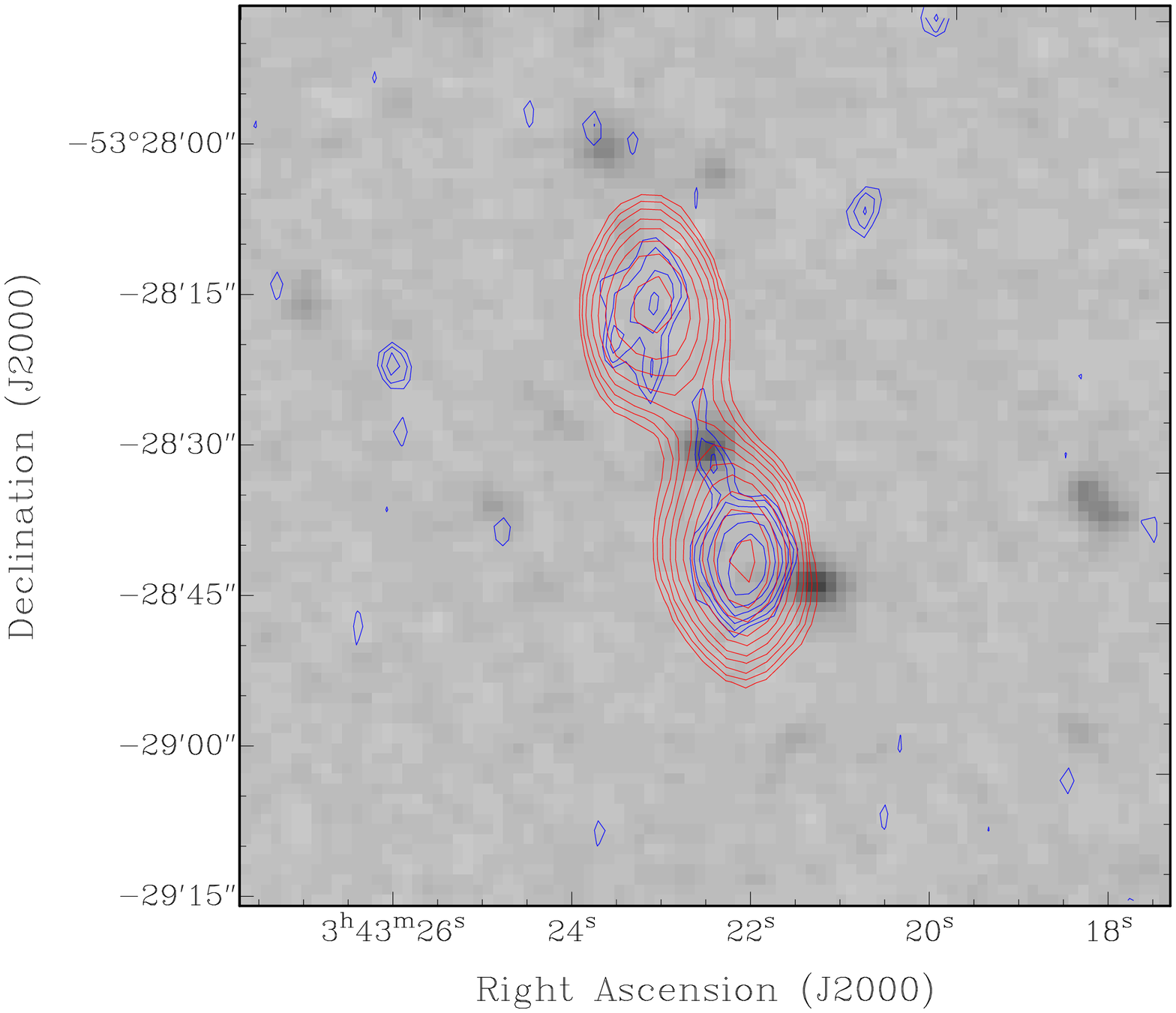}\includegraphics[angle=-90]{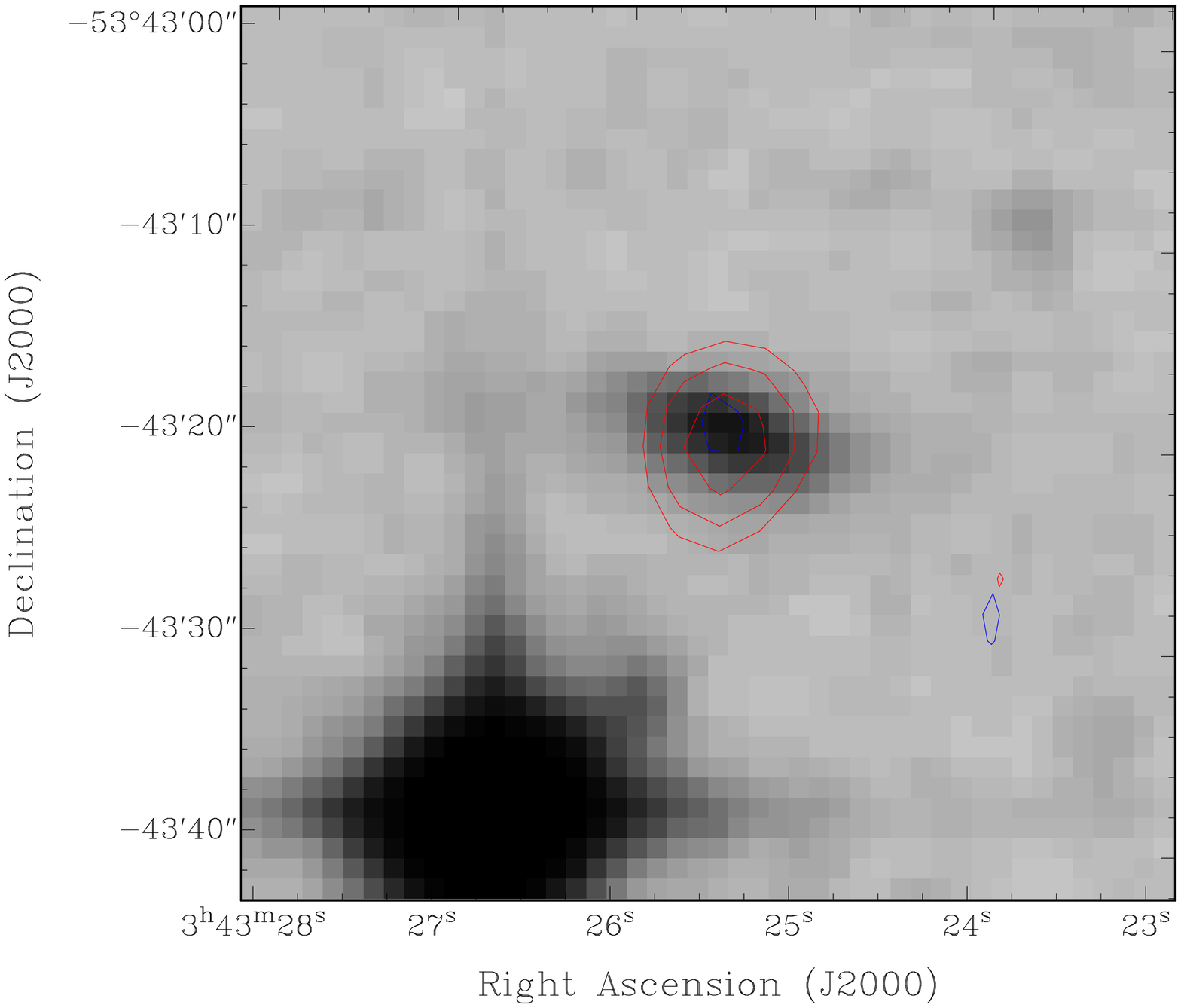}\includegraphics[angle=-90]{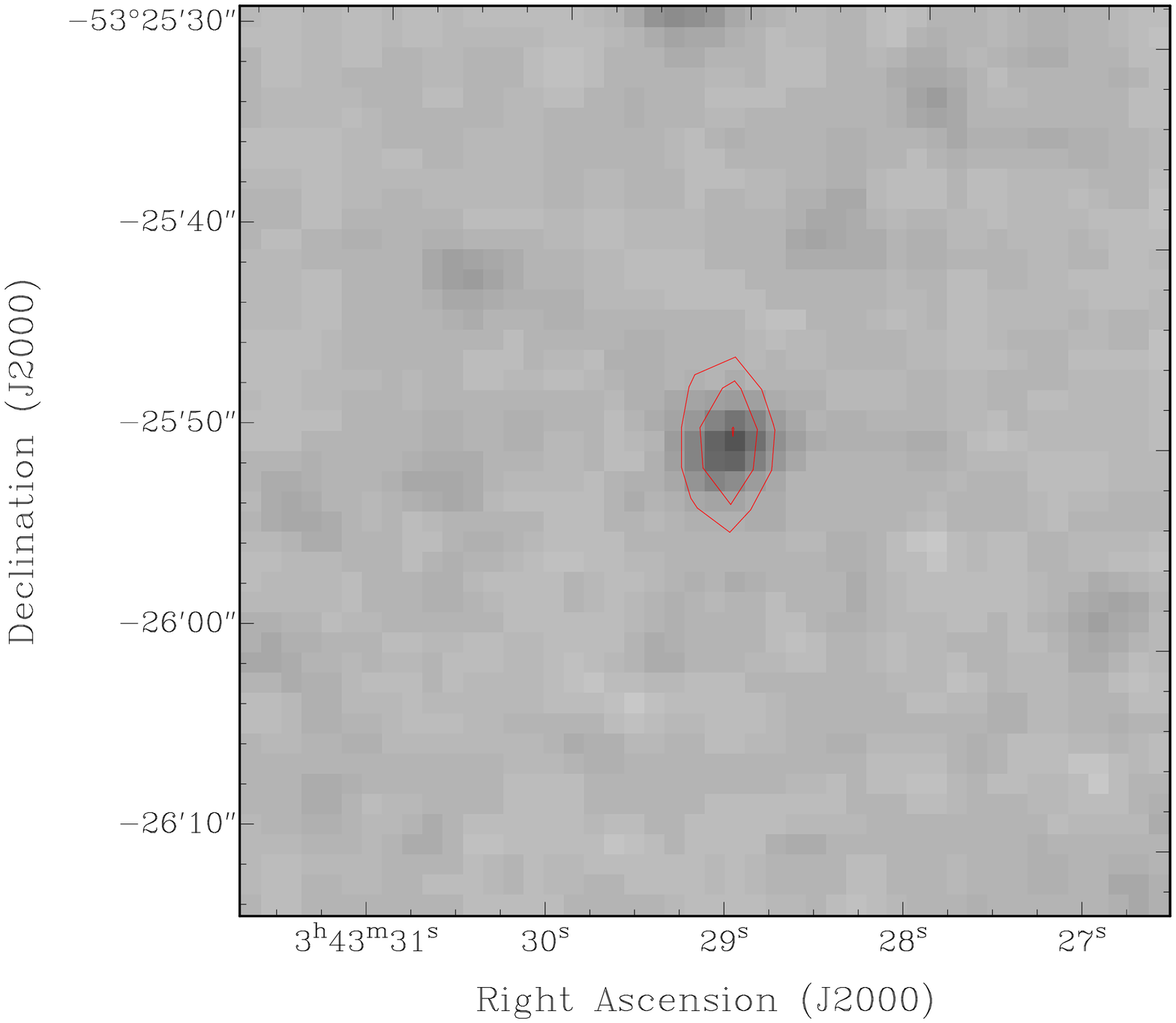}\includegraphics[angle=-90]{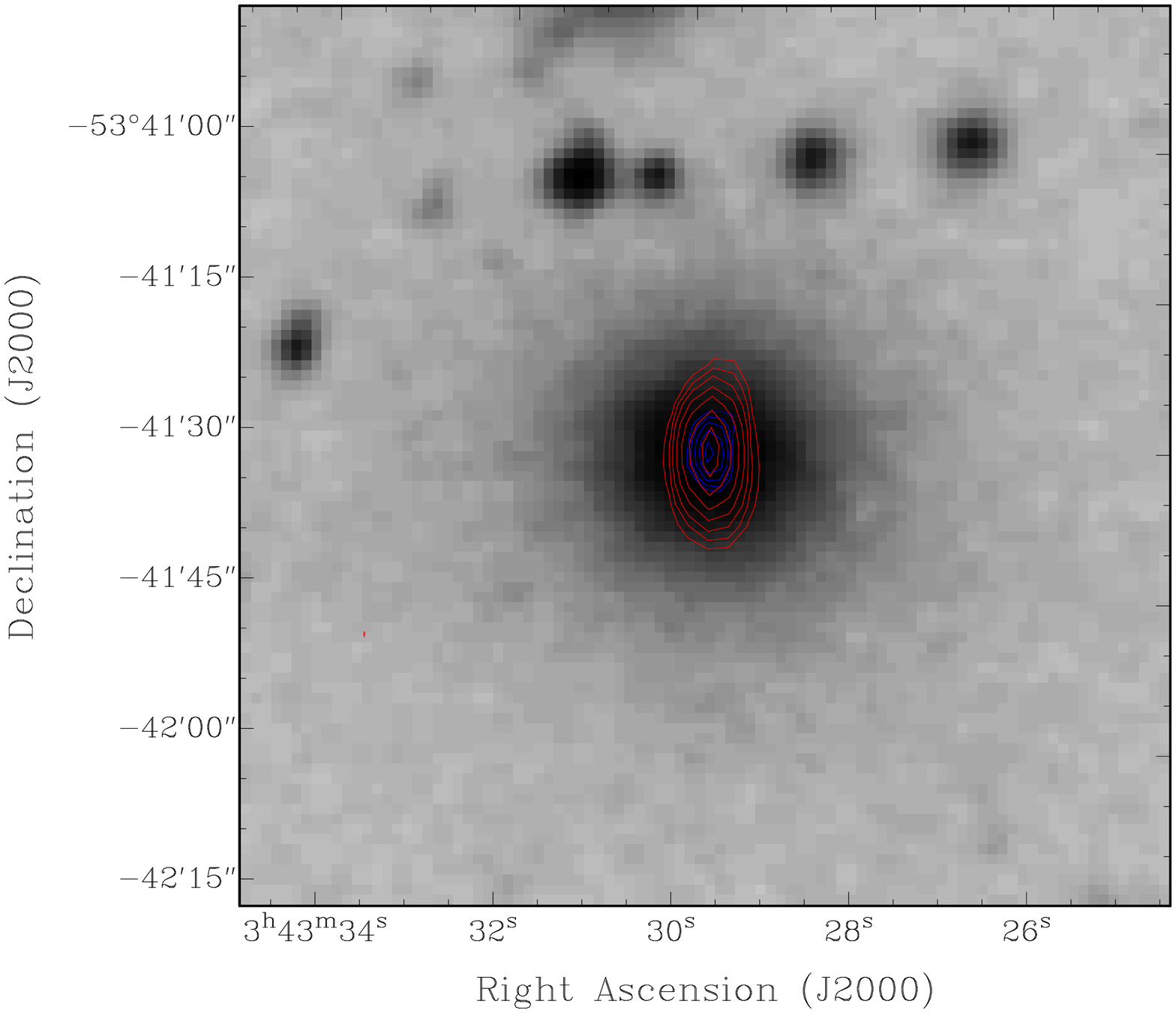}\includegraphics[angle=-90]{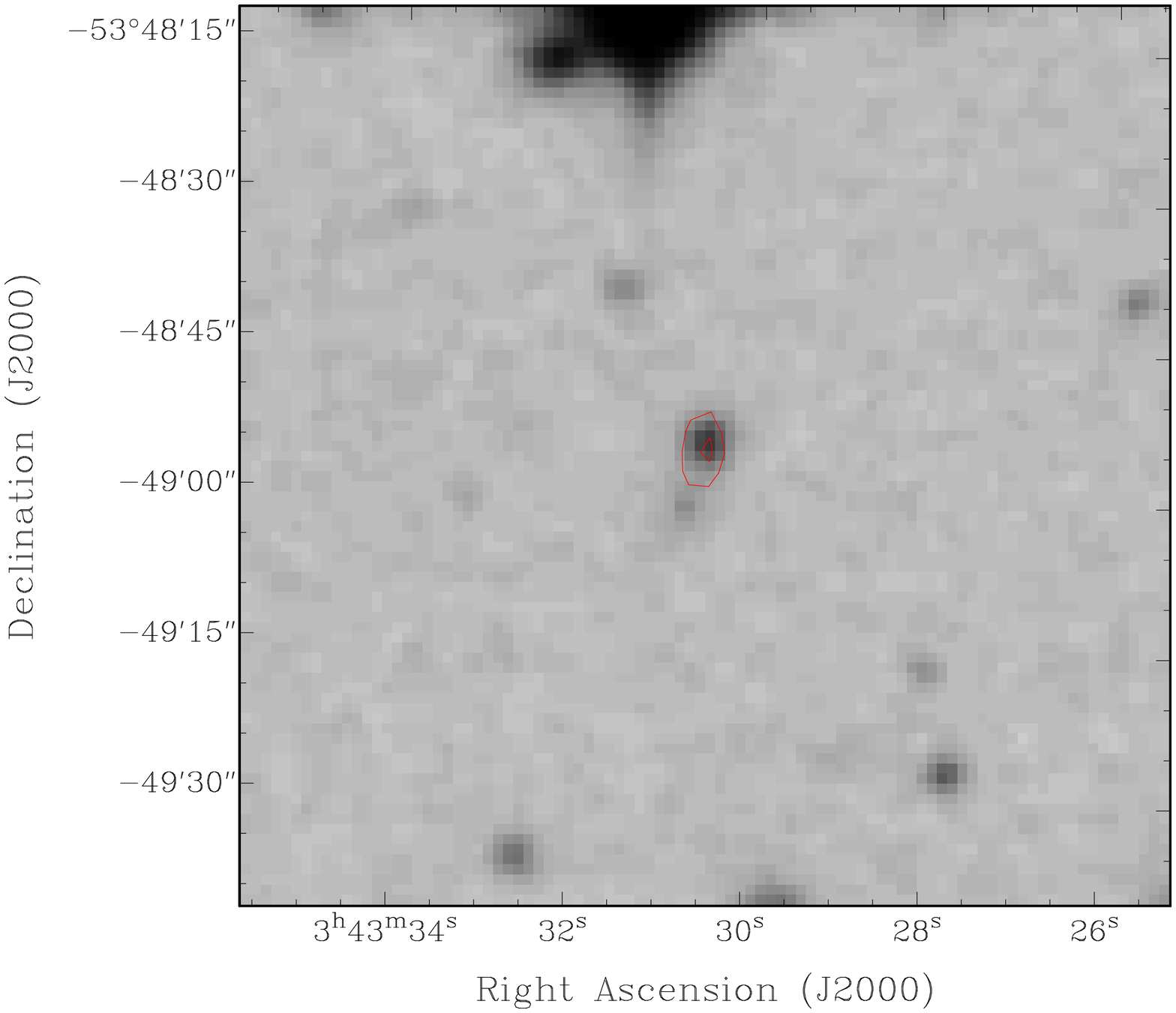}}
\resizebox{\hsize}{!}{\includegraphics[angle=-90]{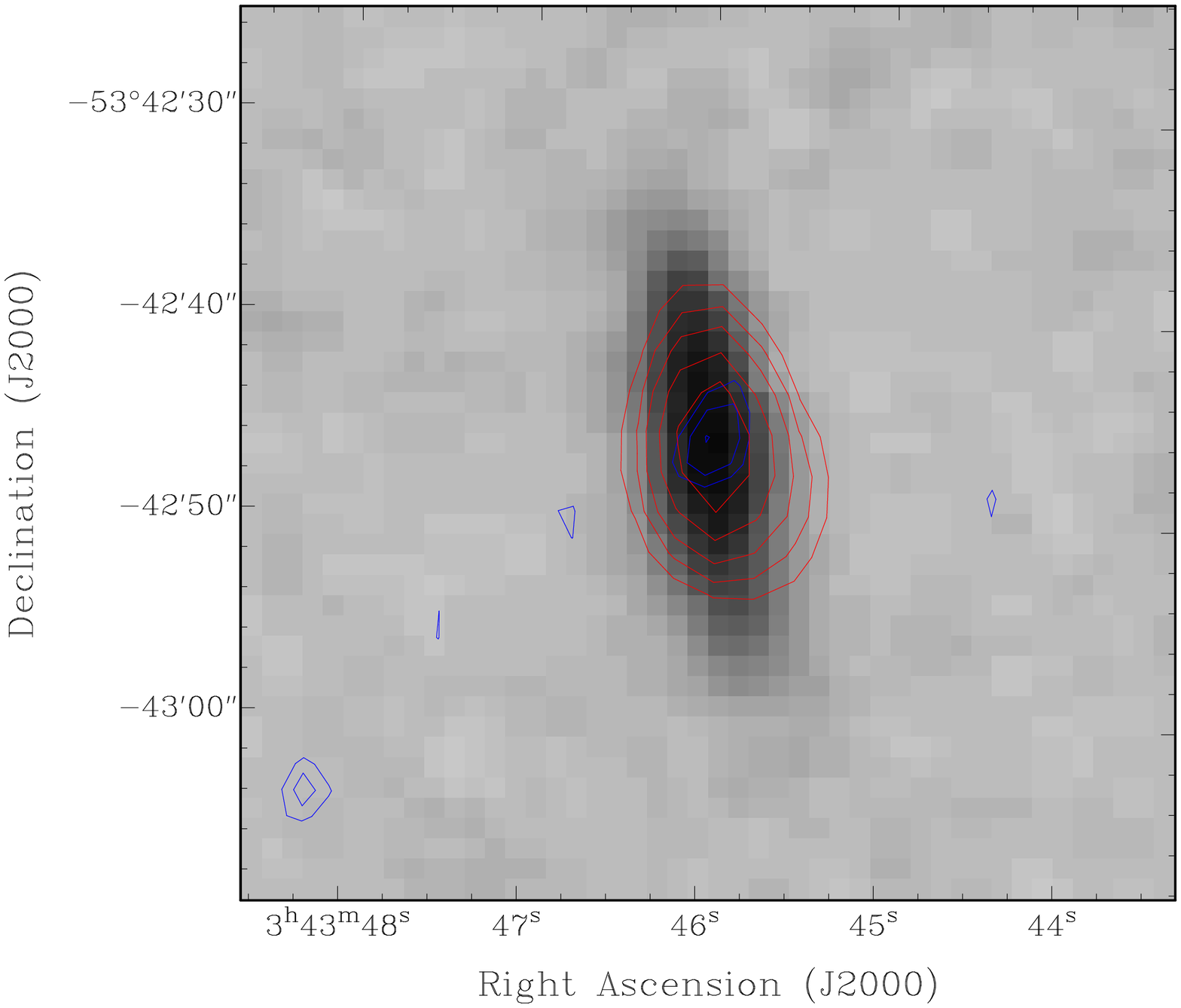}\includegraphics[angle=-90]{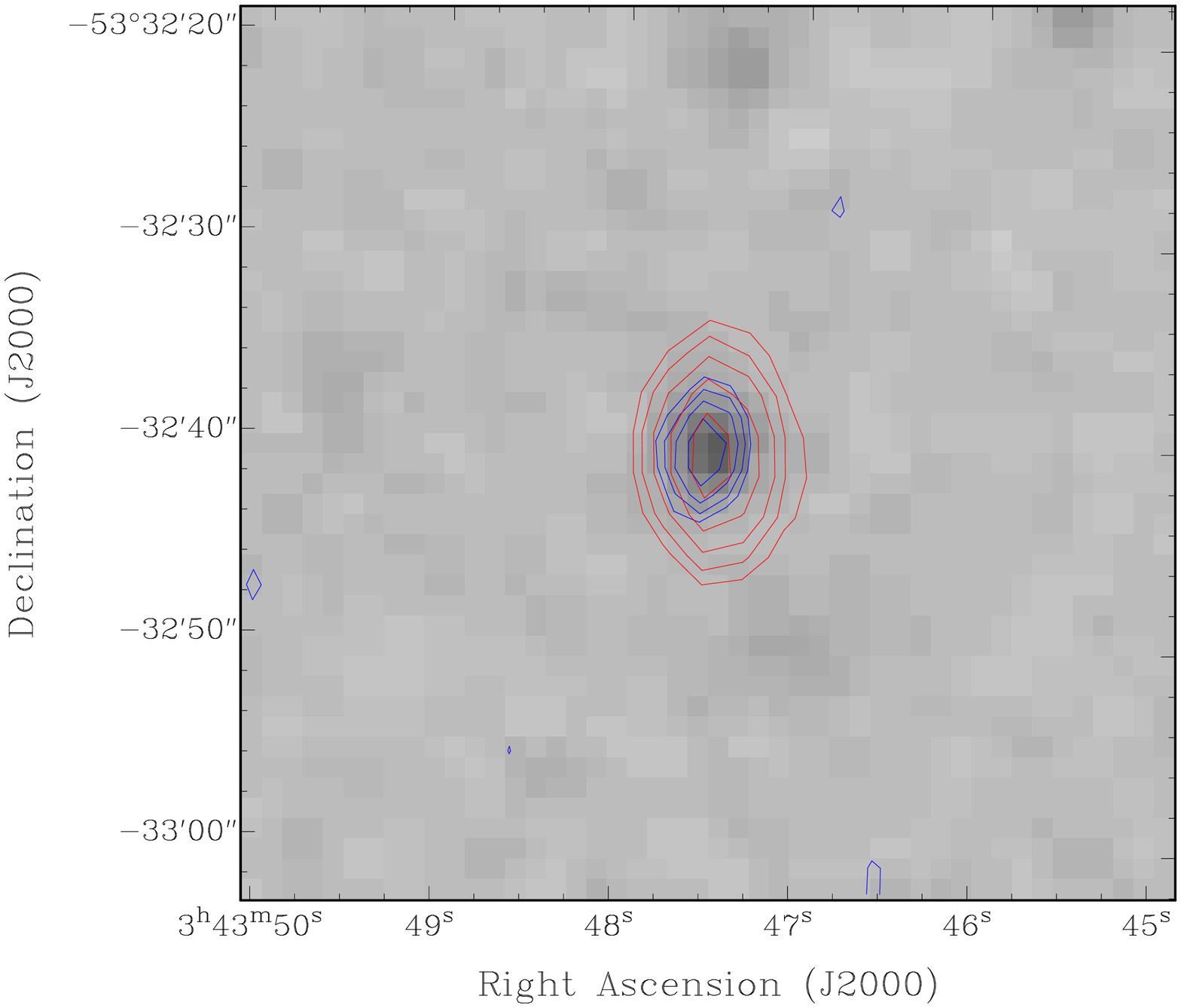}\includegraphics[angle=-90]{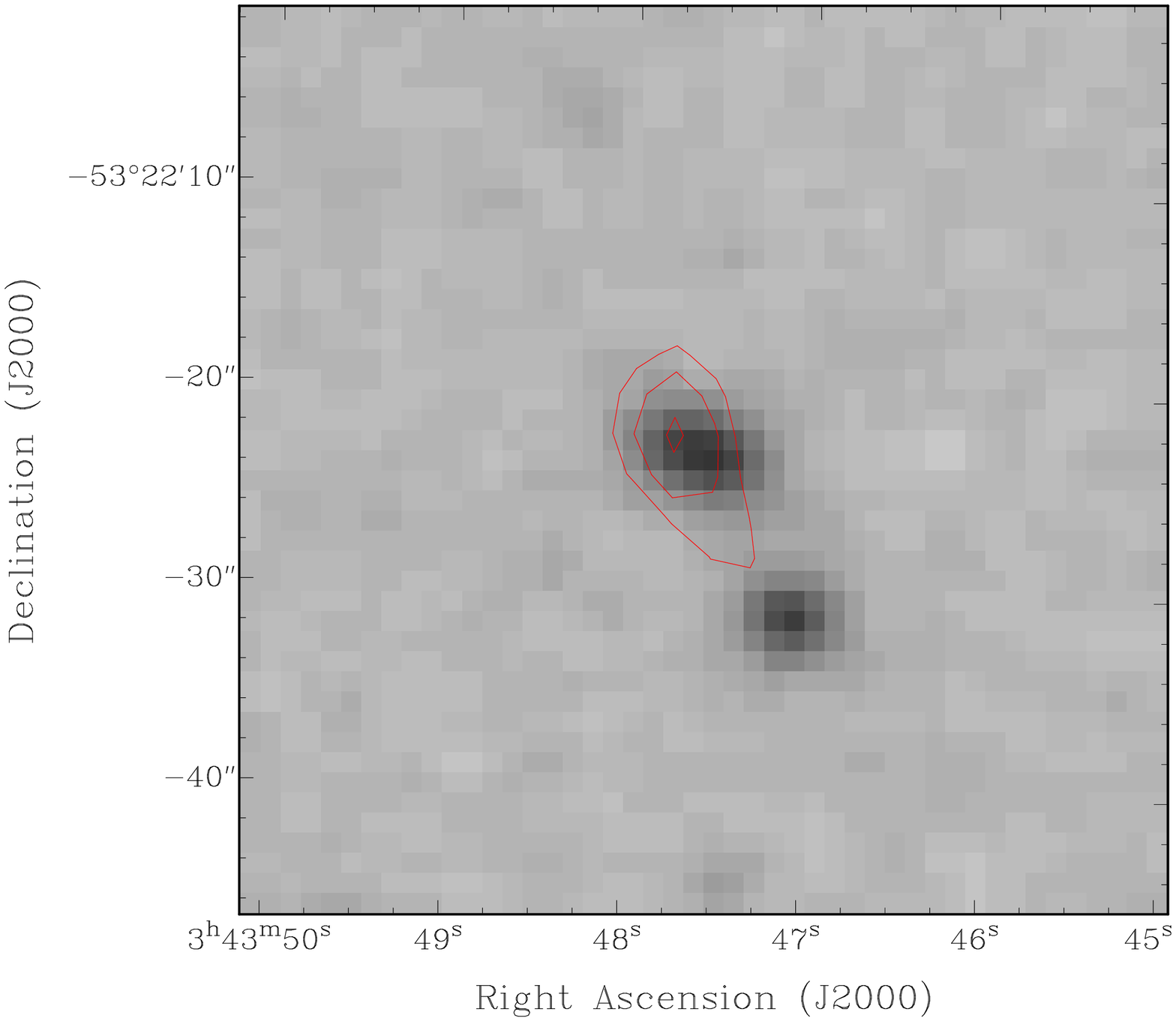}\includegraphics[angle=-90]{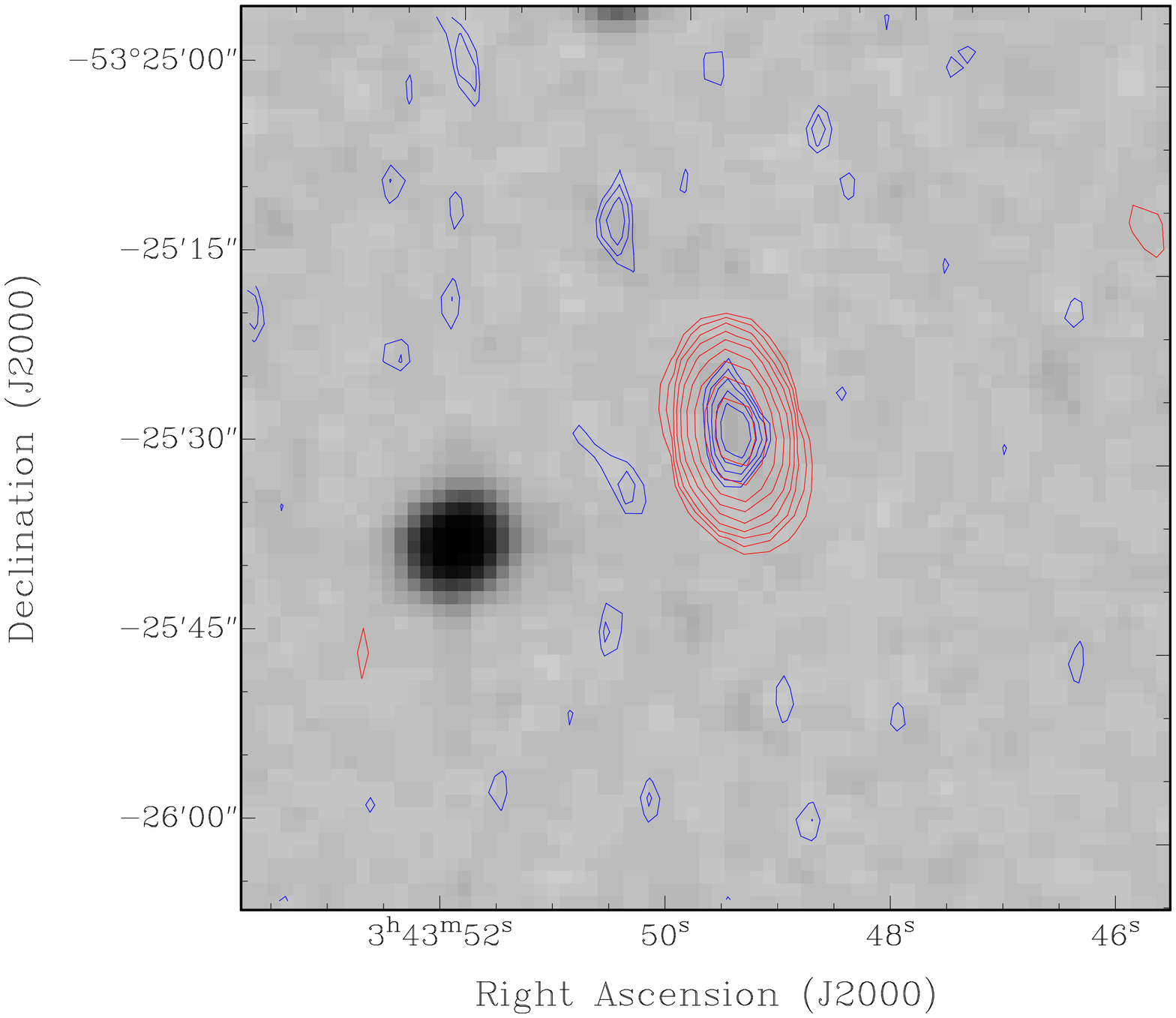}\includegraphics[angle=-90]{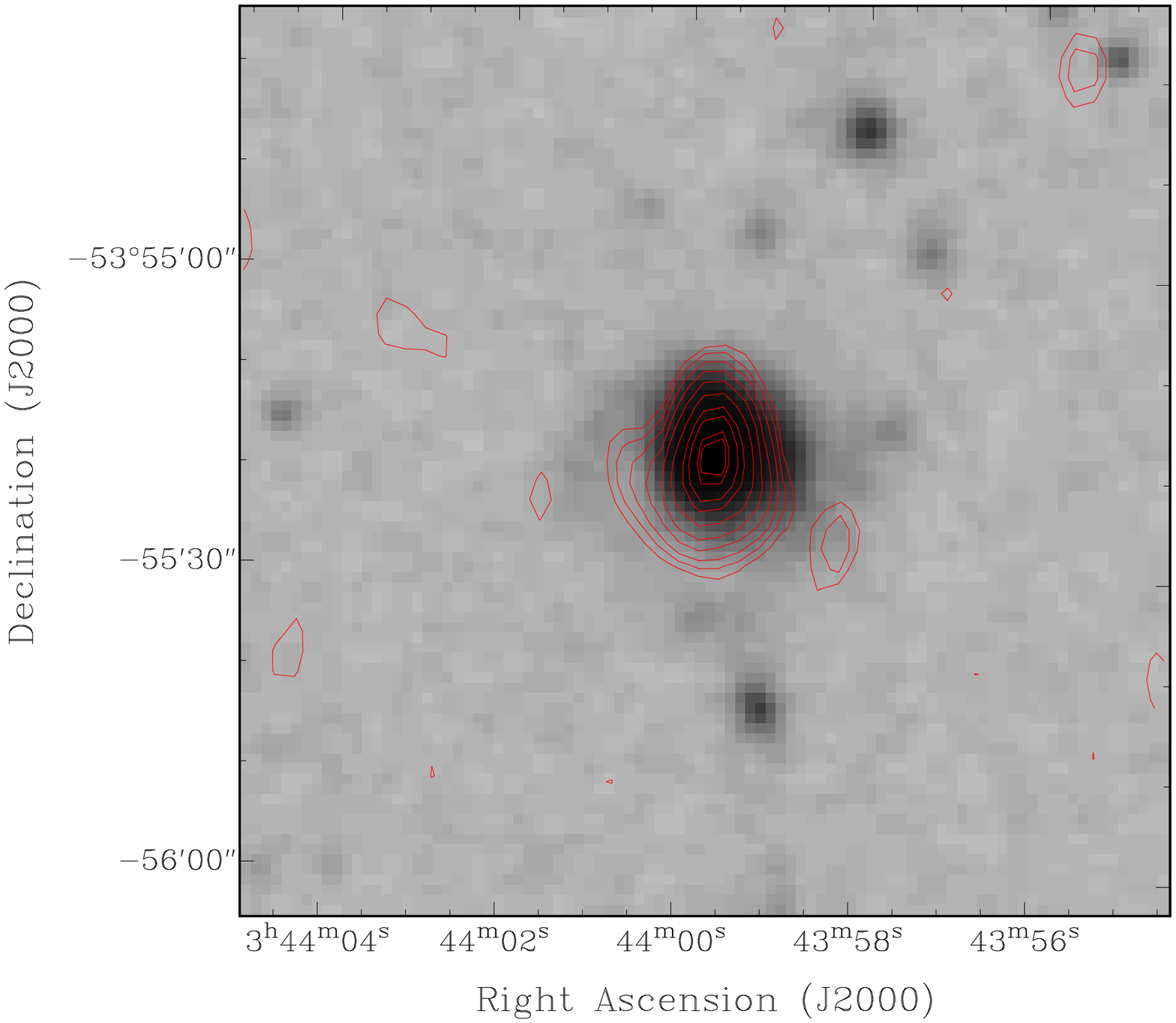}\includegraphics[angle=-90]{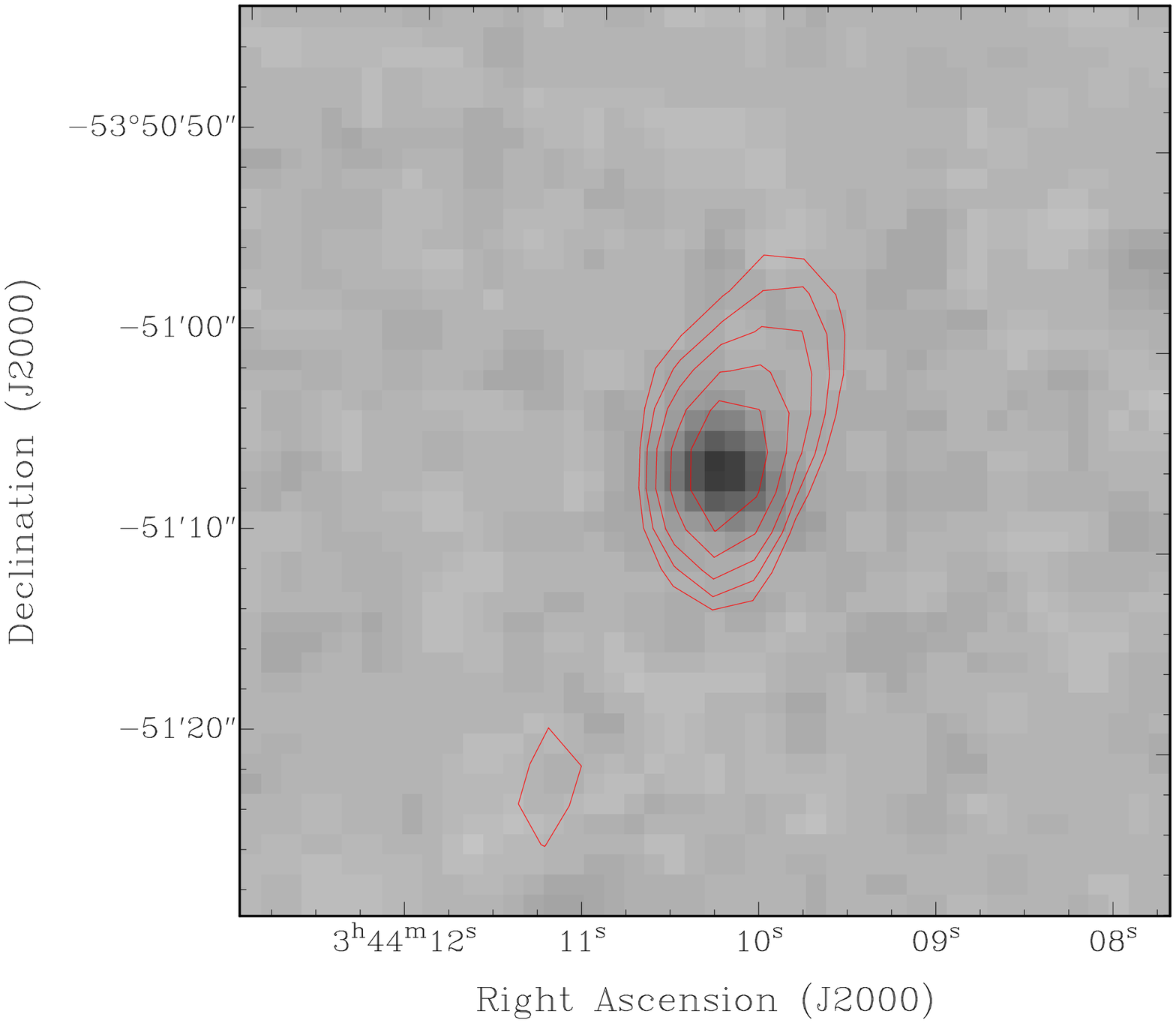}}
\resizebox{\hsize}{!}{\includegraphics[angle=-90]{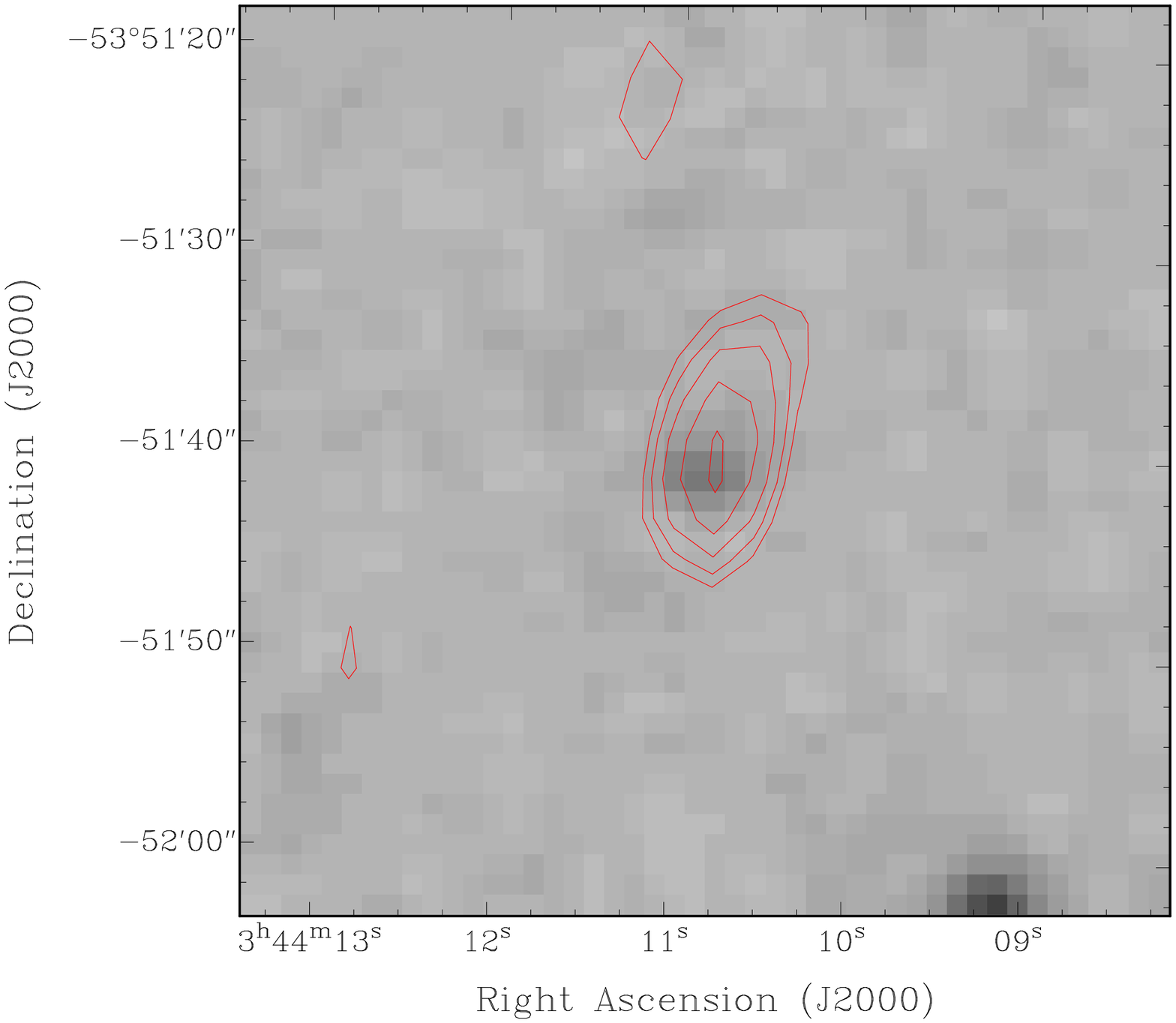}\includegraphics[angle=-90]{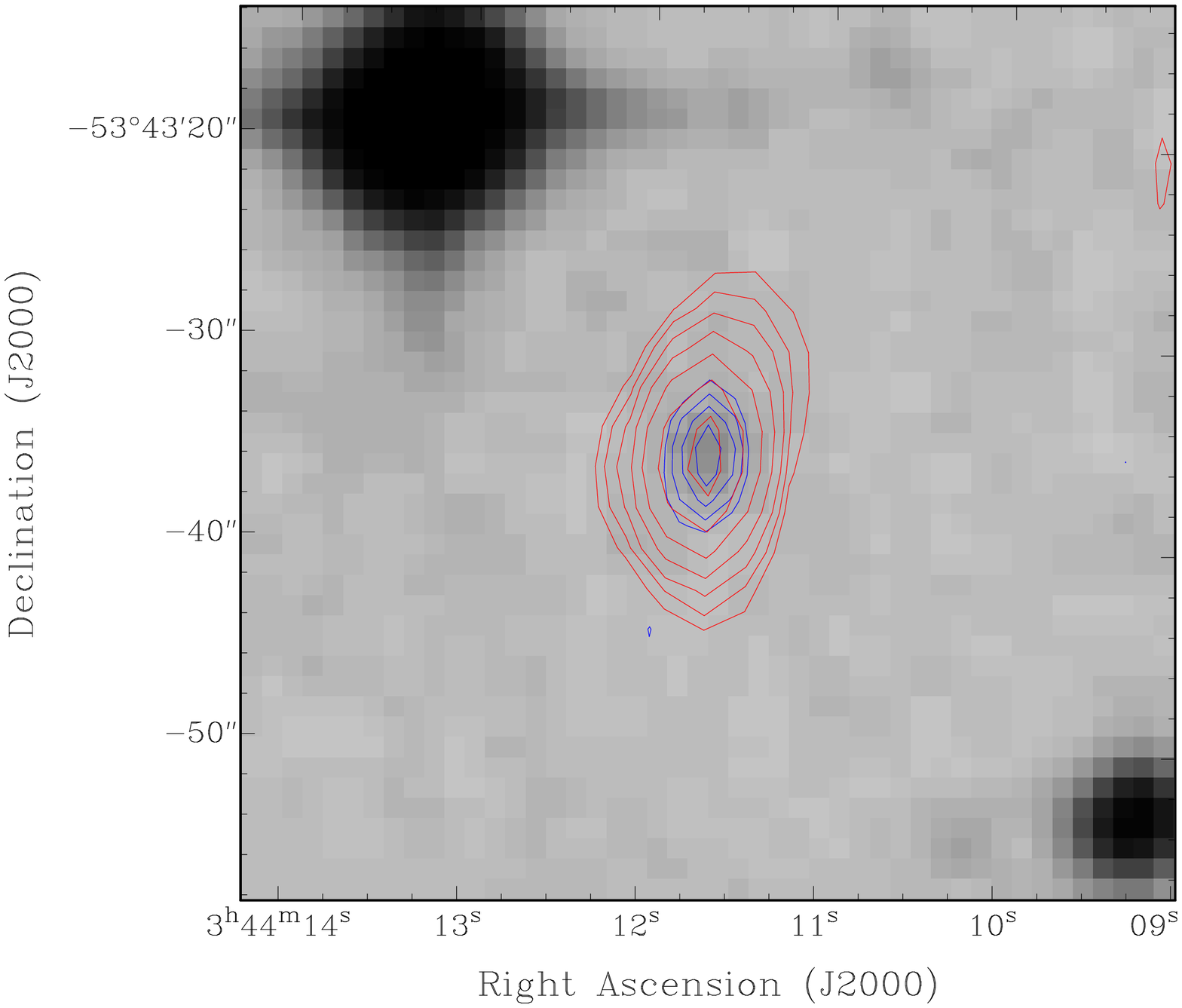}\includegraphics[angle=-90]{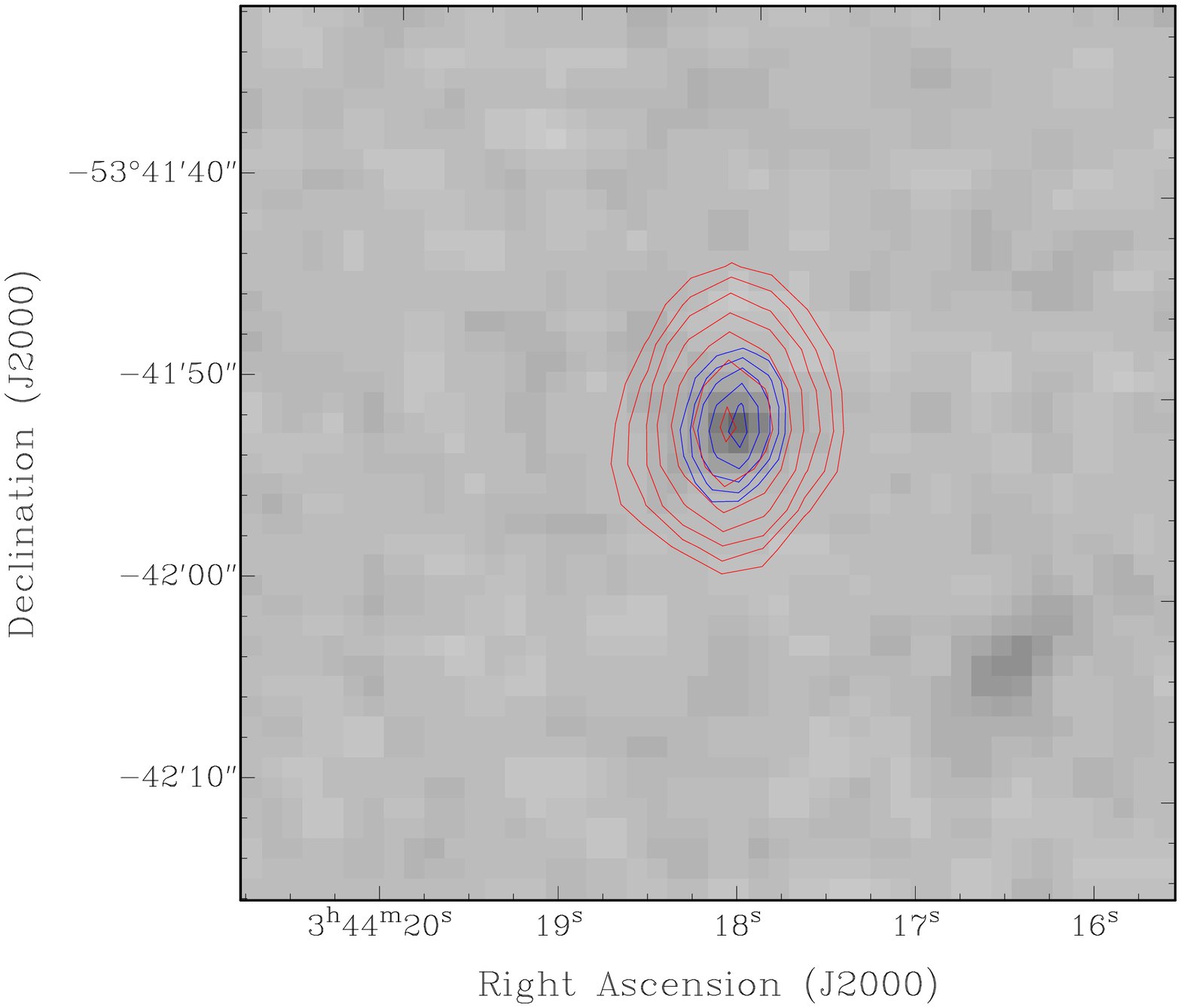}\includegraphics[angle=-90]{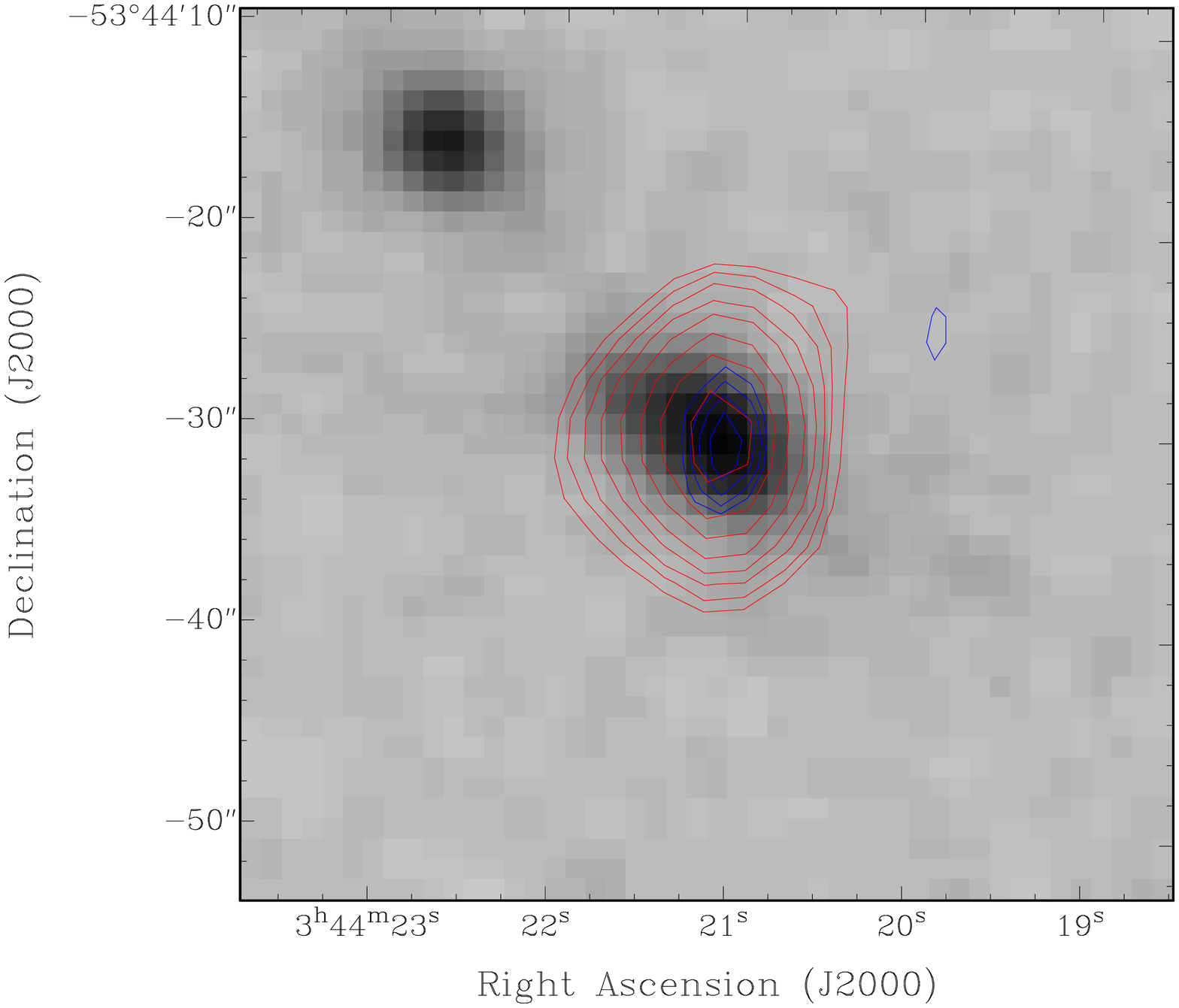}\includegraphics[angle=-90]{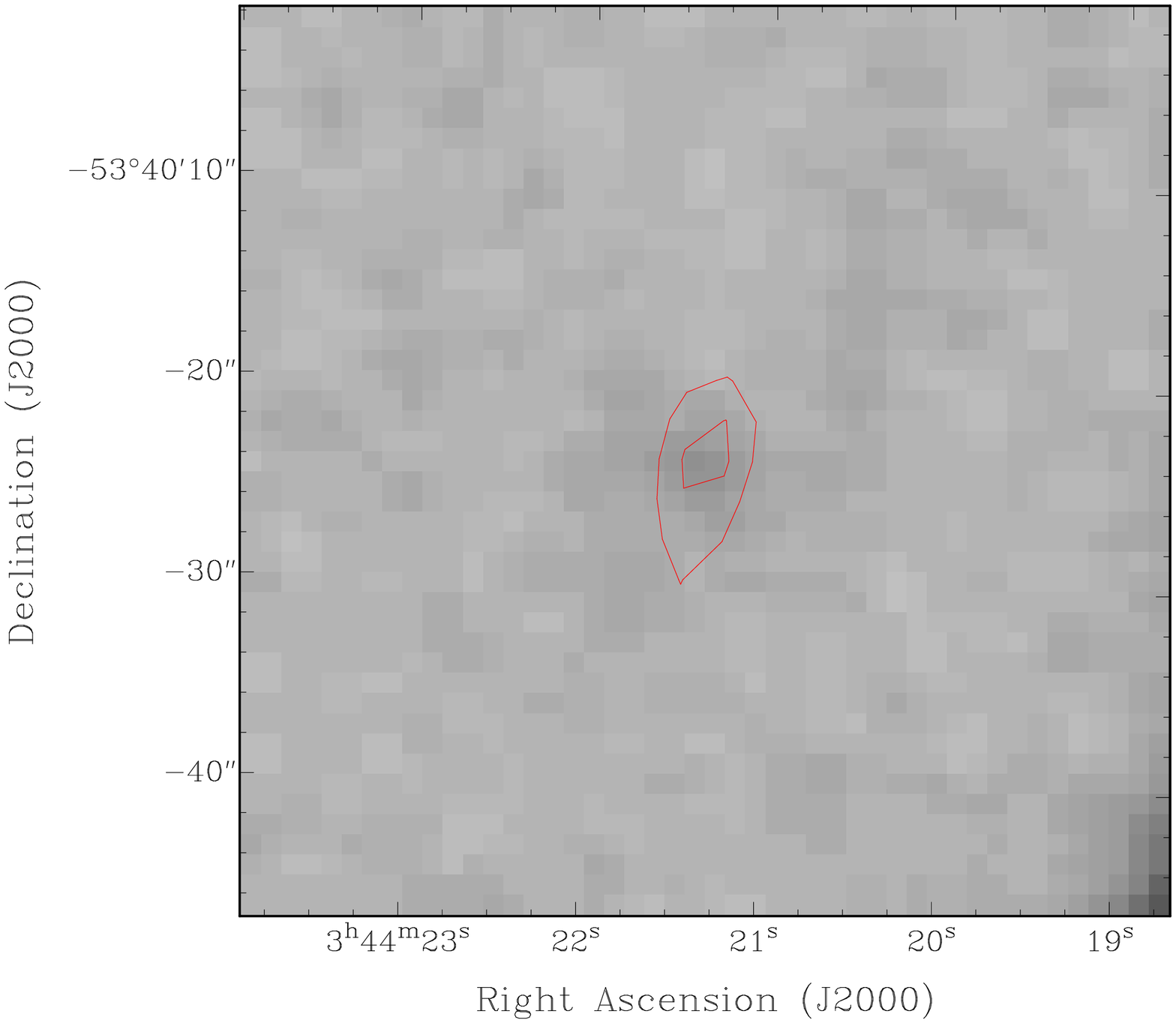}\includegraphics[angle=-90]{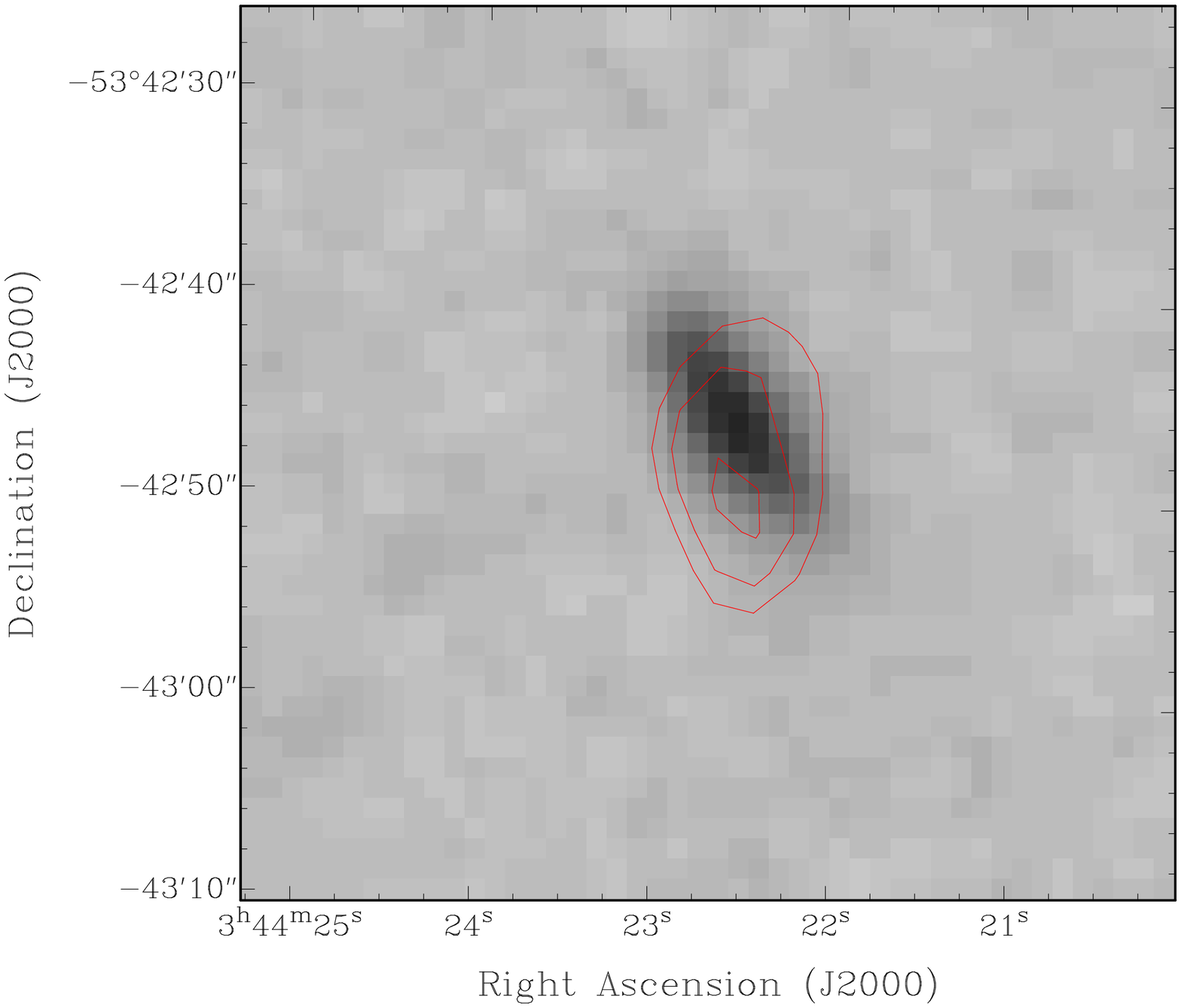}}
\resizebox{\hsize}{!}{\includegraphics[angle=-90]{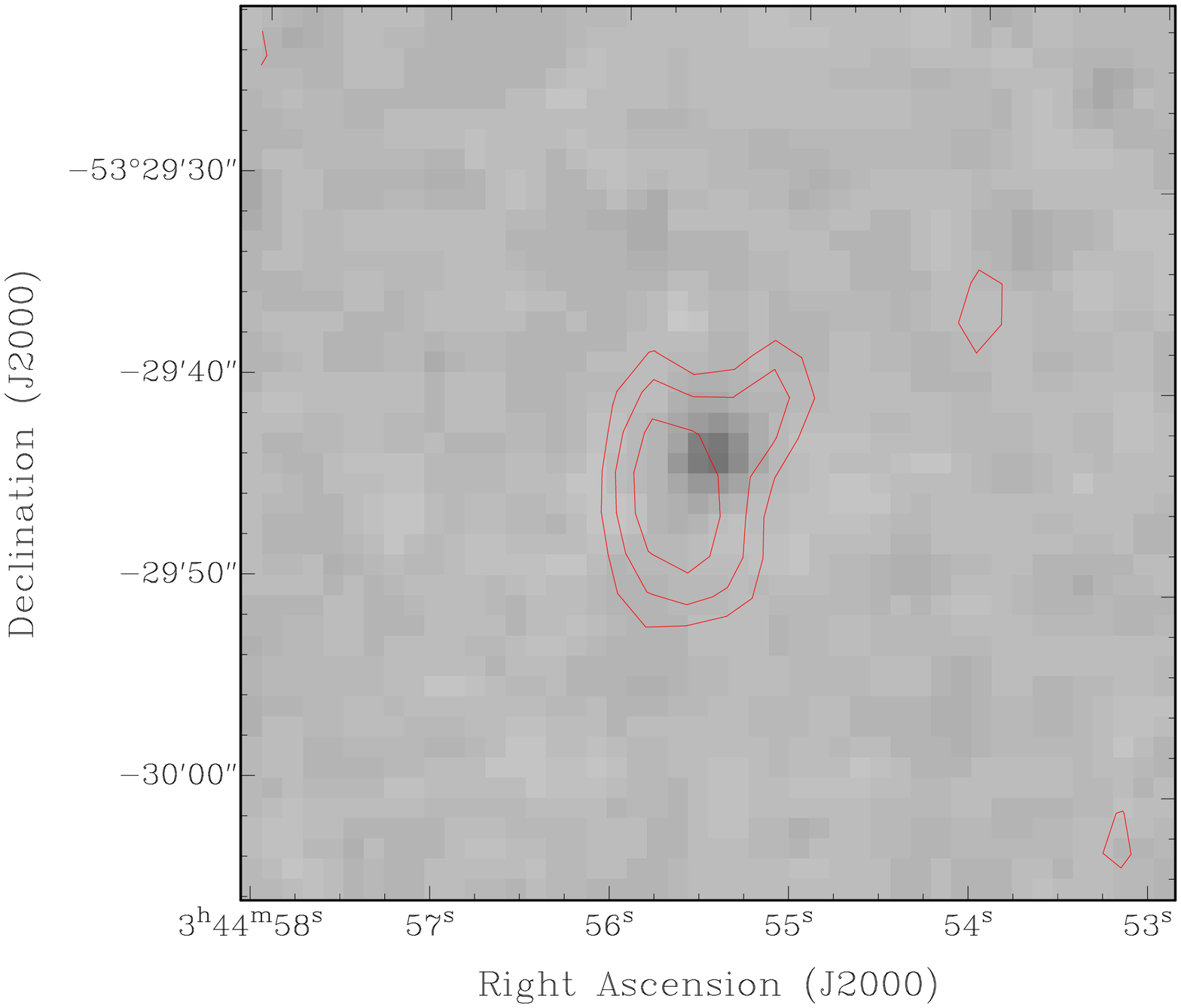}\includegraphics[angle=-90]{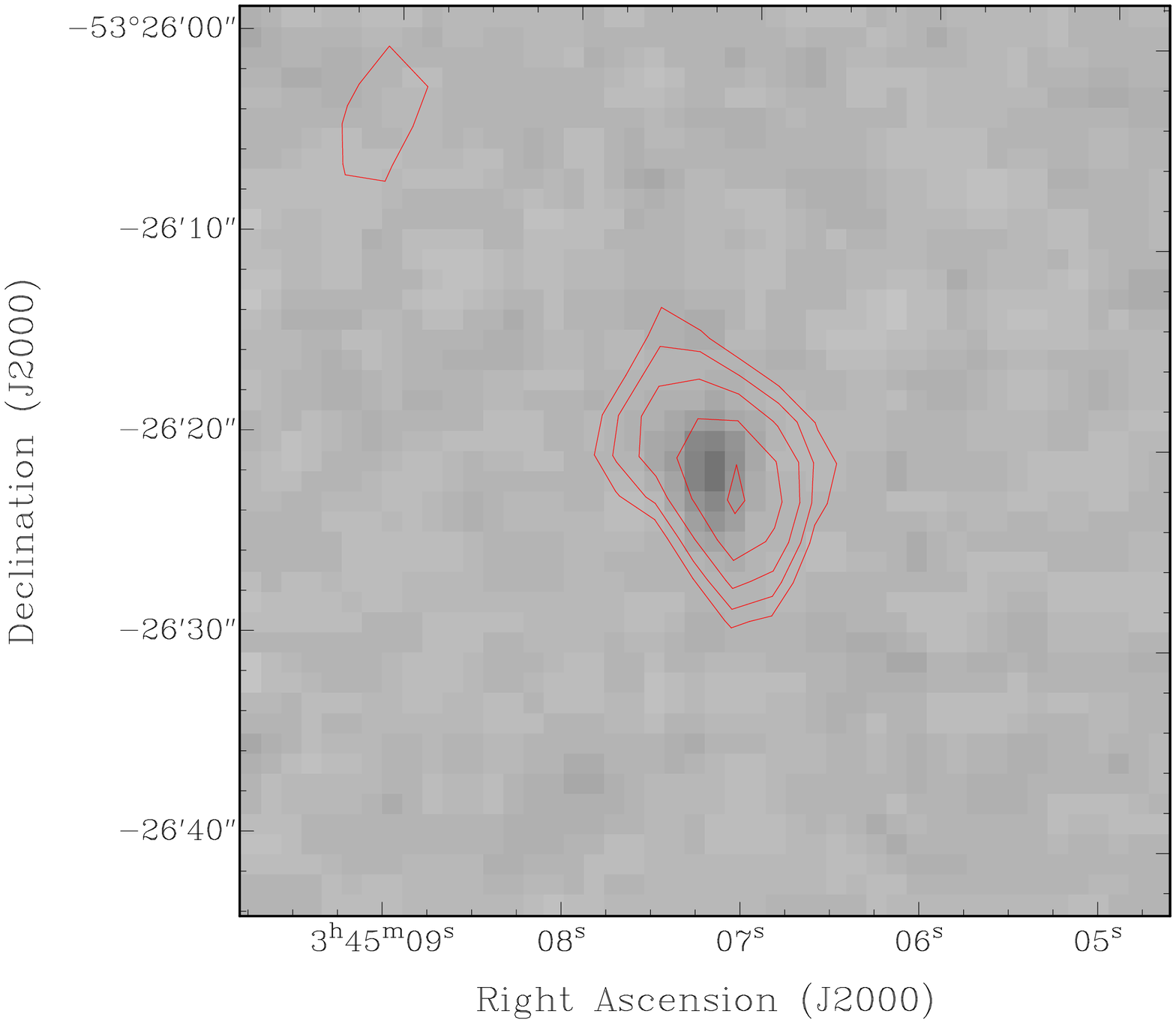}\includegraphics[angle=-90]{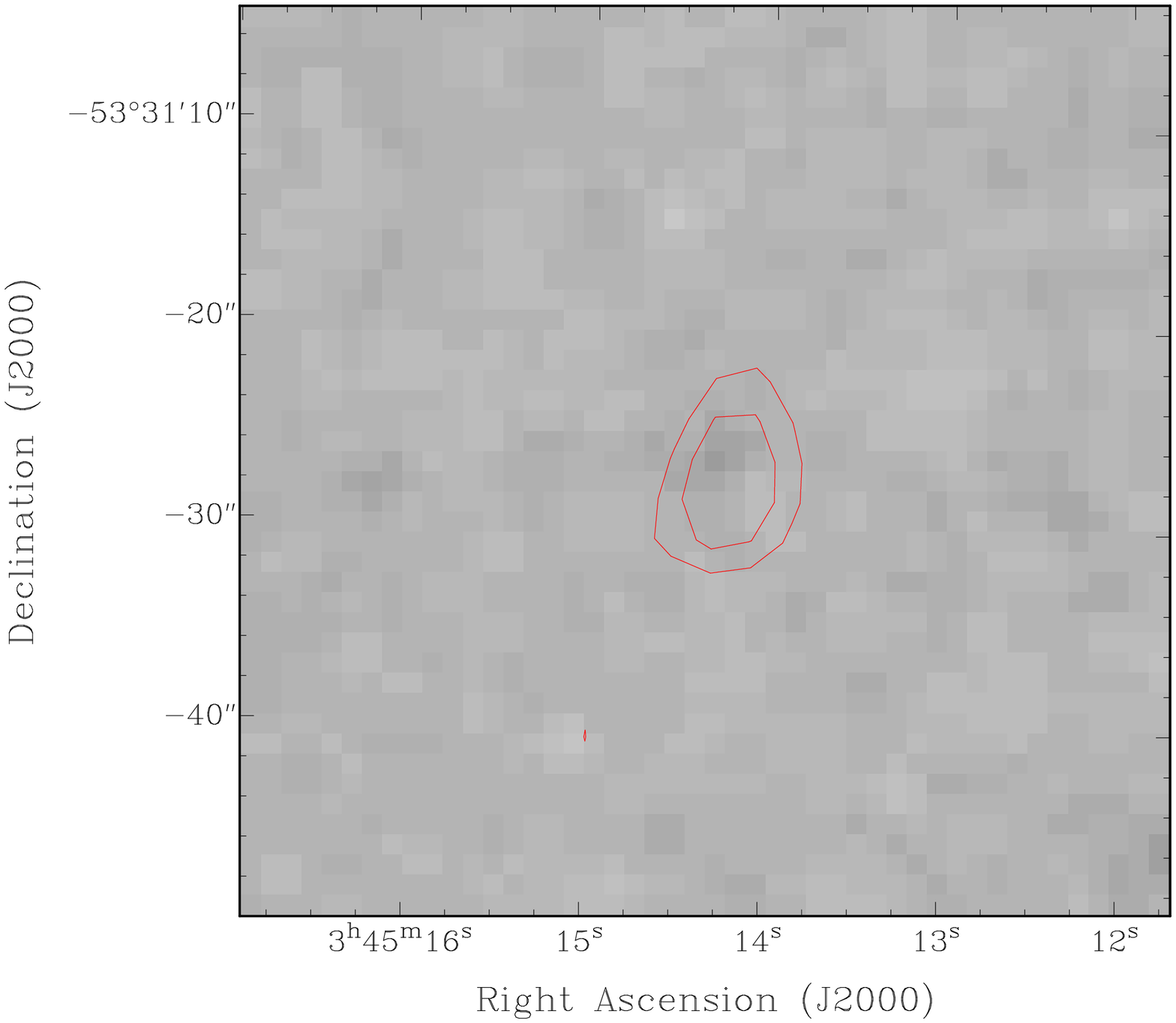}\includegraphics[angle=-90]{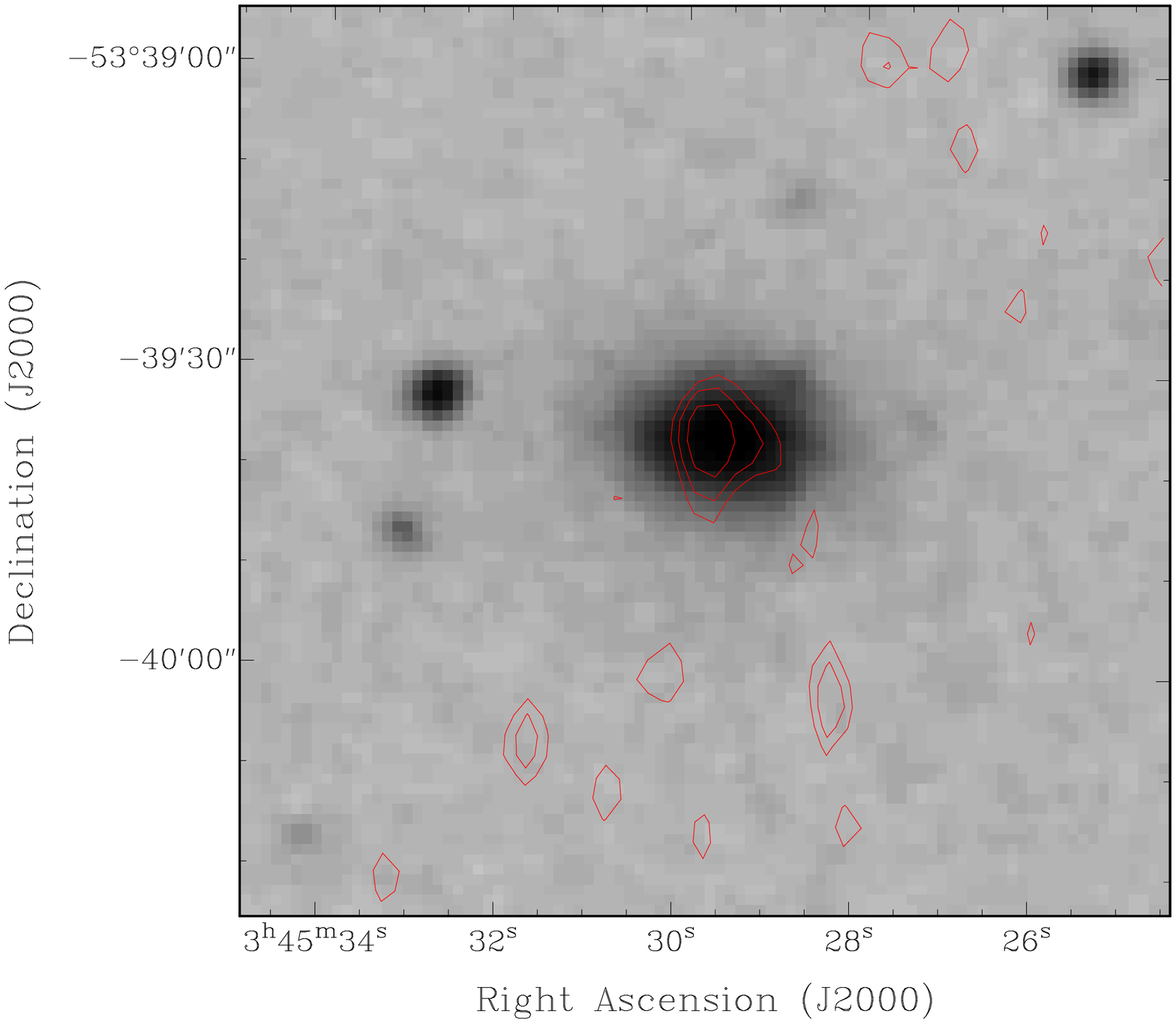}\includegraphics[angle=-90]{mjh_blank.eps}\includegraphics[angle=-90, width=22cm]{mjh_blank.eps}}
\vspace{-1.6cm}
\caption{Postage stamps of the 52 radio sources with optical counterparts detected within $23\farcm05$ of 03 42 53.00 -53 37 43.0. Red contours show the 1.4 GHz emission and blue the 2.5 GHz emission. Contours begin at
0.14 mJy beam$^{-1}$ and increase at intervals of $\sqrt 2$, except for the 3rd, 4th \& 5th images in the first row and the 4th image in row 9 which begin at 
0.2 mJy beam$^{-1}$ and the 2nd image in the first row and the first image in the 3rd row which begin at 0.28 mJy beam$^{-1}$. 
Images are presented in the order given in Table 
%\ref{tab:a}
3.}
\label{fig:rad_opt}
\end{figure*}

\section{Radio Sources}
Radio sources were found via two independent methods. In addition to searching for sources by eye, we used
a modified version of the automatic source detection algorithm ``Duchamp'' created by Matthew Whiting over 
the dataset. In both methods sources were considered to be detected if there was a total of 4 or more 
congruent pixels above 3 times the local RMS noise.

One of the difficulties in source determination is how to deal with sources which have multiple components. We have assumed that
unless sources have contiguous pixels above the local RMS noise threshold that they correspond to separate sources.

A total of 110 sources have been detected in the 1.4 GHz image, of these
30 are also detected at 2.5 GHz, 62 sources have optical counterparts, while 48 do not. 
Of the 110 radio sources, 88 were detected in the area over which the beam attenuation is
minimal and a radio luminosity function can be calculated; 52 have optical counterparts within 3 arcseconds
of the peak radio emission these are shown in Figure 
%\ref{fig:rad_opt}  
3 and the remaining
and 36, shown in Figure 
%\ref{fig:rad_only}
4, have no optical counterpart within 3 arcseconds
of the peak radio emission. Properties of the radio galaxies are given in Tables 3, 4 and 5.
%\ref{tab:a}, \ref{tab:b} and \ref{tab:c}.

In Figures 
%\ref{fig:rad_opt} and \ref{fig:rad_only} 
3 and 4 we show the Digitized Sky Survey (DSS) image overlaid with the 
radio contours for the 88 sources within the area of interest (the remaining 22 sources found in the image
but outside of the region are not shown). 56\% of the total 110 sources and 59\% of those within
$23\farcm05$ of the centre have optical counterparts detected in the DSS (b$_{\rm{J}} \leq$ 22.7) within 3 arcseconds of
the brightest radio position, or in the case of extended sources, within 3 arcseconds of the radio 
core of the object. Venturi et al.~(2000) found
26\% of radio sources in Shapley had optical counterparts brighter than b$_{\rm{J}}$= 19.5. At this limit we obtain
31\% for both the whole sample and the reduced subsample corresponding to the area with little attenuation.

\begin{table}
   \begin{center}

 \begin{tabular}{l|l|l|l|l}
        \hline

  Radio            & Coordinates  &  S$_{1.4}$ & S$_{2.5}$ & $\alpha^{1.4}_{2.5}$ \\
  RA$_{\rm{J}2000}$     & DEC$_{\rm{J}2000}$&  mJy           & mJy           & \\
         \hline 								
03 40 59.16&	-53 34 19.7&	0.84 $\pm$ 0.03 &		 &\\				
03 41 11.17&	-53 42 50.7&	0.34 $\pm$ 0.03 &       	 &\\				
03 41 30.82&	-53 48 07.6&	0.69 $\pm$ 0.03 & 2.14 $\pm$ 0.04& 2.10\\				
03 41 37.60&	-53 48 00.1&	0.63 $\pm$ 0.03 &		 &\\				
03 41 40.55&	-53 39 38.9&	1.12 $\pm$ 0.03 &		 &\\			
03 41 41.23&	-53 23 35.6&	1.44 $\pm$ 0.03 &		 &\\				
03 41 41.44&	-53 47 58.3&	0.46 $\pm$ 0.03 & 0.85 $\pm$ 0.03&1.14\\				
03 41 48.26&	-53 19 23.1&	6.33 $\pm$ 0.07 &		 &\\				
03 42 01.45&	-53 29 45.5&	0.25 $\pm$ 0.03 &		 &\\				
03 42 07.71&	-53 30 35.9&	0.30 $\pm$ 0.03 &		 &\\				
03 42 07.74&	-53 16 29.3&	1.69 $\pm$ 0.03 &		 &\\
03 42 09.48&	-53 18 49.9&	1.45 $\pm$ 0.03 &		 &\\
03 42 19.43&	-53 26 11.7&	3.43 $\pm$ 0.05 & 1.75 $\pm$ 0.04& -1.25\\	
03 42 34.15&    -53 33 20.7&    1.62 $\pm$ 0.03 & 0.77 $\pm$ 0.03& -1.38\\
03 42 41.33&    -53 32 52.8&    0.95 $\pm$ 0.03 & 0.43 $\pm$ 0.03& -1.47\\
03 42 44.50&	-53 25 32.0&	0.44 $\pm$ 0.03 & 0.41 $\pm$ 0.03& -0.13\\				
03 42 51.66&	-53 25 16.0&	0.22 $\pm$ 0.03 &		 &\\				
03 42 51.87&	-53 44 56.7&	0.45 $\pm$ 0.03 & 0.29 $\pm$ 0.03& -0.82\\				
03 42 52.09&	-53 59 47.9&	2.33 $\pm$ 0.04 &		 &\\				
03 42 54.80&	-53 43 24.8&	0.26 $\pm$ 0.03 &		 &\\				
03 42 56.80&	-53 24 45.9&	0.48 $\pm$ 0.03 &		 &\\				
03 42 58.37&	-53 23 59.8&	0.34 $\pm$ 0.03 &		 &\\				
03 43 10.67&	-53 56 00.7&	1.30 $\pm$ 0.03 &		 &\\				
03 43 16.70&	-53 47 36.2&	0.53 $\pm$ 0.03 &		 &\\				
03 43 38.73&	-53 42 38.0&	0.30 $\pm$ 0.03 &		 &\\				
03 44 04.87&	-53 42 48.8&	0.74 $\pm$ 0.03 &		 &\\				
03 44 07.32&	-53 41 56.7&	0.32 $\pm$ 0.03 &		 &\\
03 44 29.90&	-53 37 33.2&	0.30 $\pm$ 0.03 &		 &\\				
03 44 32.32&	-53 53 17.5&	176.03 $\pm$ 1.76 & 84.73 $\pm$ 0.85& -1.36\\			
03 44 40.90&	-53 42 22.2&	1.75 $\pm$ 0.04 &		  &\\					
03 44 44.93&	-53 41 45.9&	21.45 $\pm$ 0.24 & 11.40 $\pm$ 0.13& -1.17\\%here%	
03 44 49.41&	-53 36 31.5&	0.64 $\pm$ 0.03 &		 &\\	
03 45 19.13&	-53 37 22.4&	0.77 $\pm$ 0.03 &		 &\\	
03 45 26.35&	-53 41 27.5&	57.89 $\pm$ 0.58 &	         &\\
03 45 27.46&	-53 33 57.3&	0.38 $\pm$ 0.03 &		 &\\	
03 45 36.86&	-53 40 20.2&	3.69 $\pm$ 0.05 &		 &\\	
        \hline

 \end{tabular}
 \caption{Details of radio sources detected within $23\farcm05$ of 03 42 53.00 -53 37 43.0 for which no optical counterpart is found.
Column 1 and 2 are the J2000 coordinates of the radio source; column 3 and 4 are the total flux at 1.4 GHz and 2.5 GHz 
respectively, and column 5 is the spectral index between 1.4 and 2.5 GHz.}
 \end{center}
 \label{tab:b}
\end{table}

Of the 110 sources detected the vast majority of 
sources are unresolved. There are 15 extended sources (13.7\%) of which only four are 
double radio galaxies. This compares well
with the ATCA observations of the A3558 complex in which around 11\% of the 263 radio galaxies found at
1.4 GHz and the same resolution (6 arcseconds) are extended (Venturi et al.~2000 \& 2001).

\begin{figure*}
\centering
\resizebox{\hsize}{!}{\includegraphics[angle=-90]{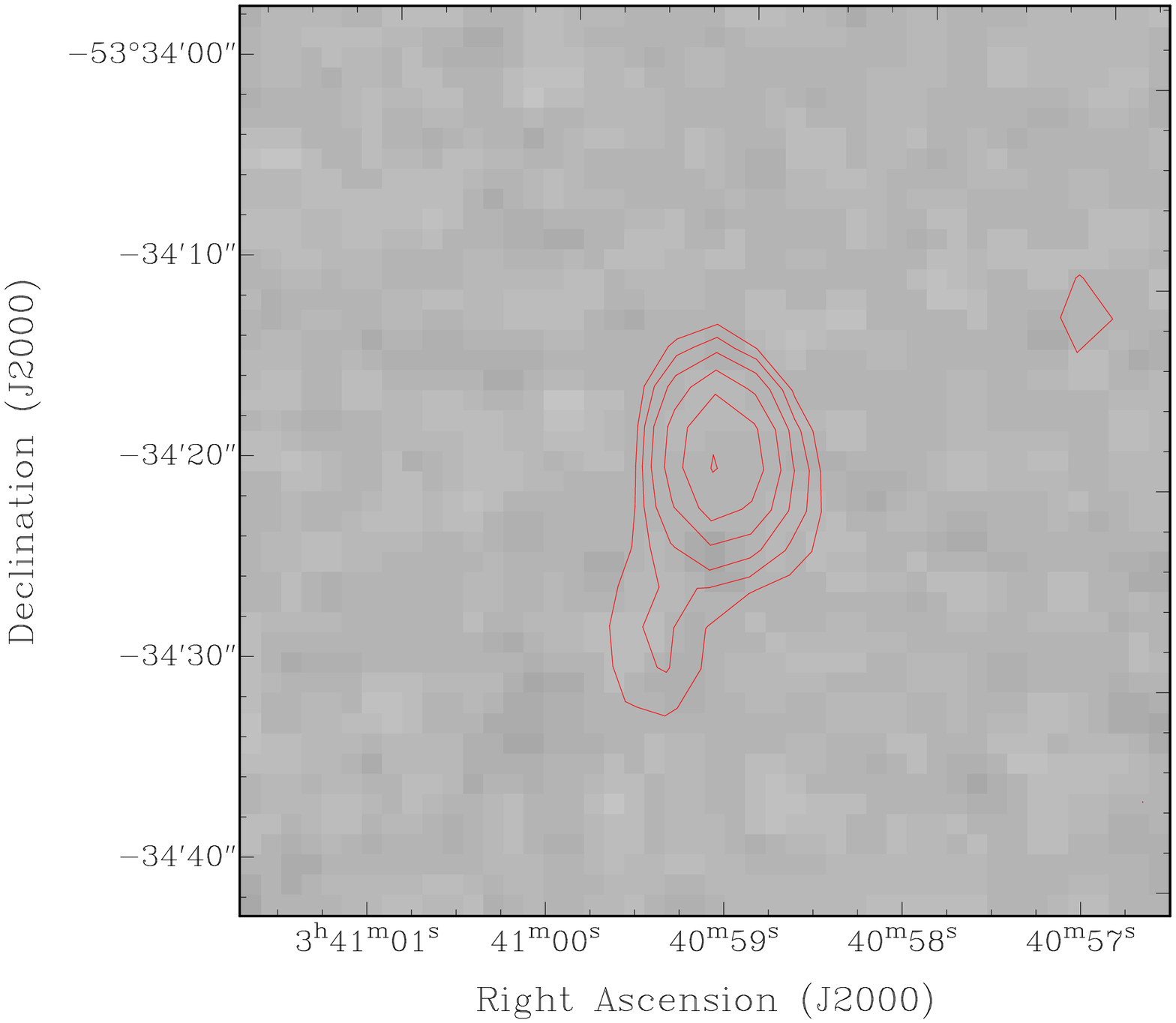}\includegraphics[angle=-90]{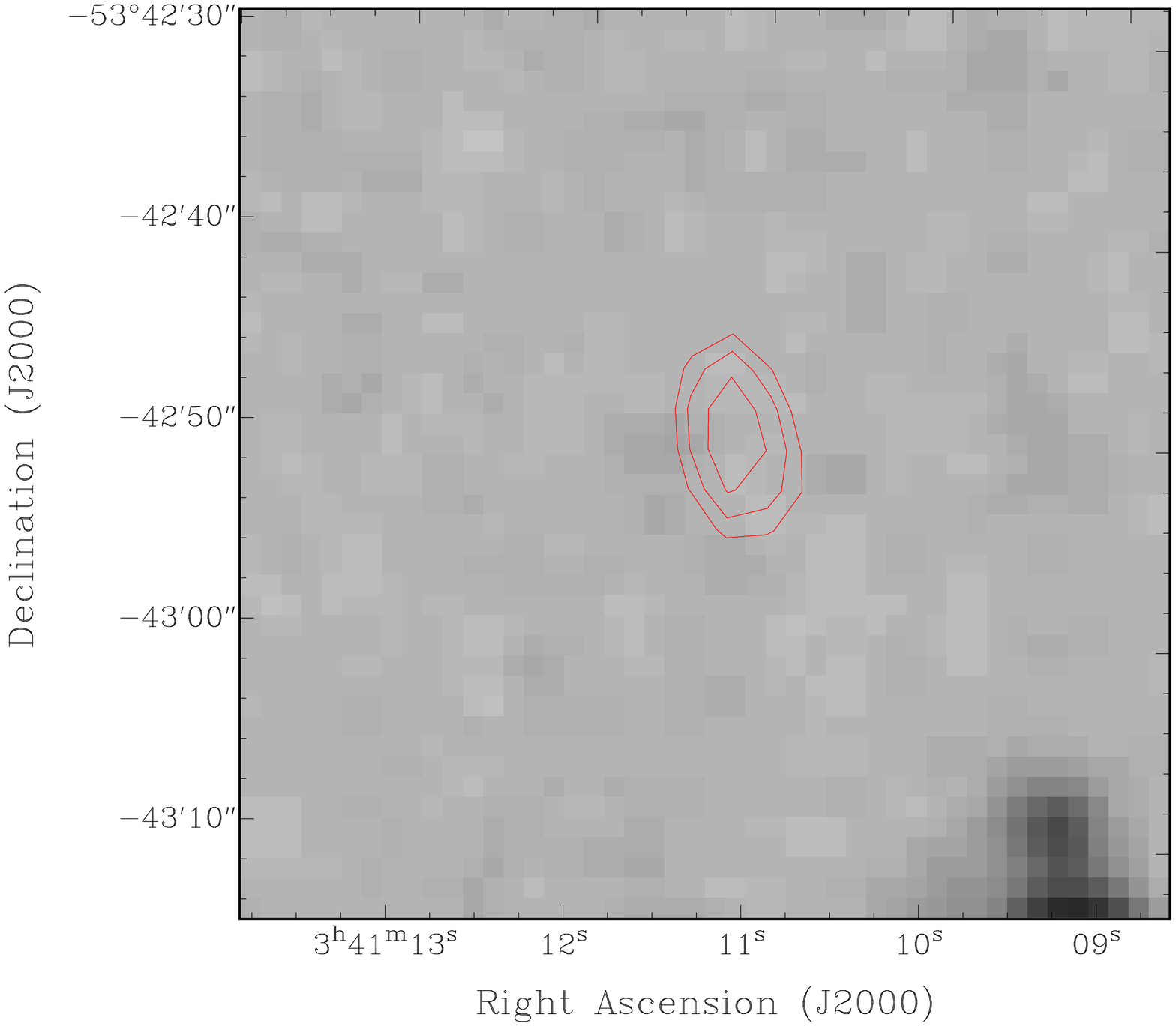}\includegraphics[angle=-90]{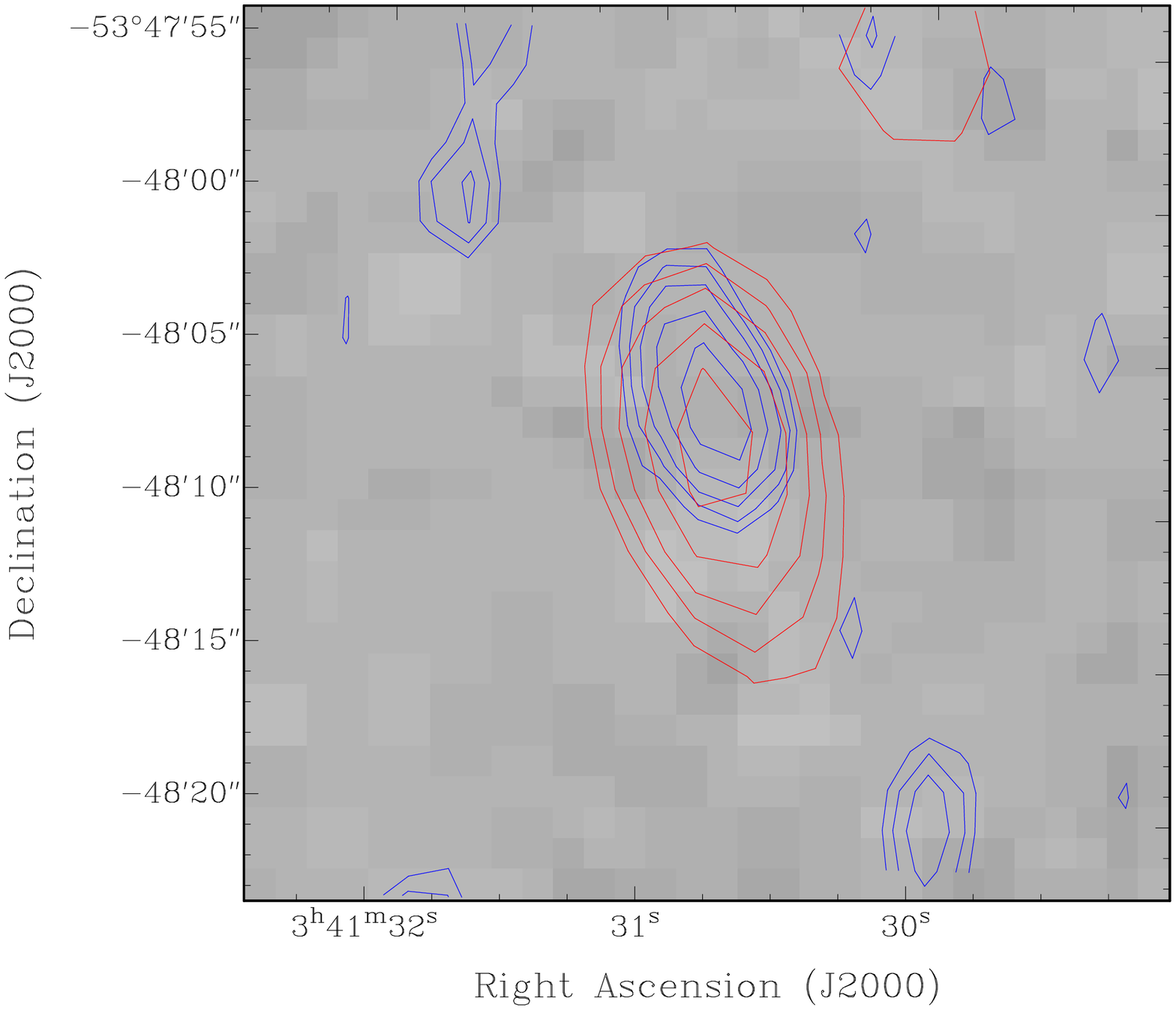}\includegraphics[angle=-90]{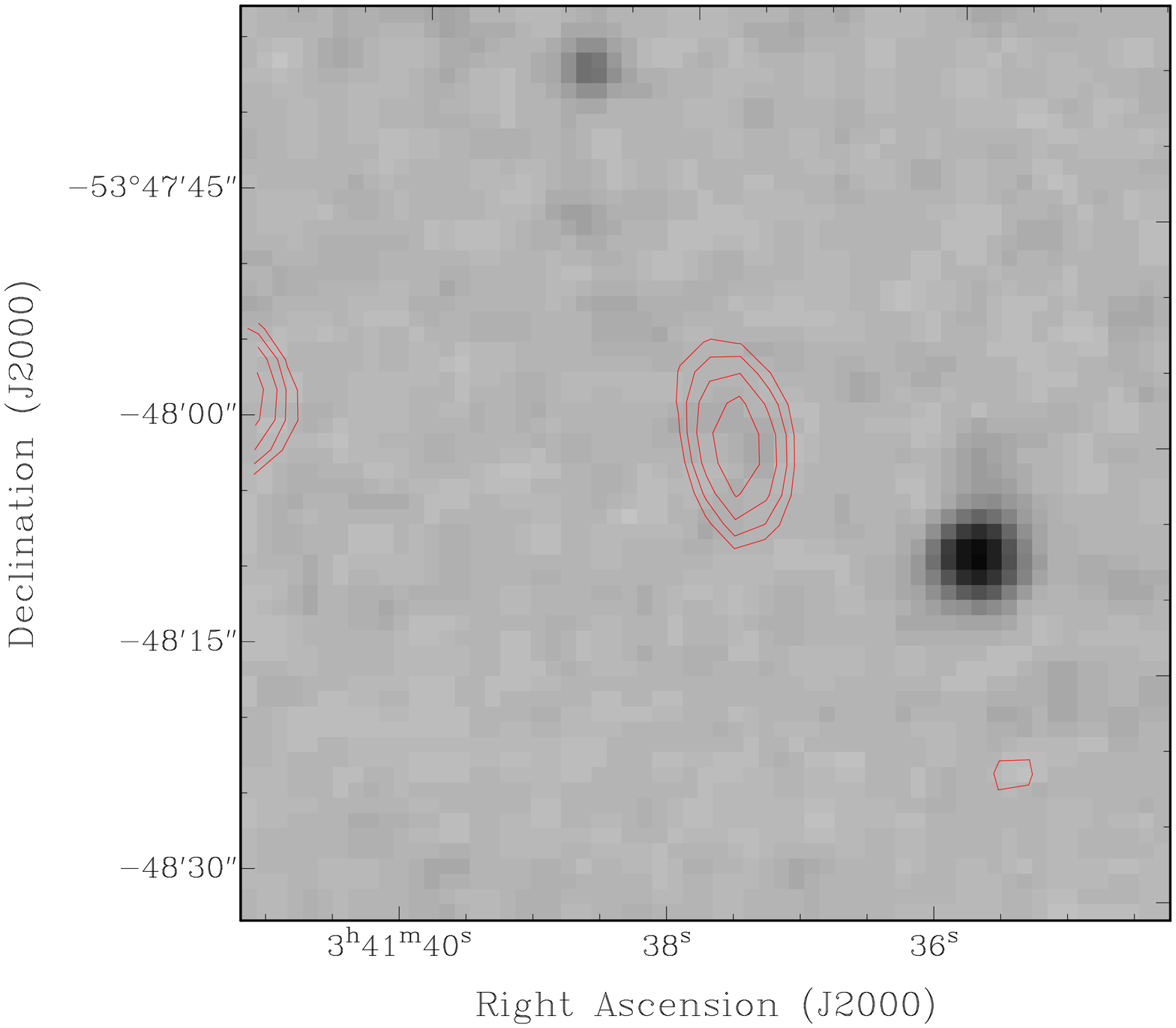}\includegraphics[angle=-90]{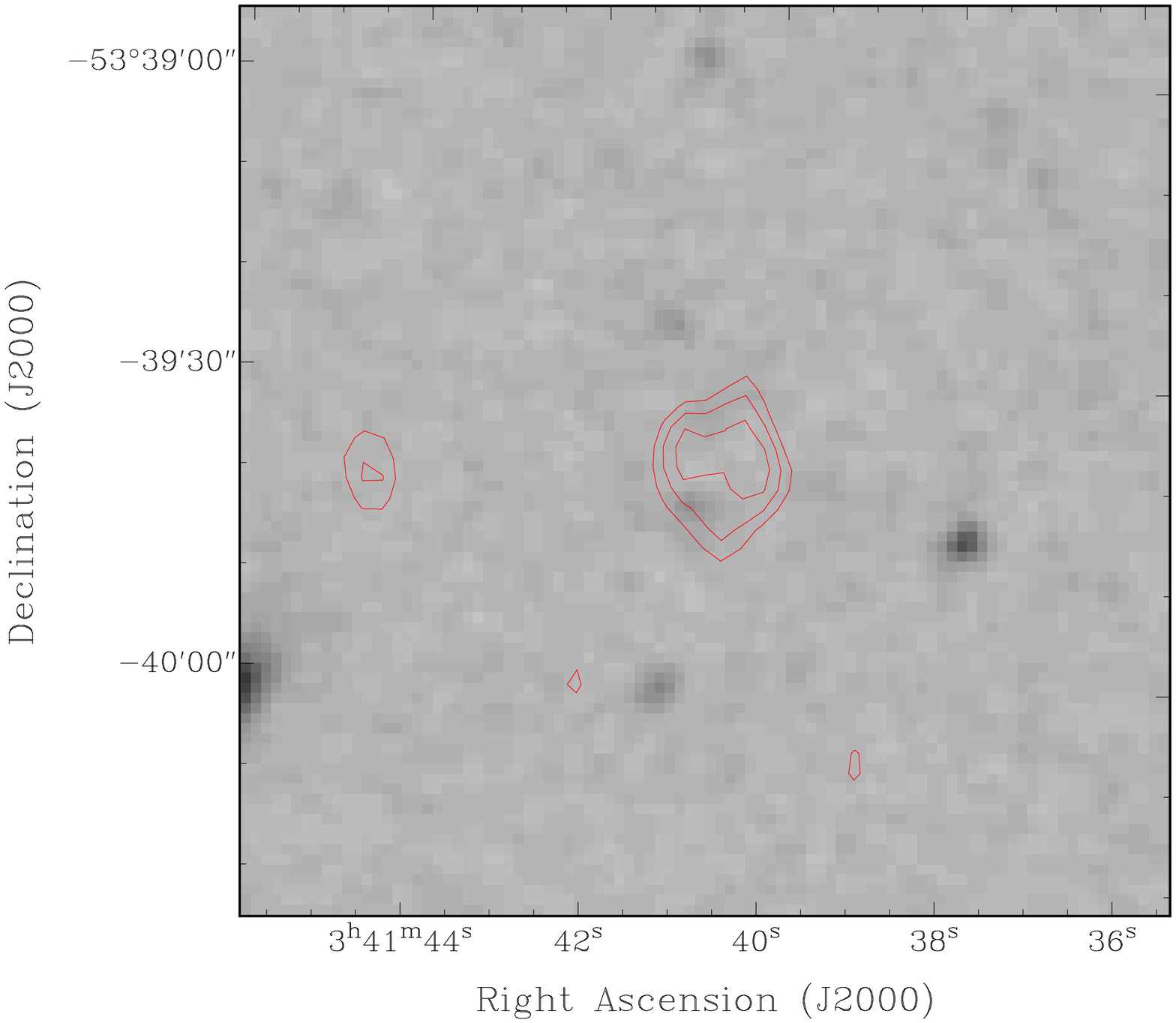}\includegraphics[angle=-90]{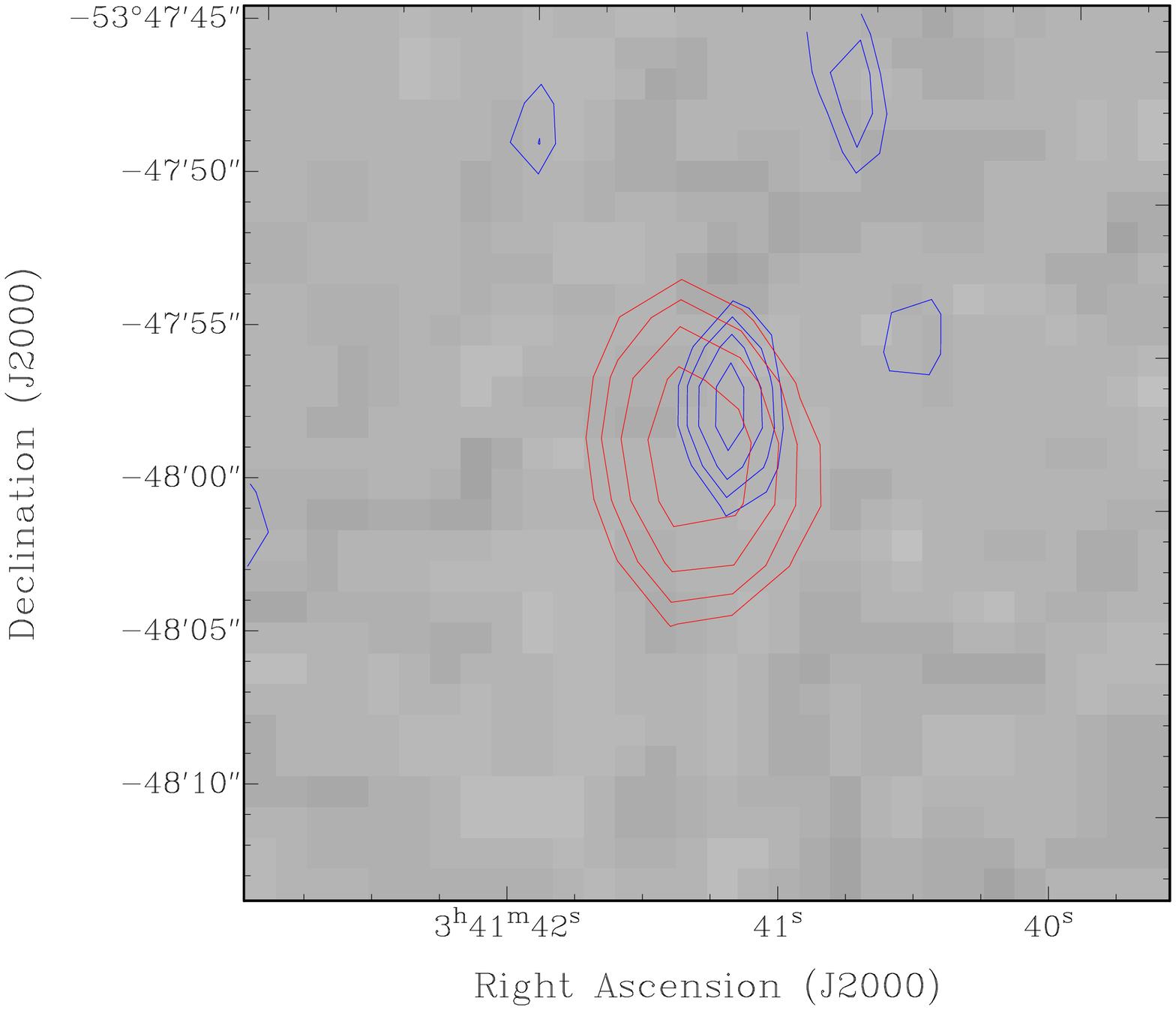}}
\centering
\resizebox{\hsize}{!}{\includegraphics[angle=-90]{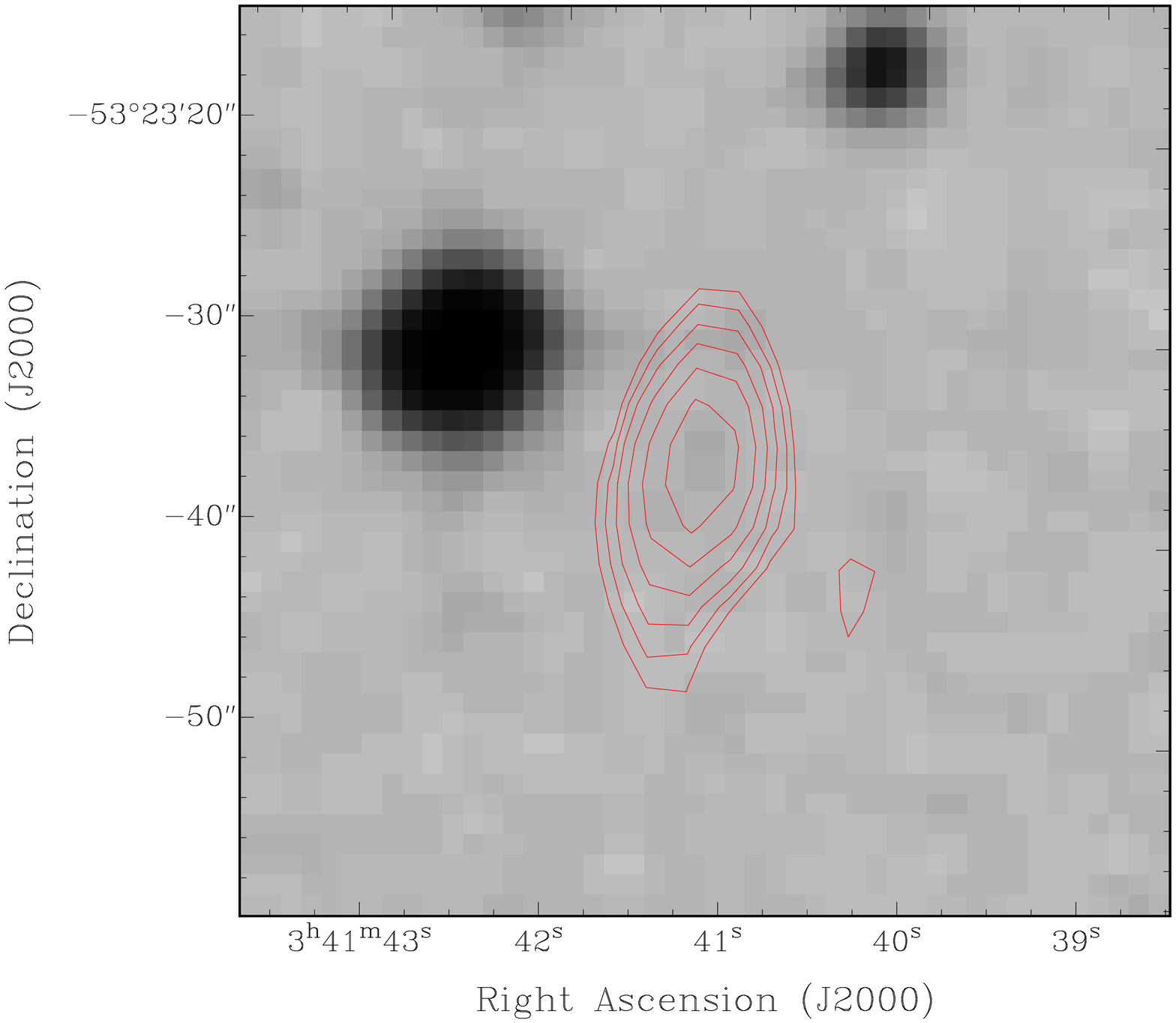}\includegraphics[angle=-90]{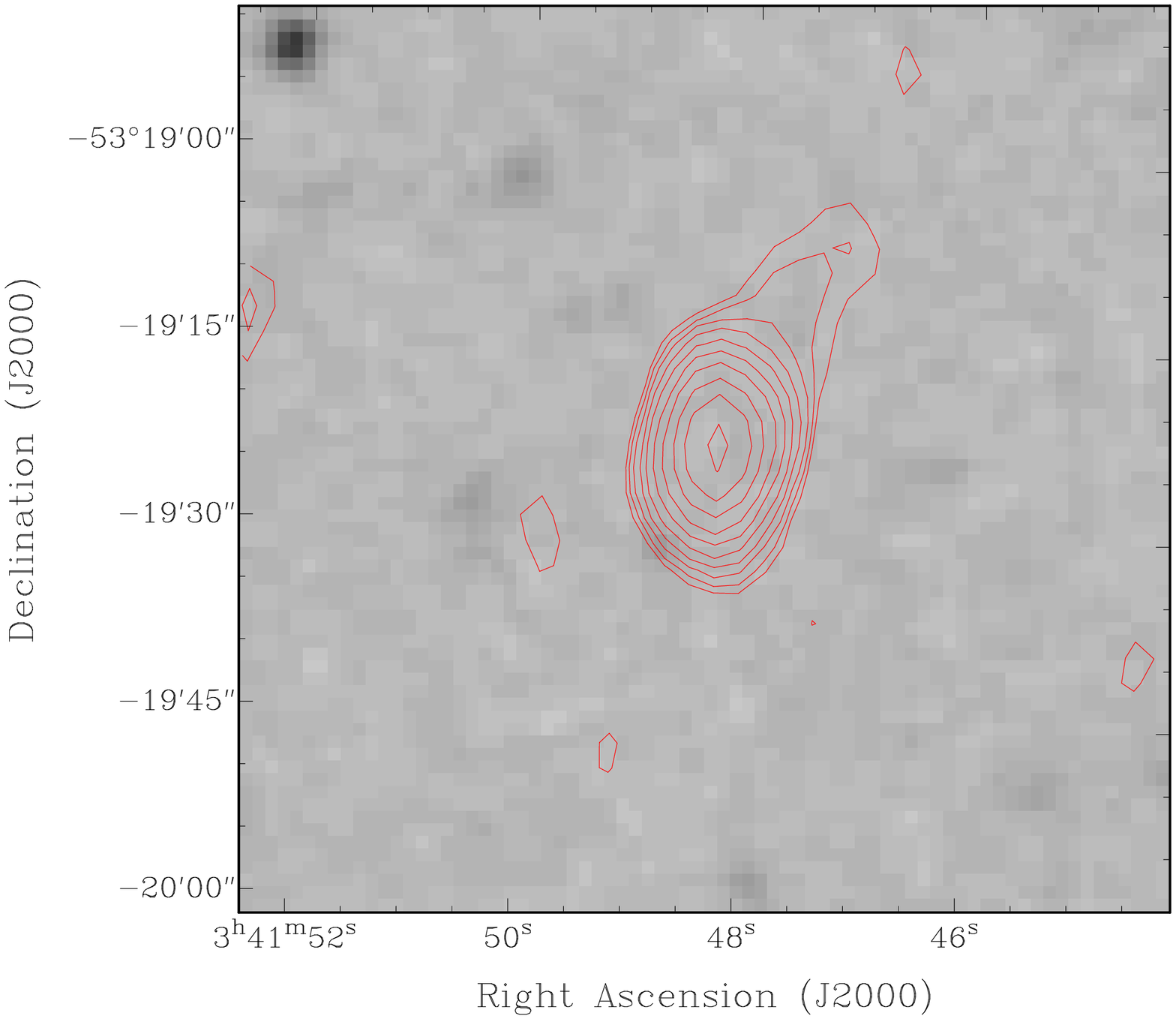}\includegraphics[angle=-90]{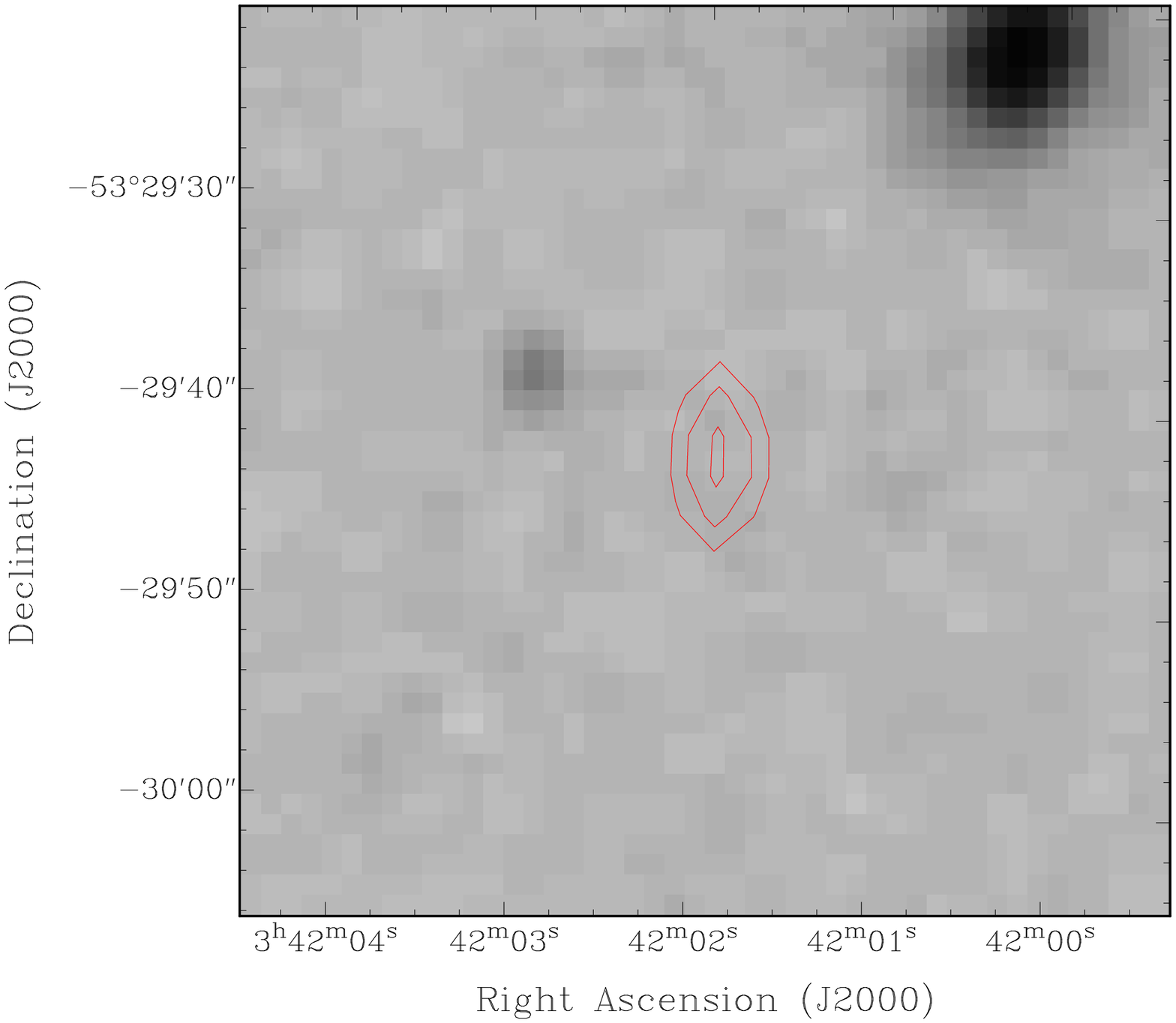}\includegraphics[angle=-90]{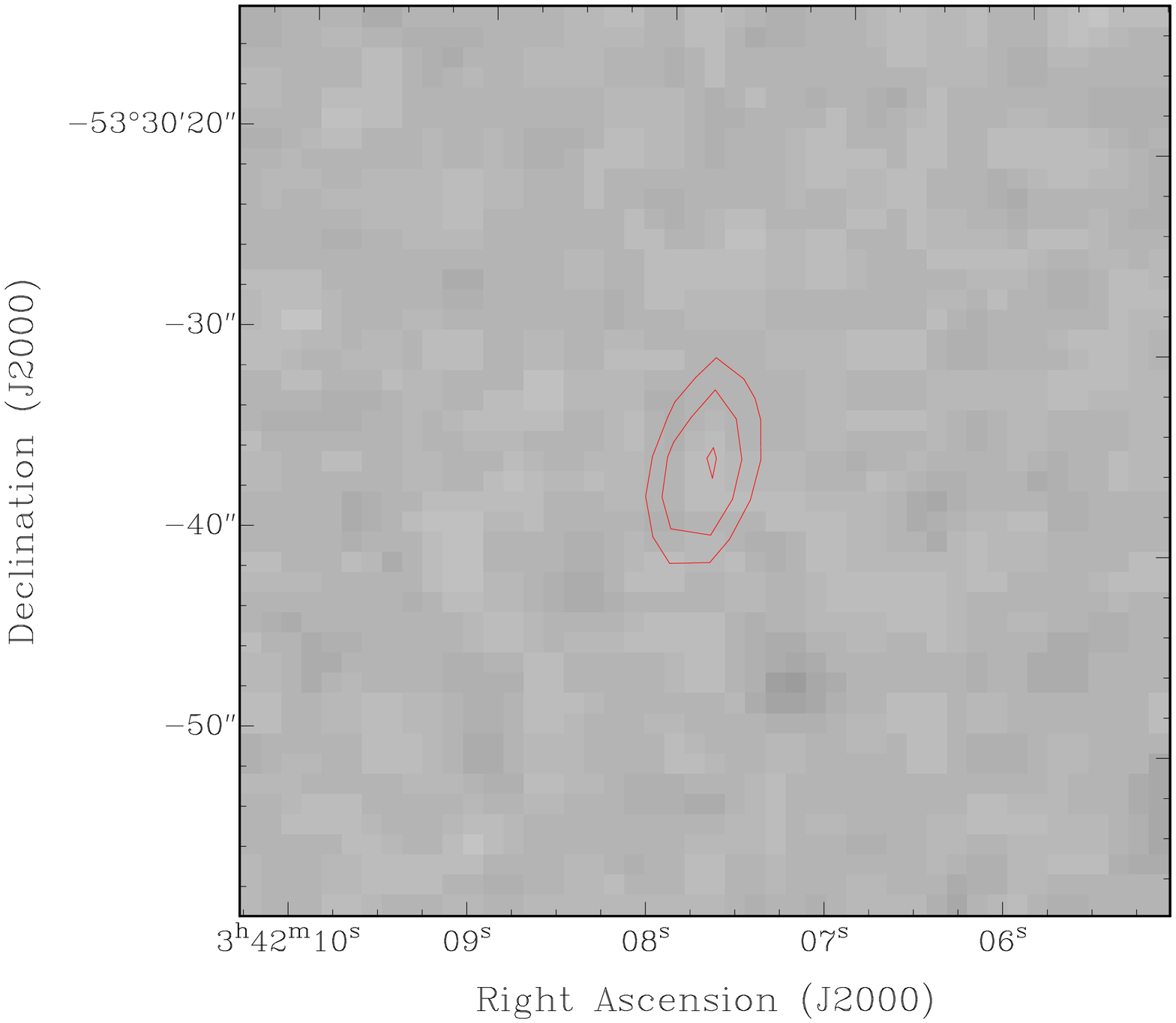}\includegraphics[angle=-90]{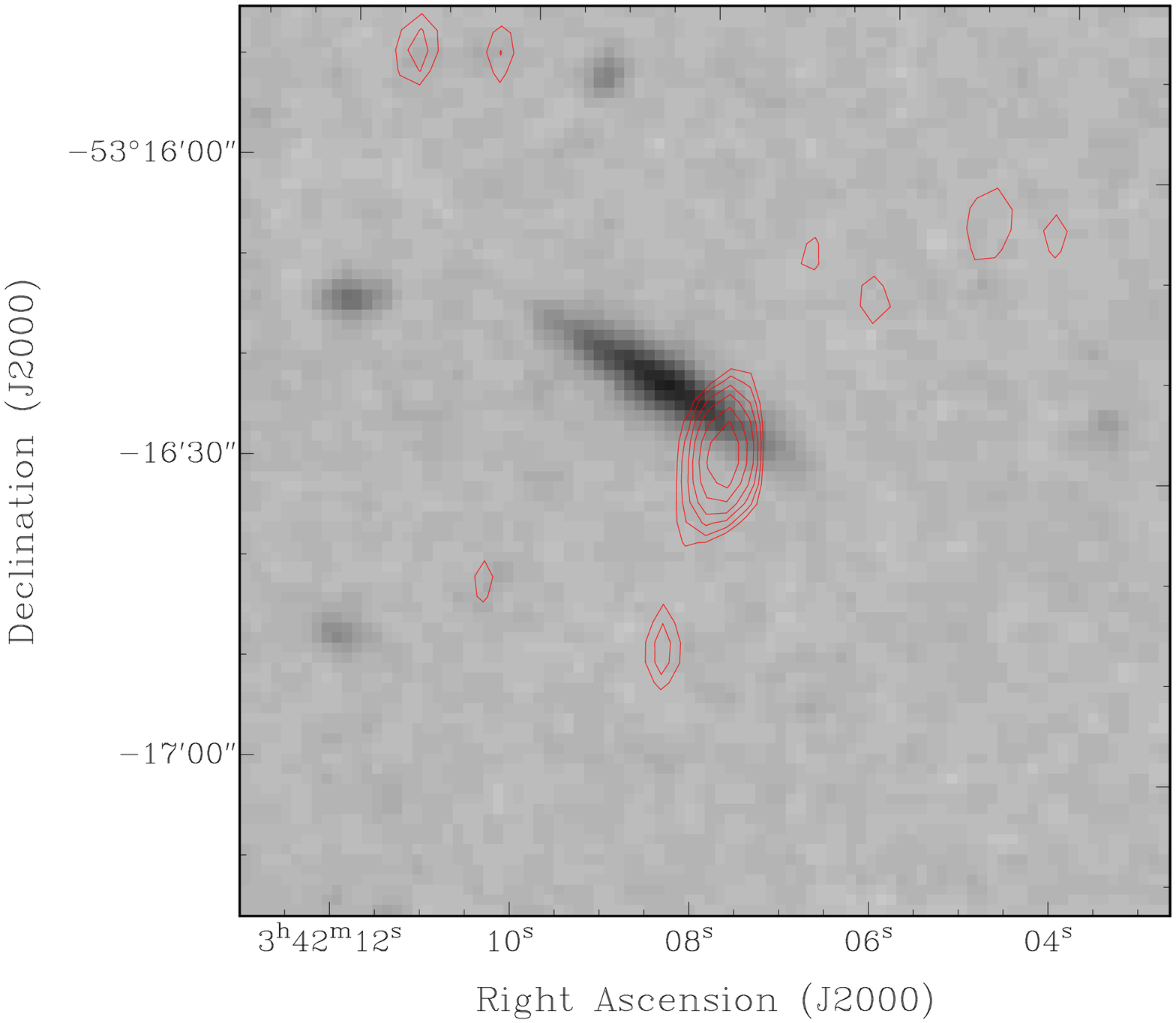}\includegraphics[angle=-90]{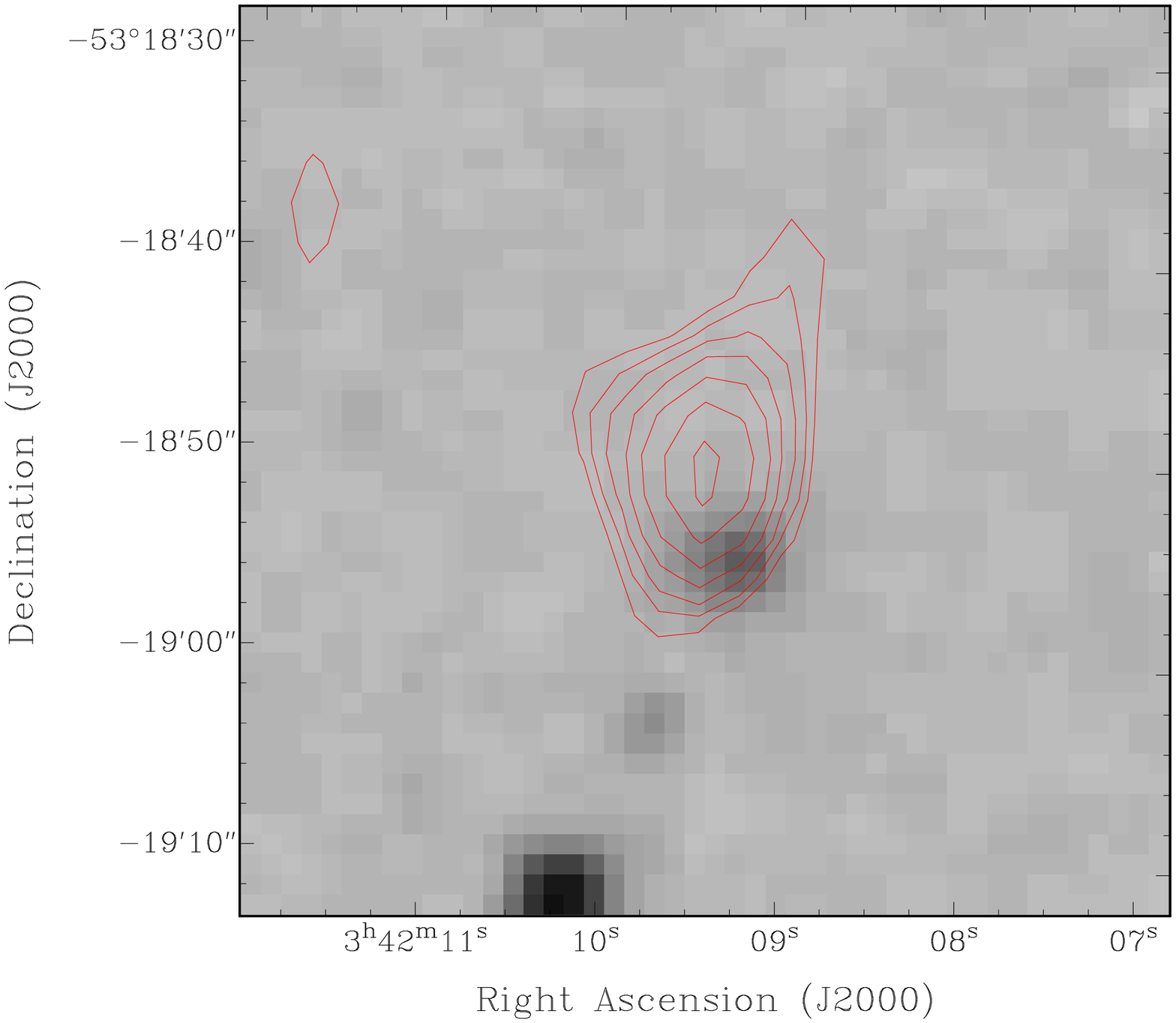}}
\resizebox{\hsize}{!}{\includegraphics[angle=-90]{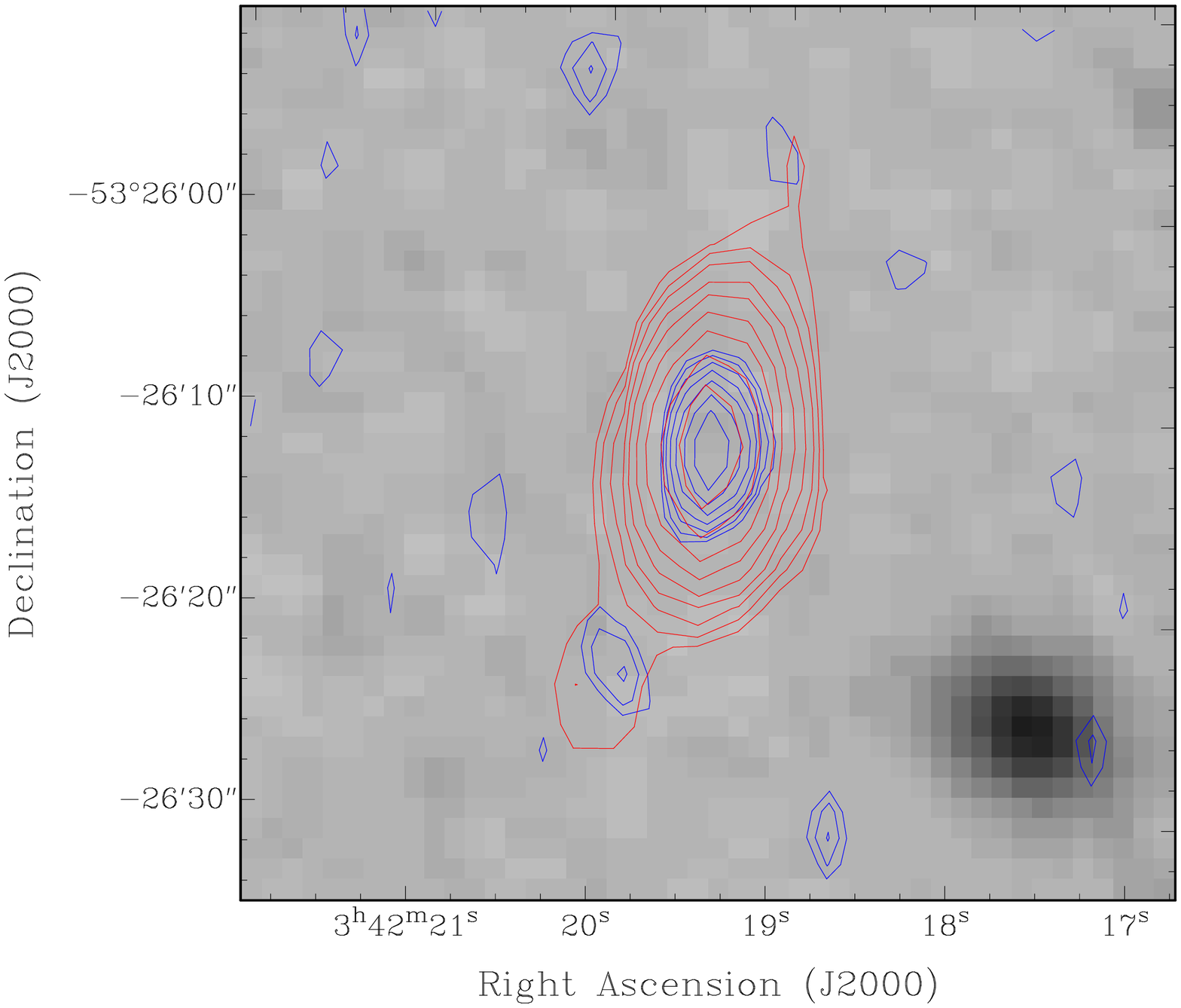}\includegraphics[angle=-90]{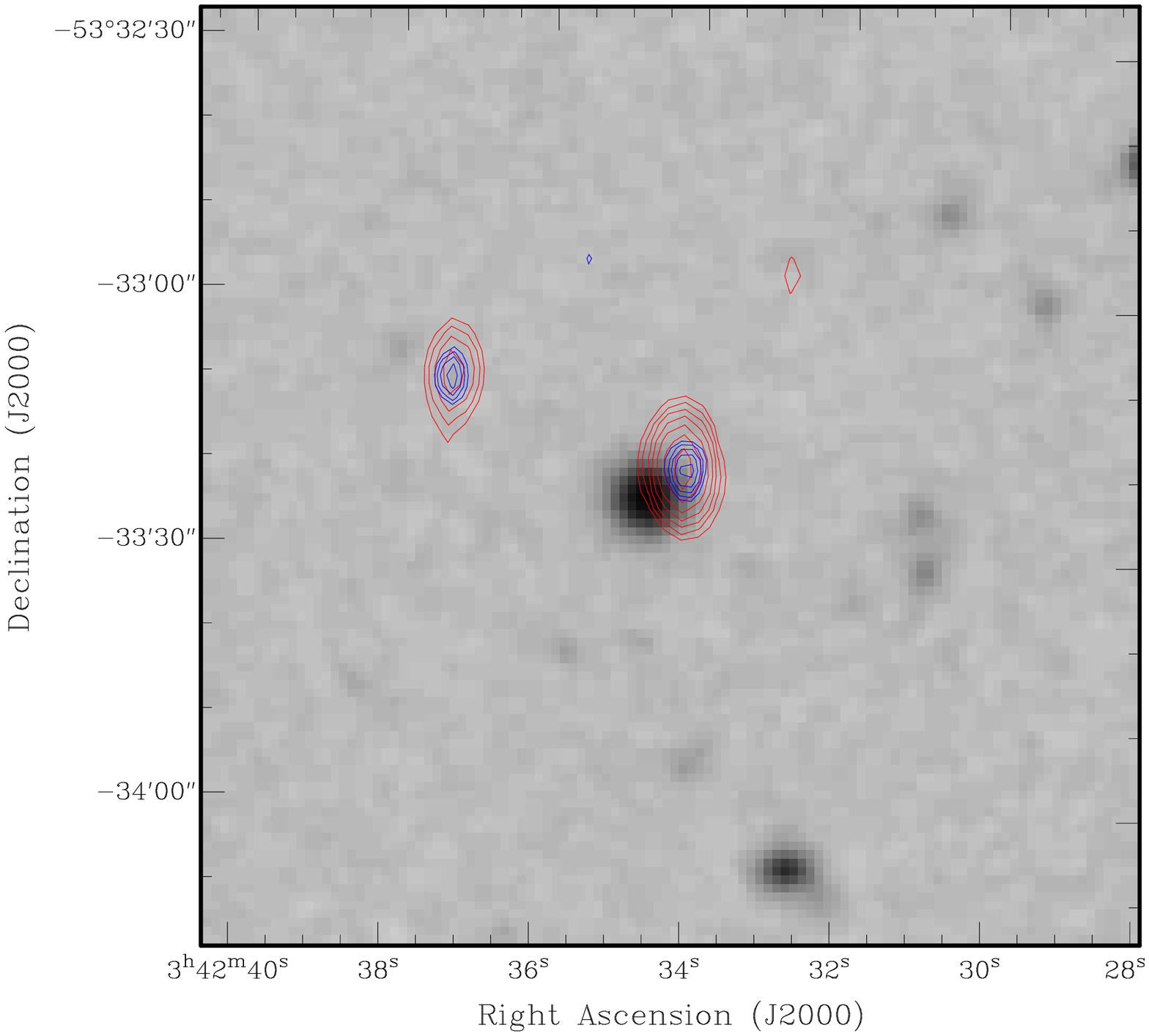}\includegraphics[angle=-90]{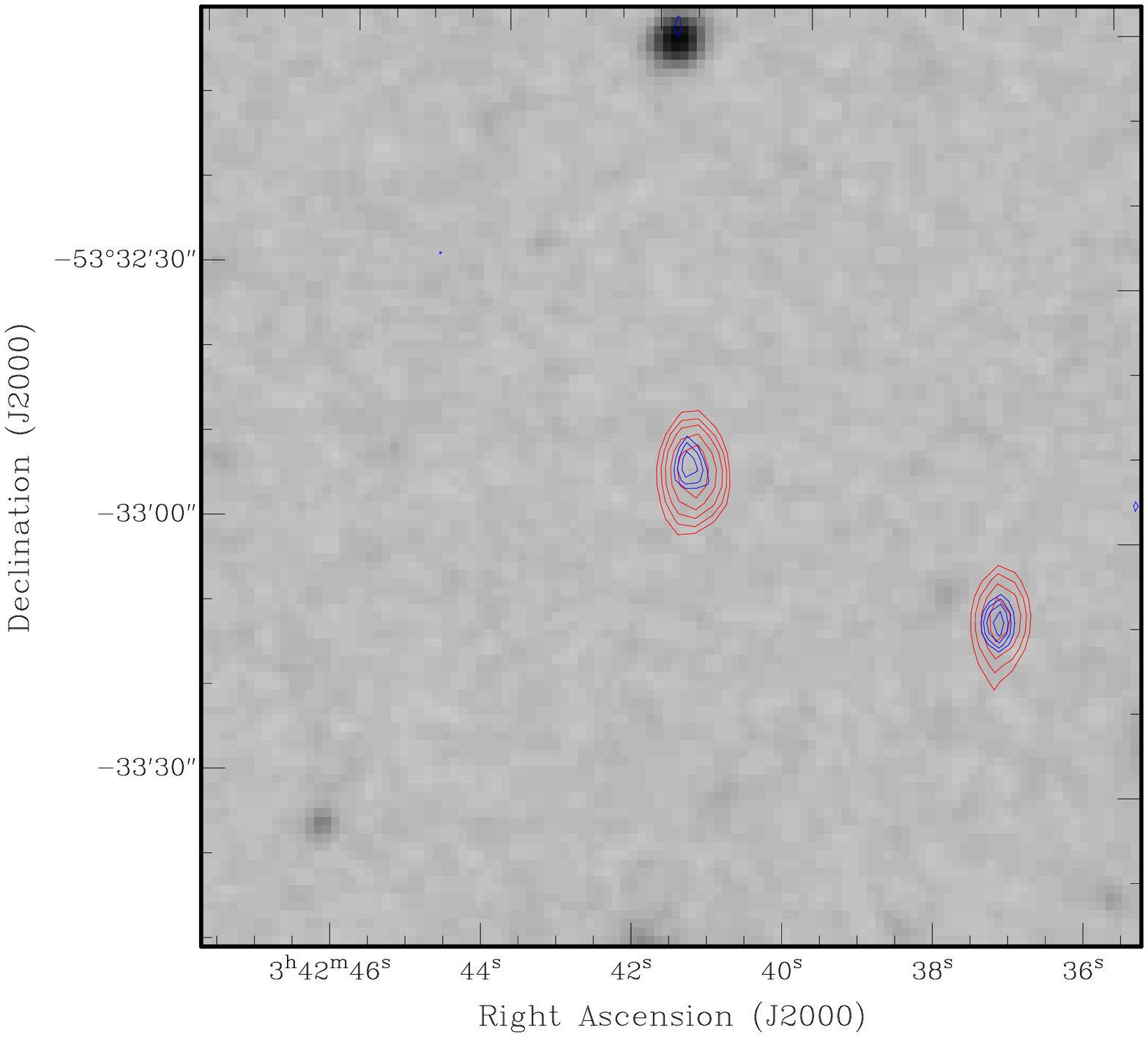}\includegraphics[angle=-90]{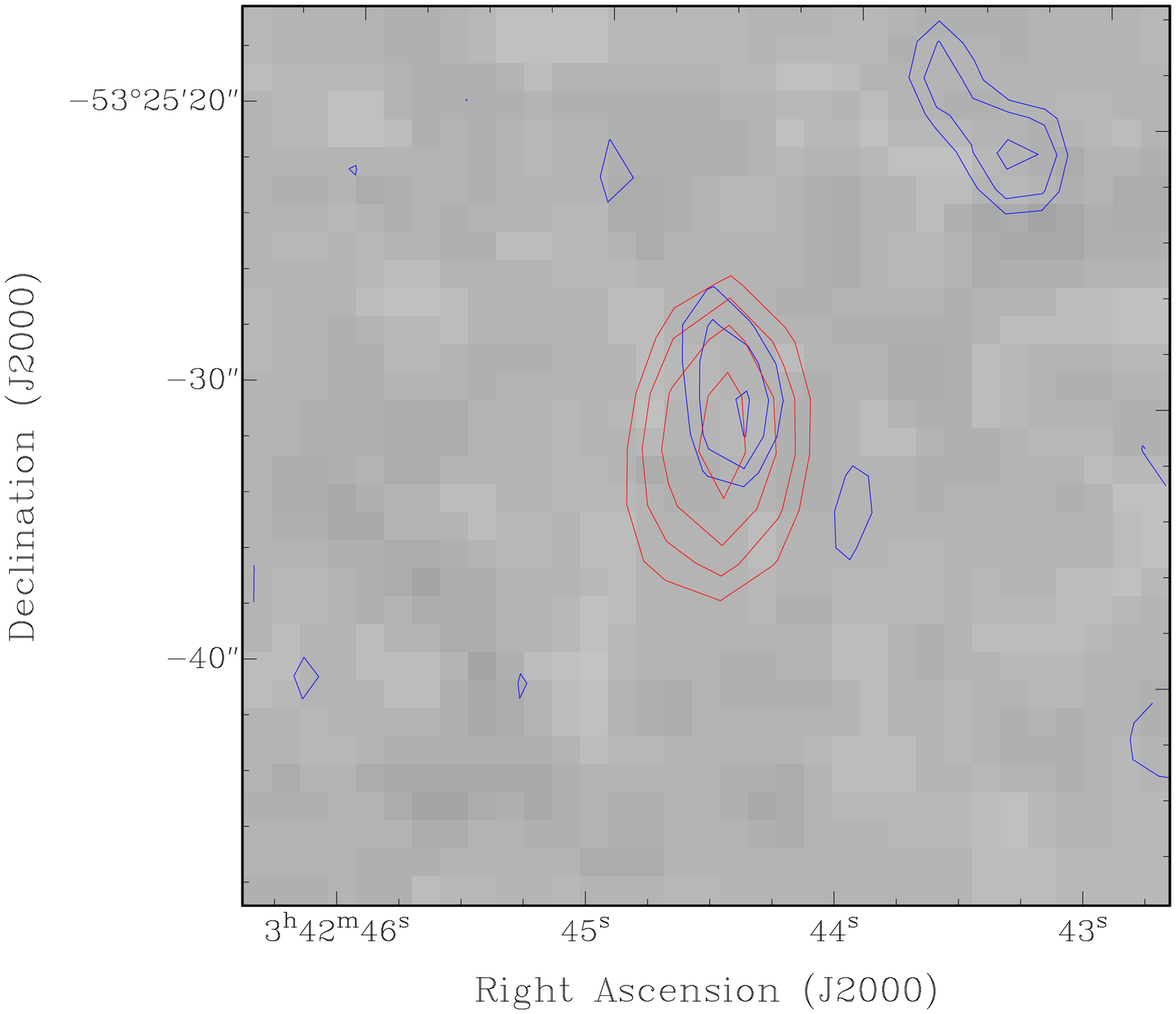}\includegraphics[angle=-90]{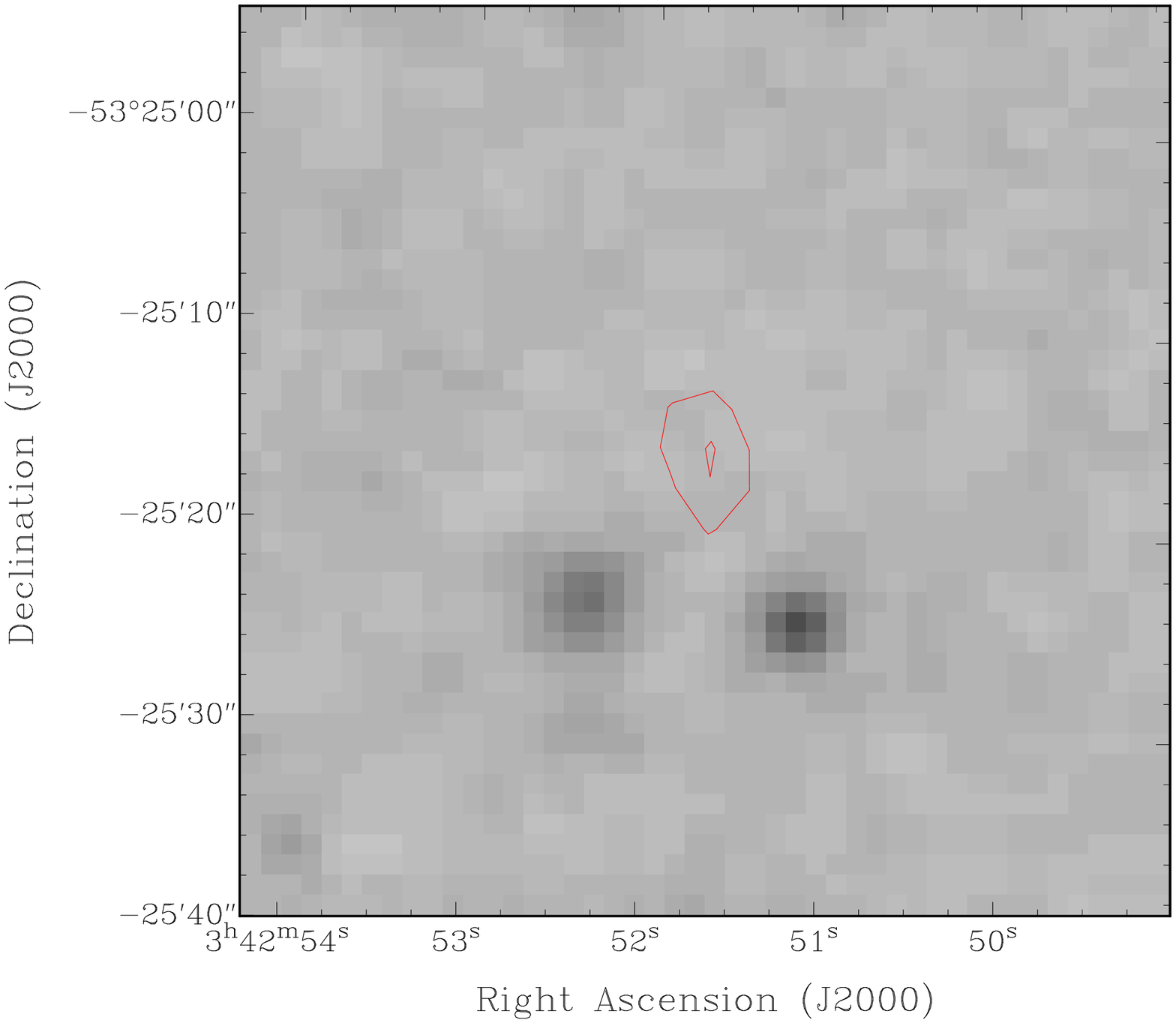}\includegraphics[angle=-90]{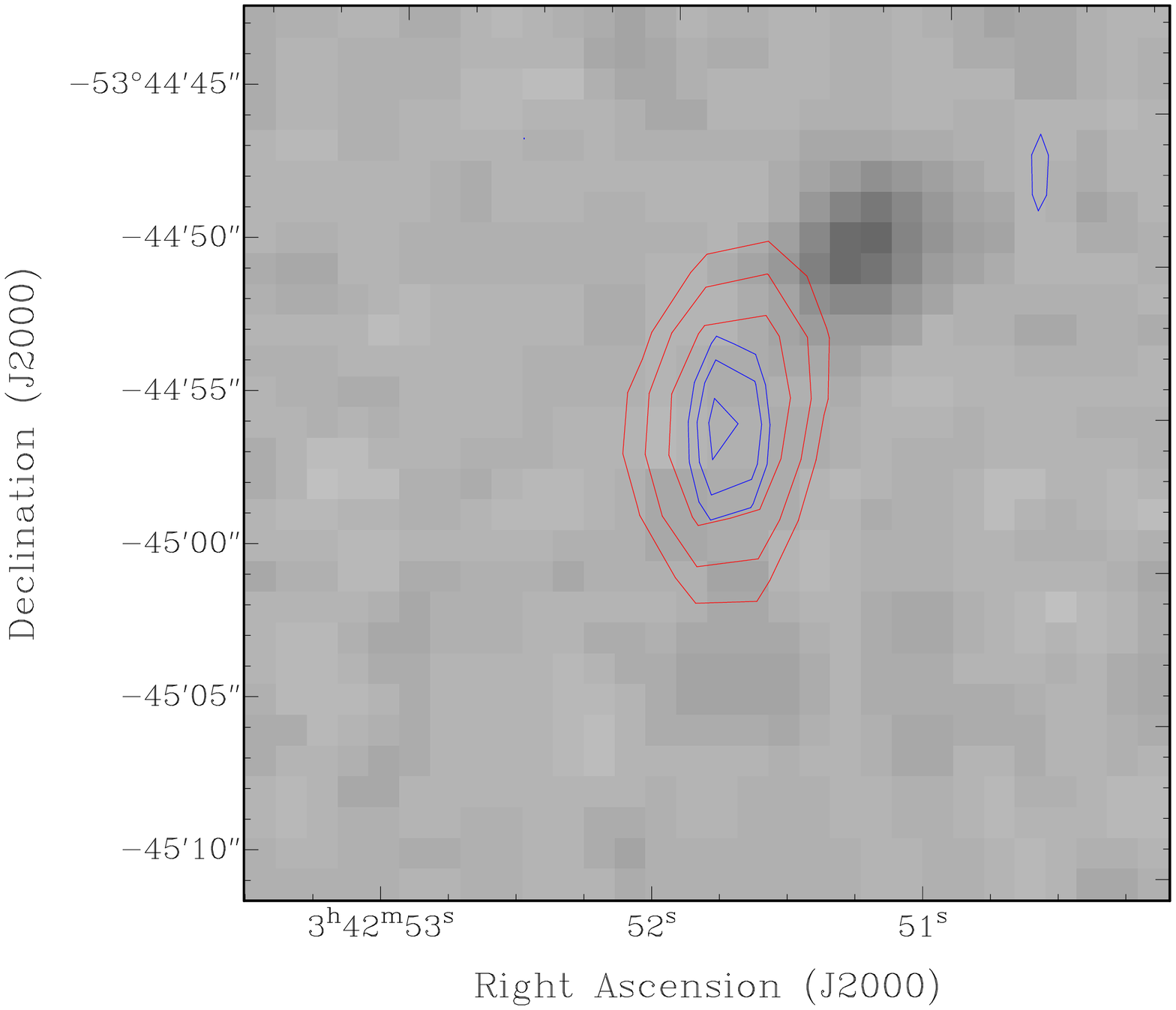}}
\resizebox{\hsize}{!}{\includegraphics[angle=-90]{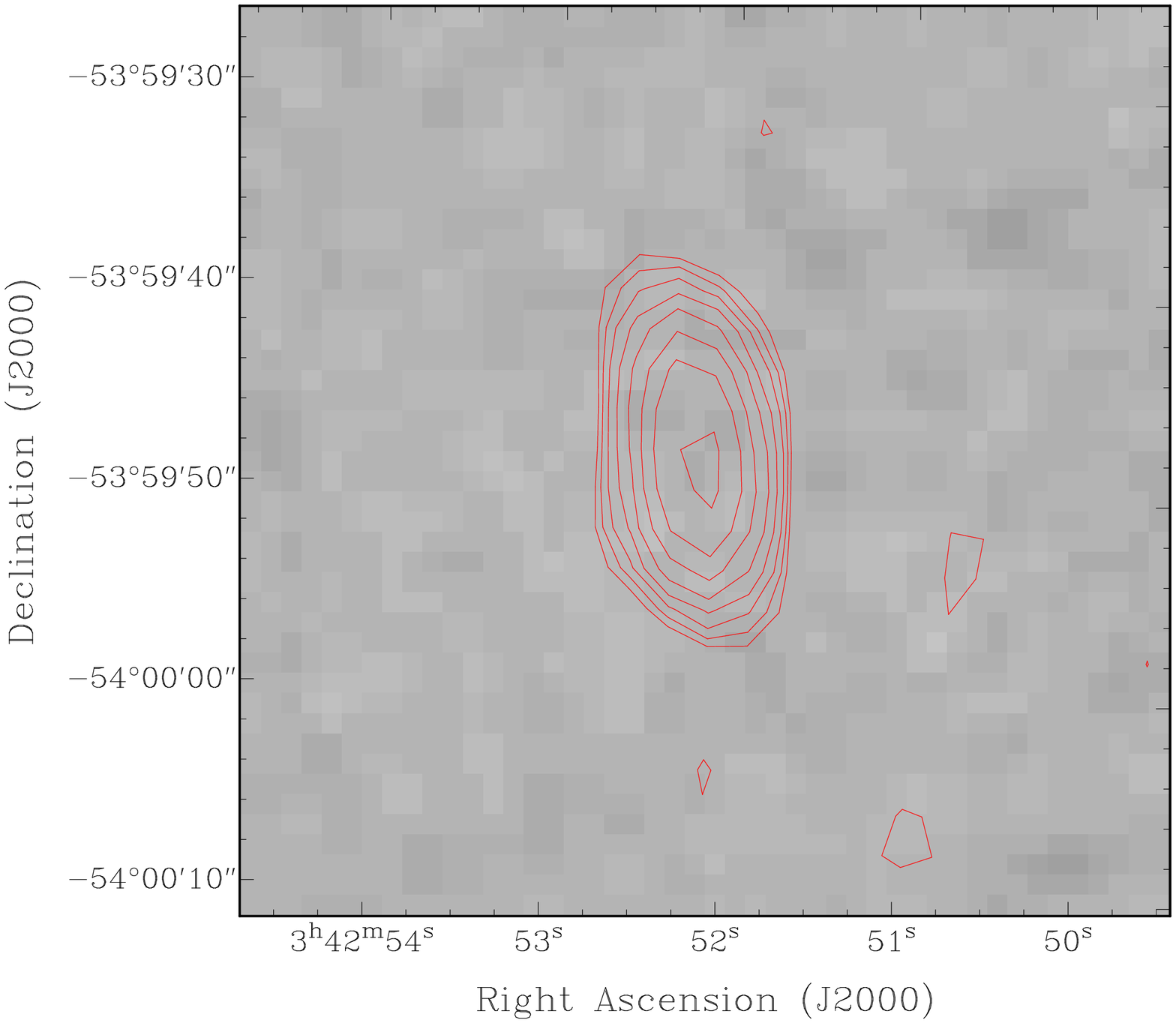}\includegraphics[angle=-90]{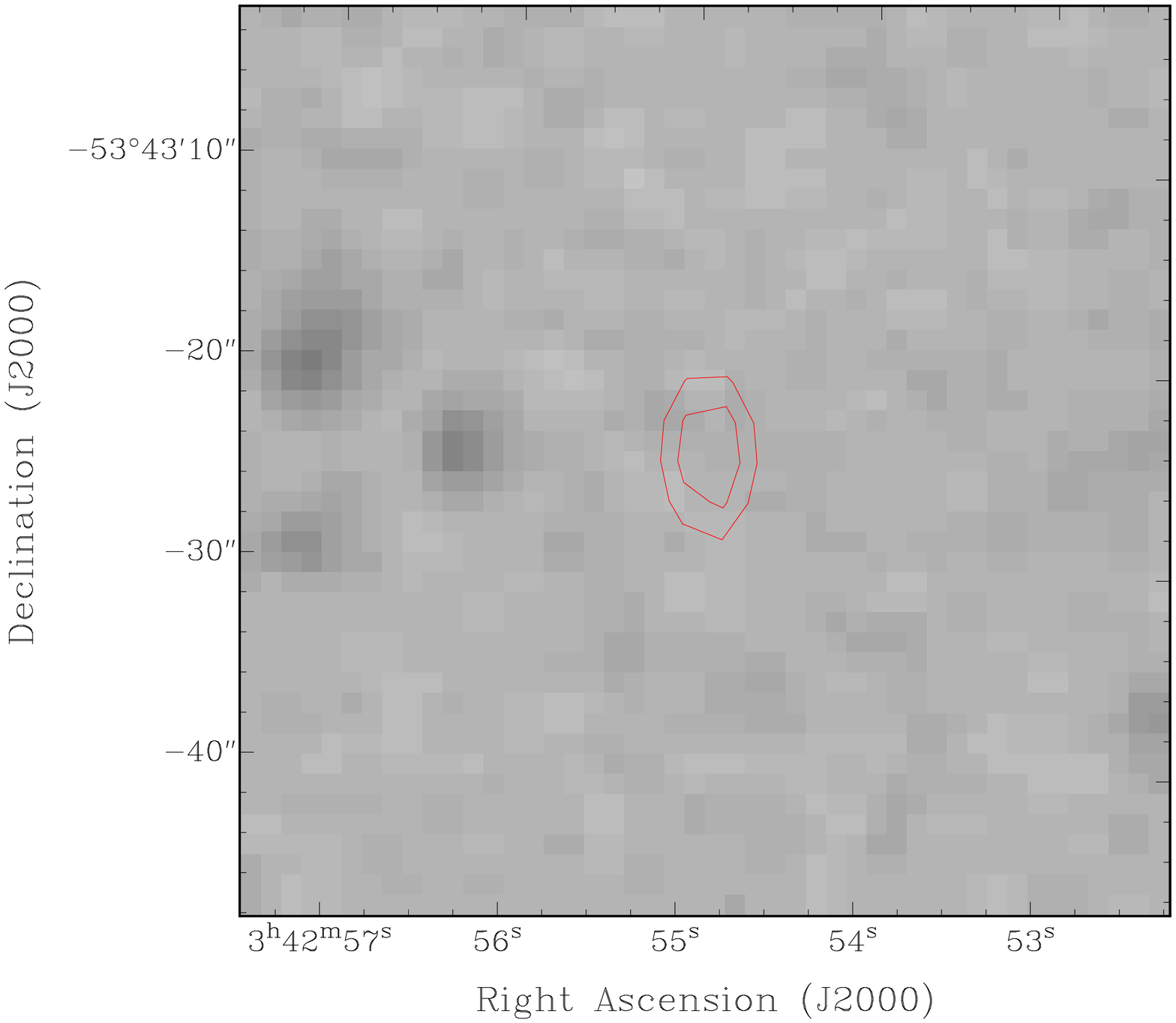}\includegraphics[angle=-90]{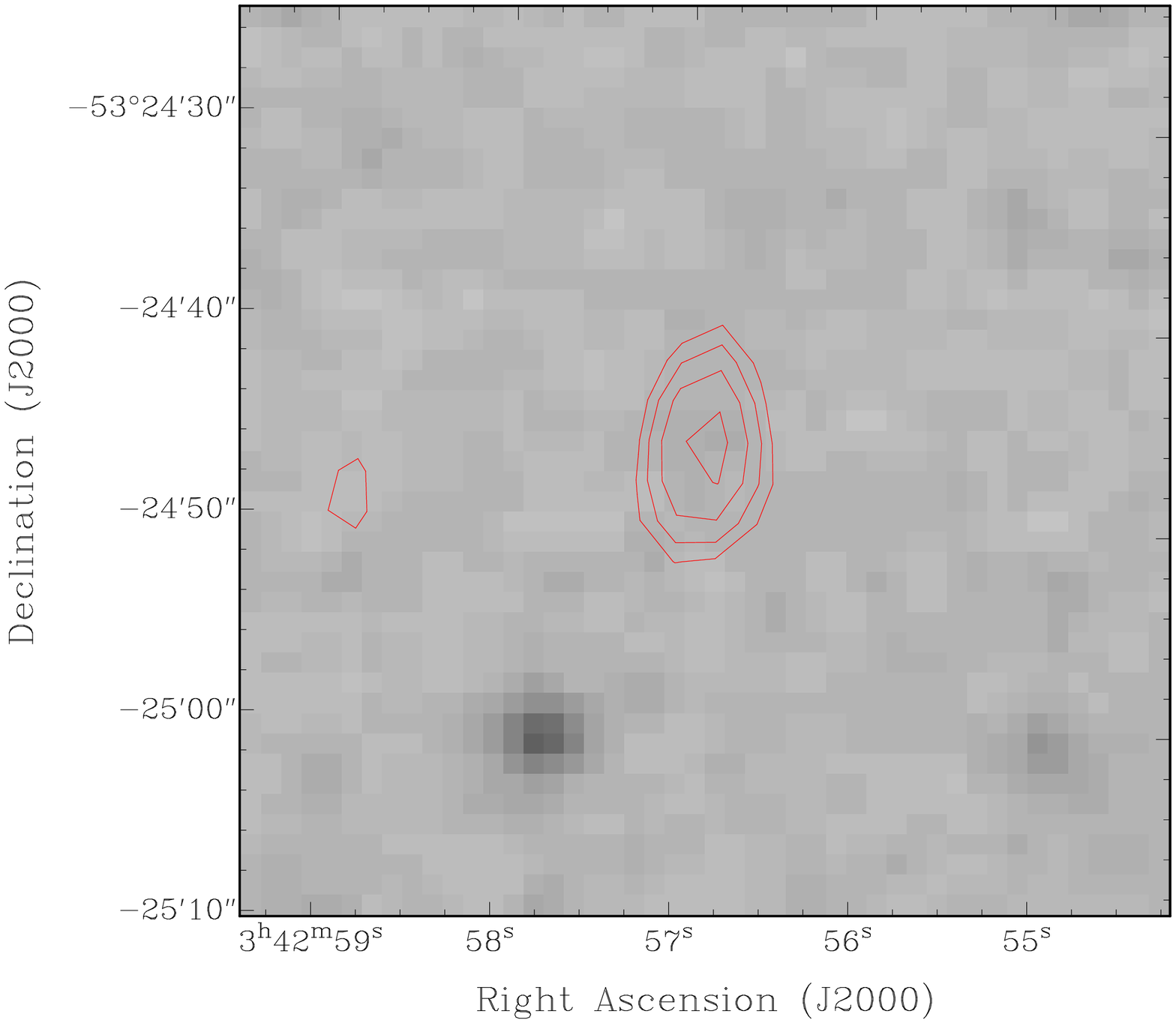}\includegraphics[angle=-90]{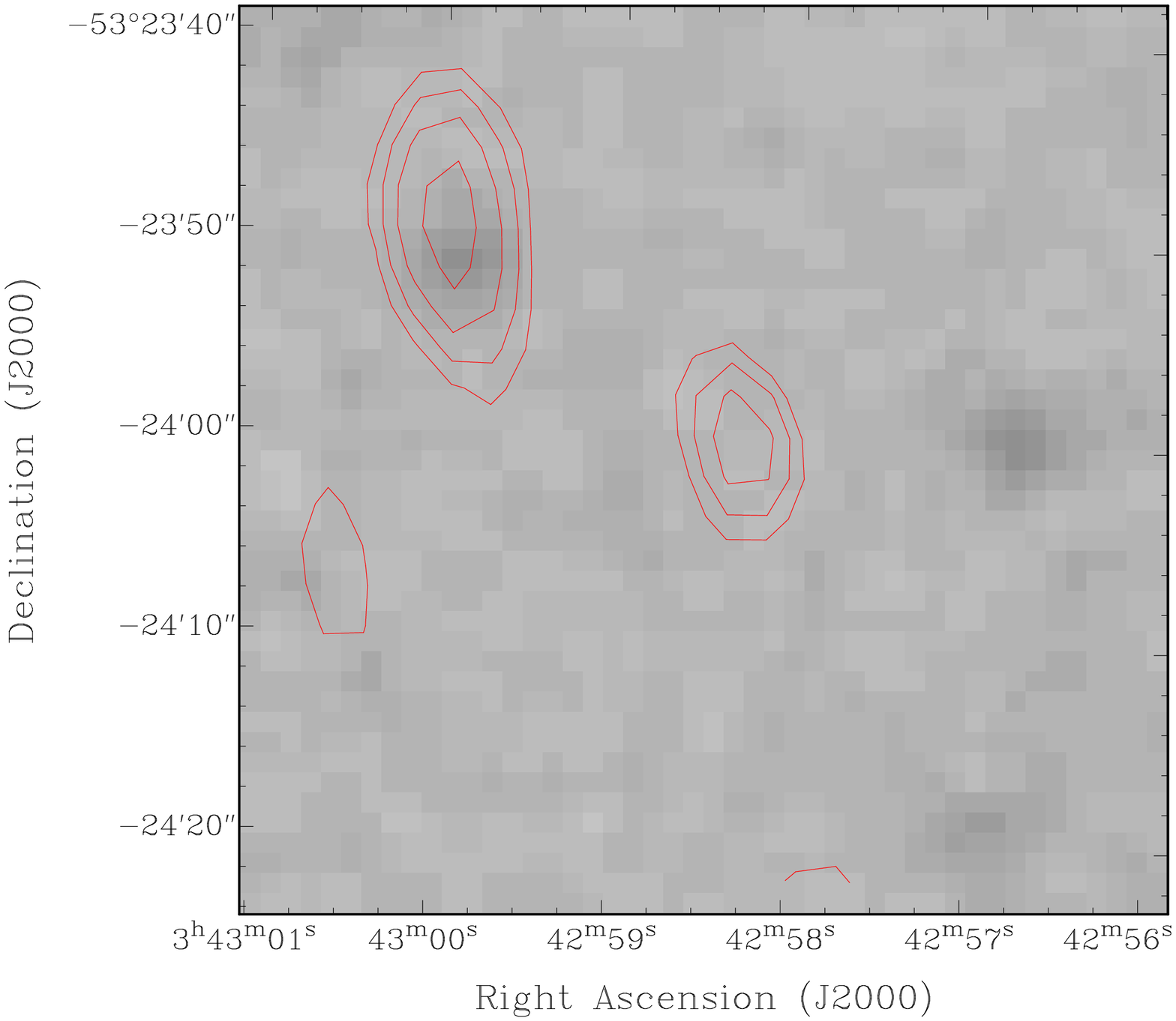}\includegraphics[angle=-90]{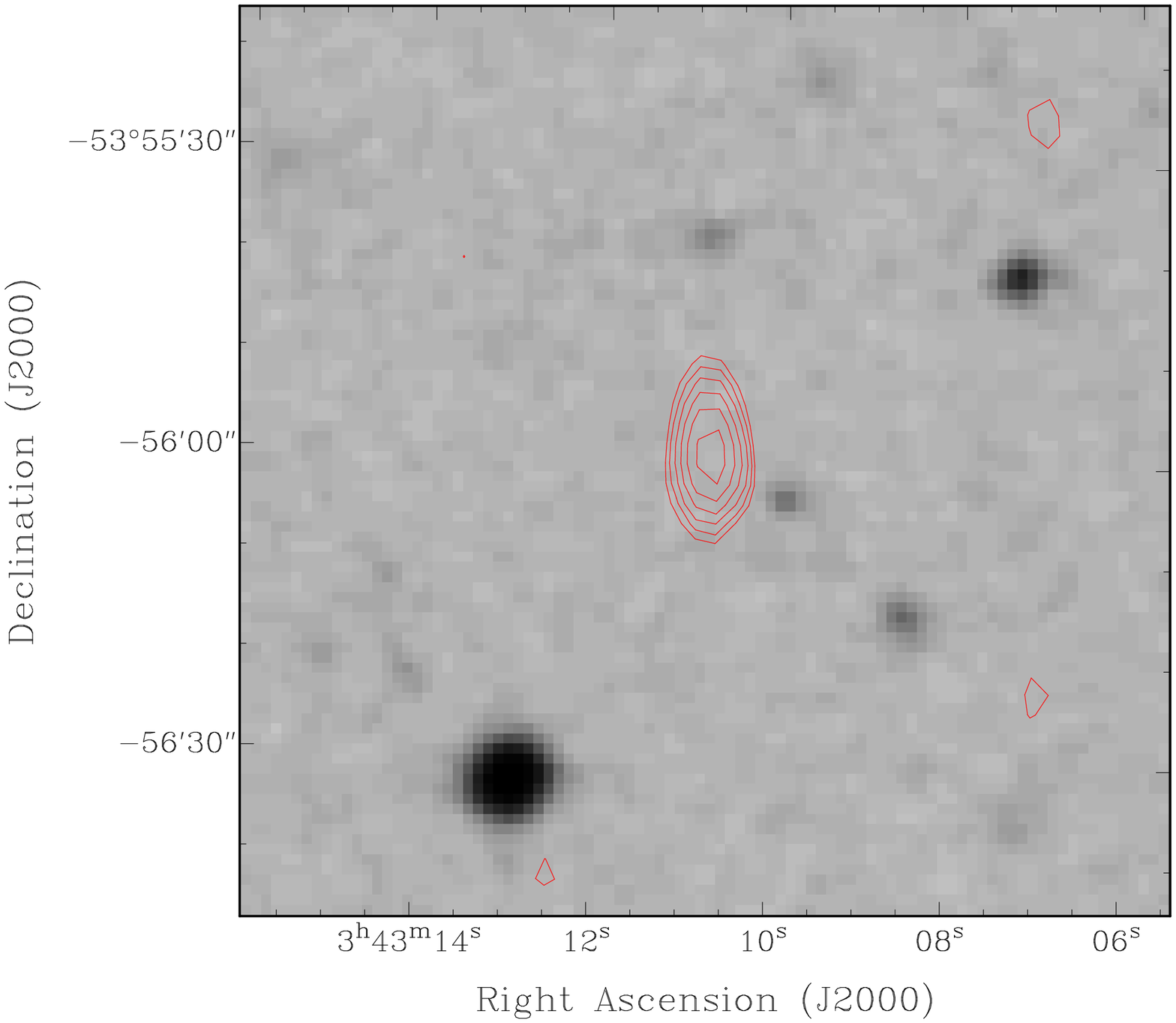}\includegraphics[angle=-90]{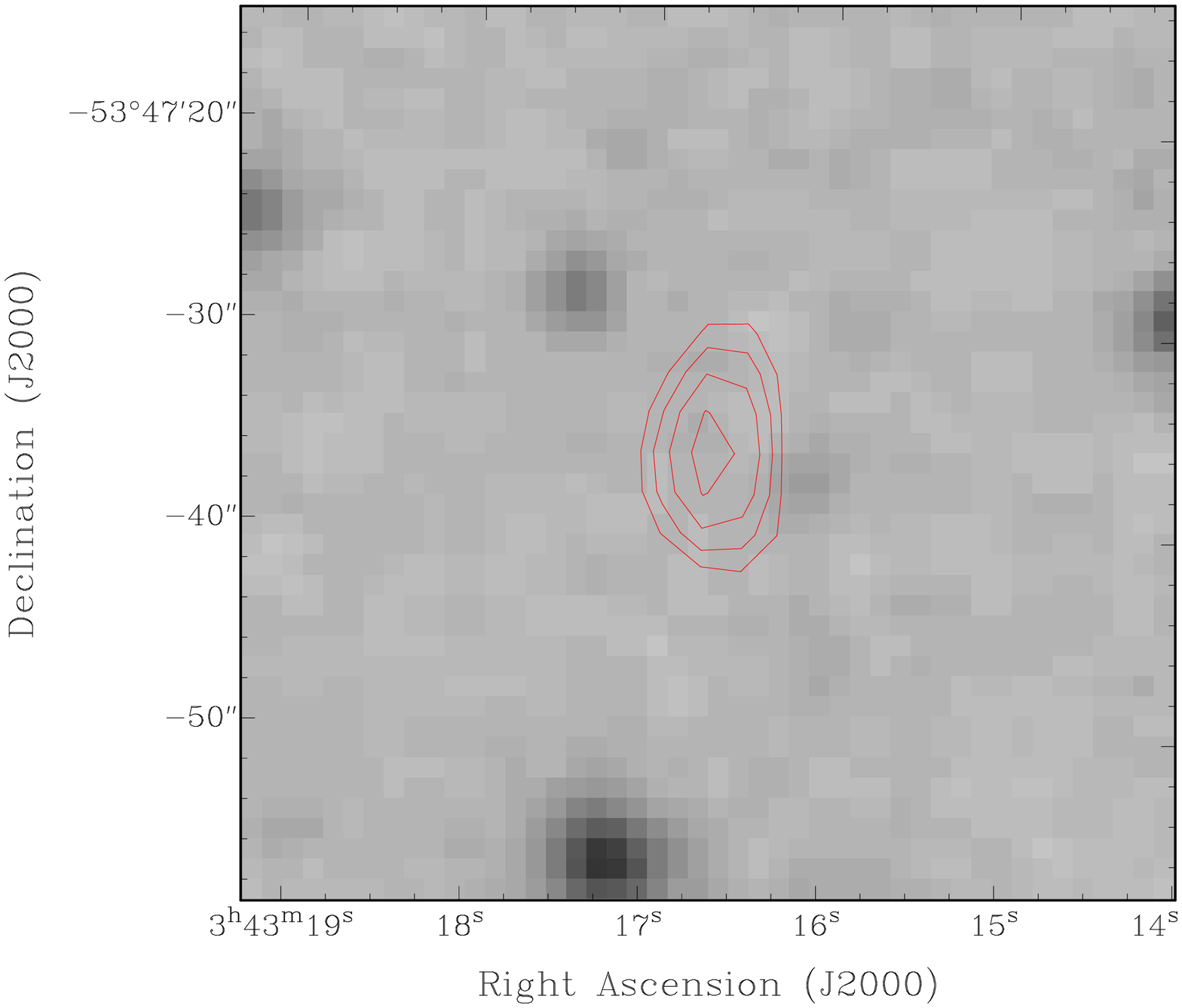}}
\resizebox{\hsize}{!}{\includegraphics[angle=-90]{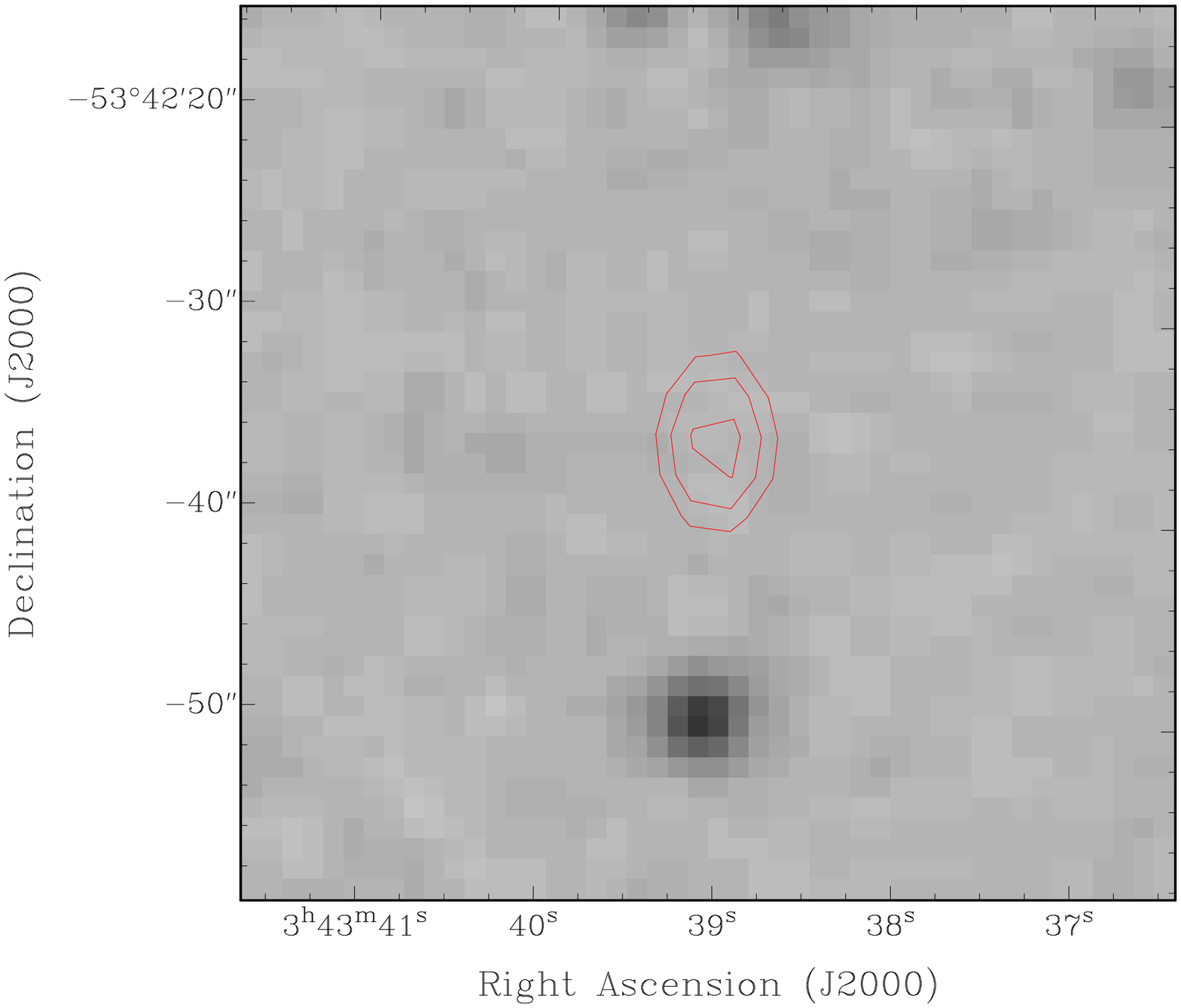}\includegraphics[angle=-90]{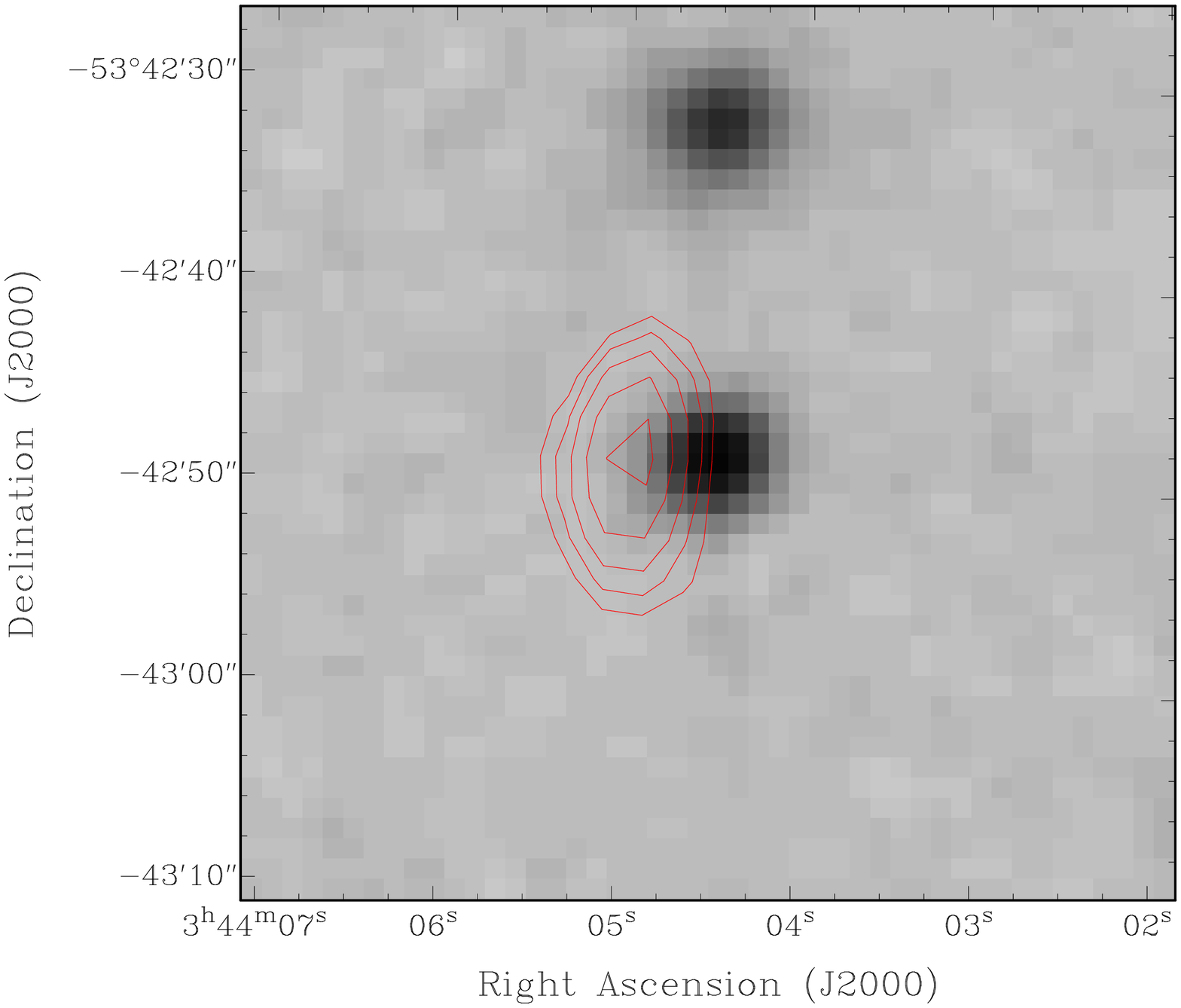}\includegraphics[angle=-90]{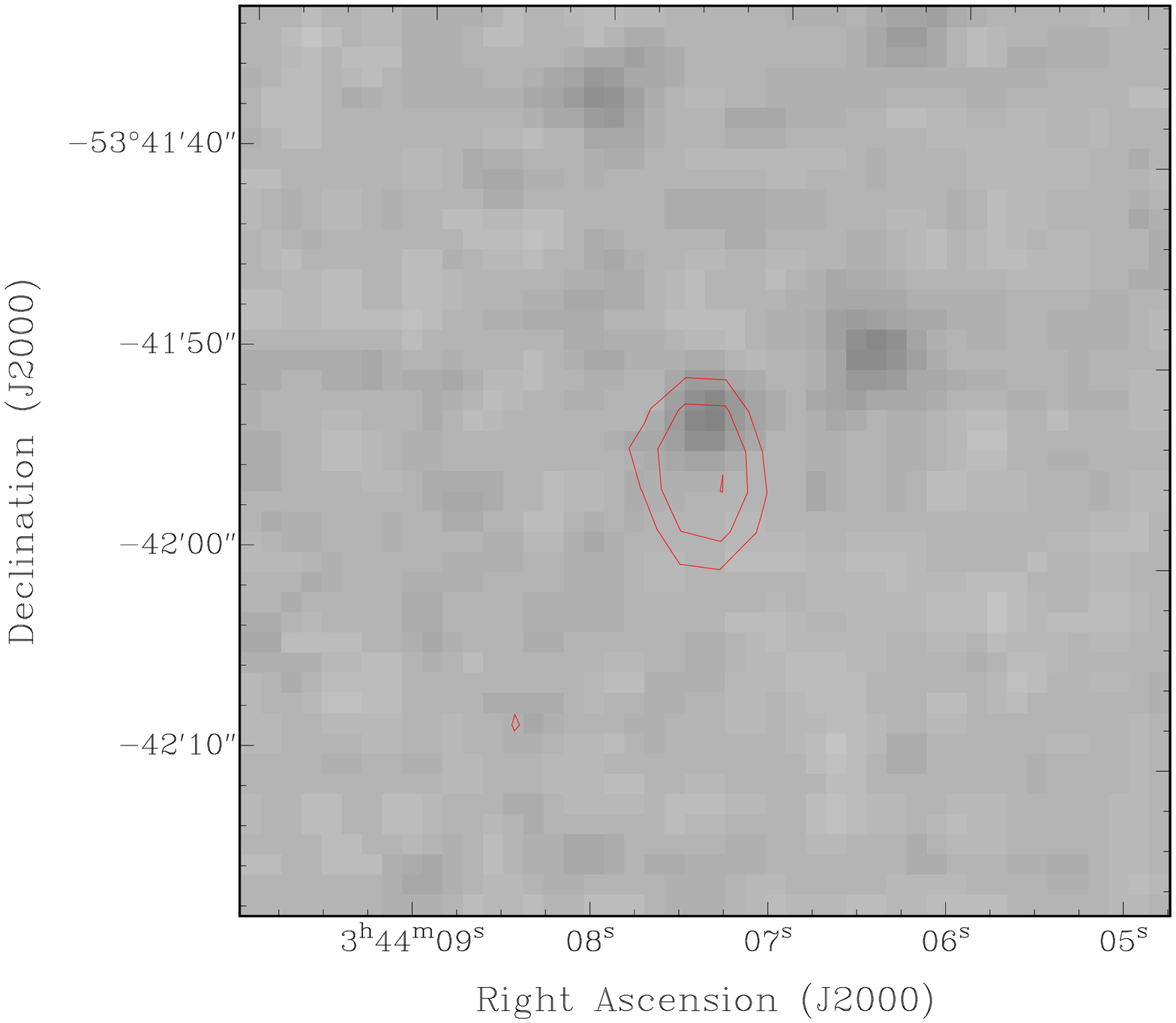}\includegraphics[angle=-90]{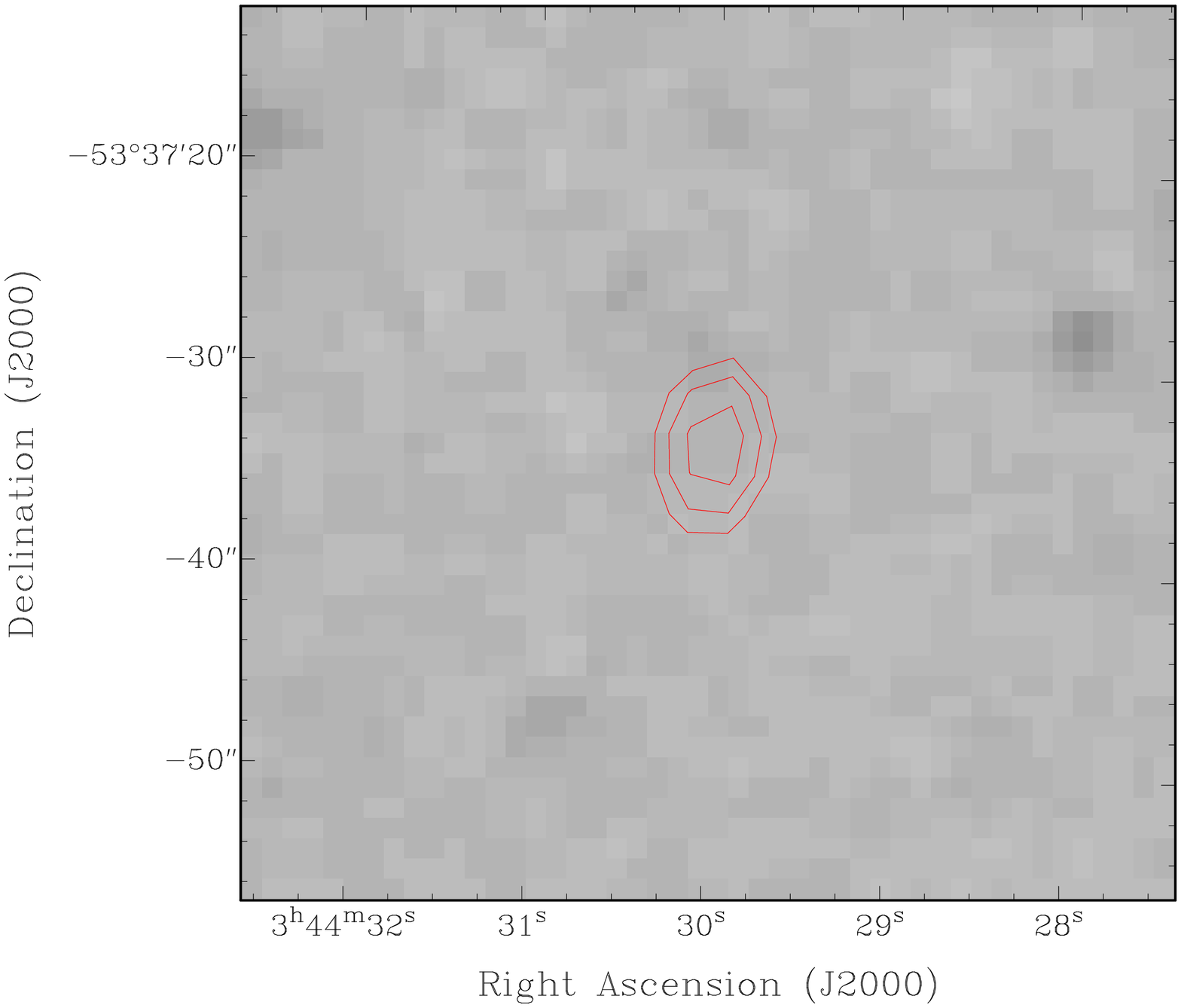}\includegraphics[angle=-90]{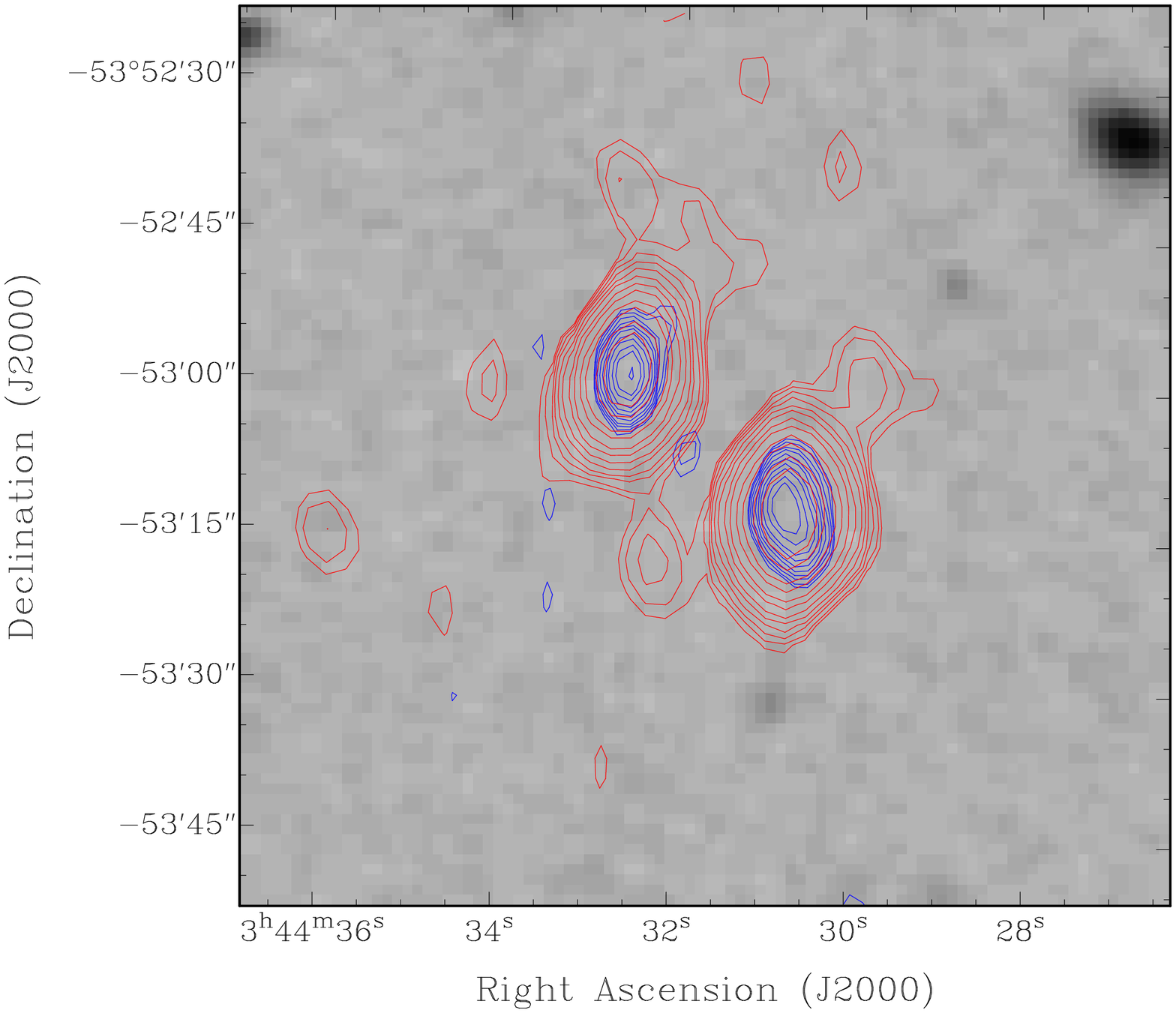}\includegraphics[angle=-90]{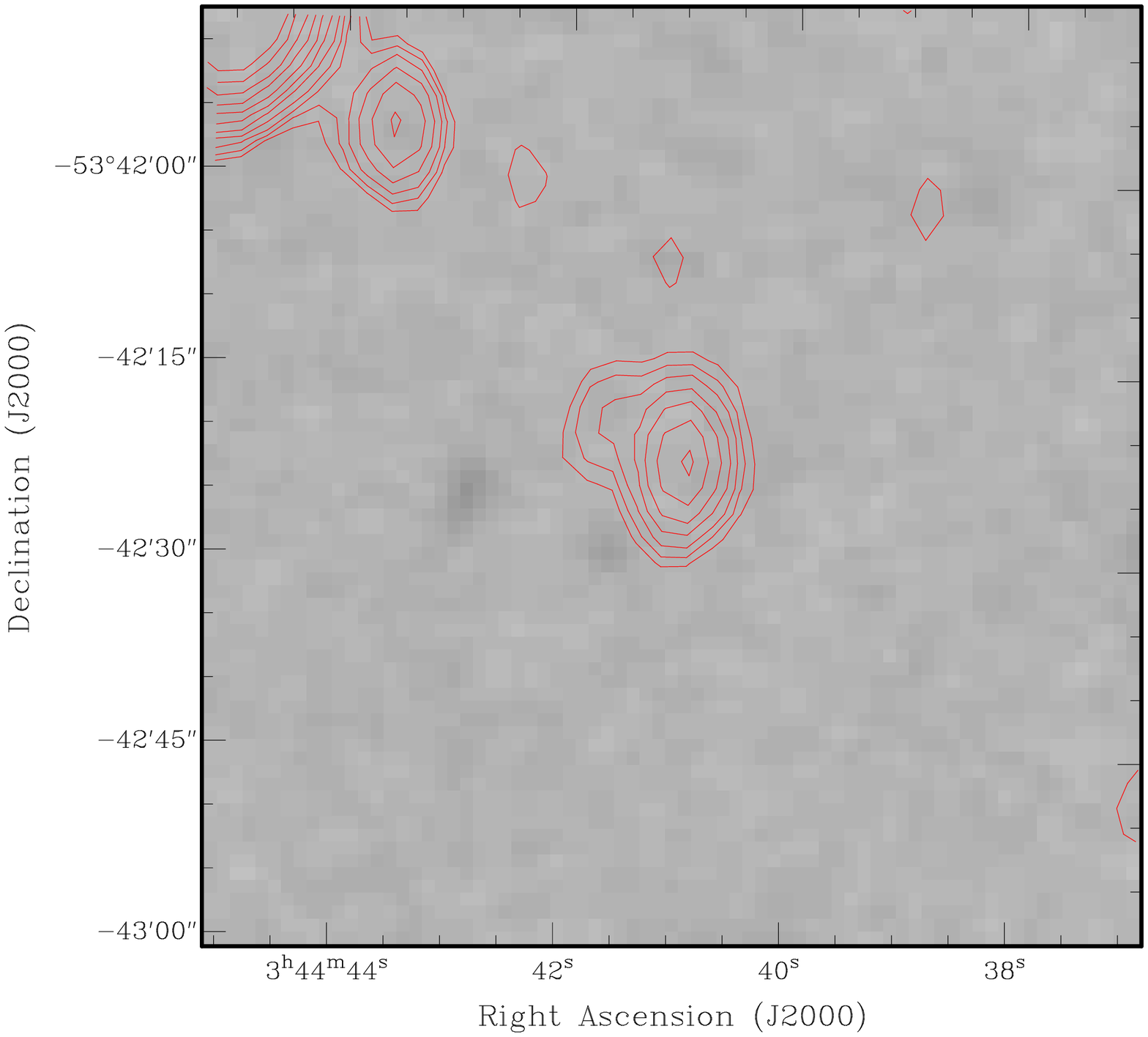}}
\resizebox{\hsize}{!}{\includegraphics[angle=-90]{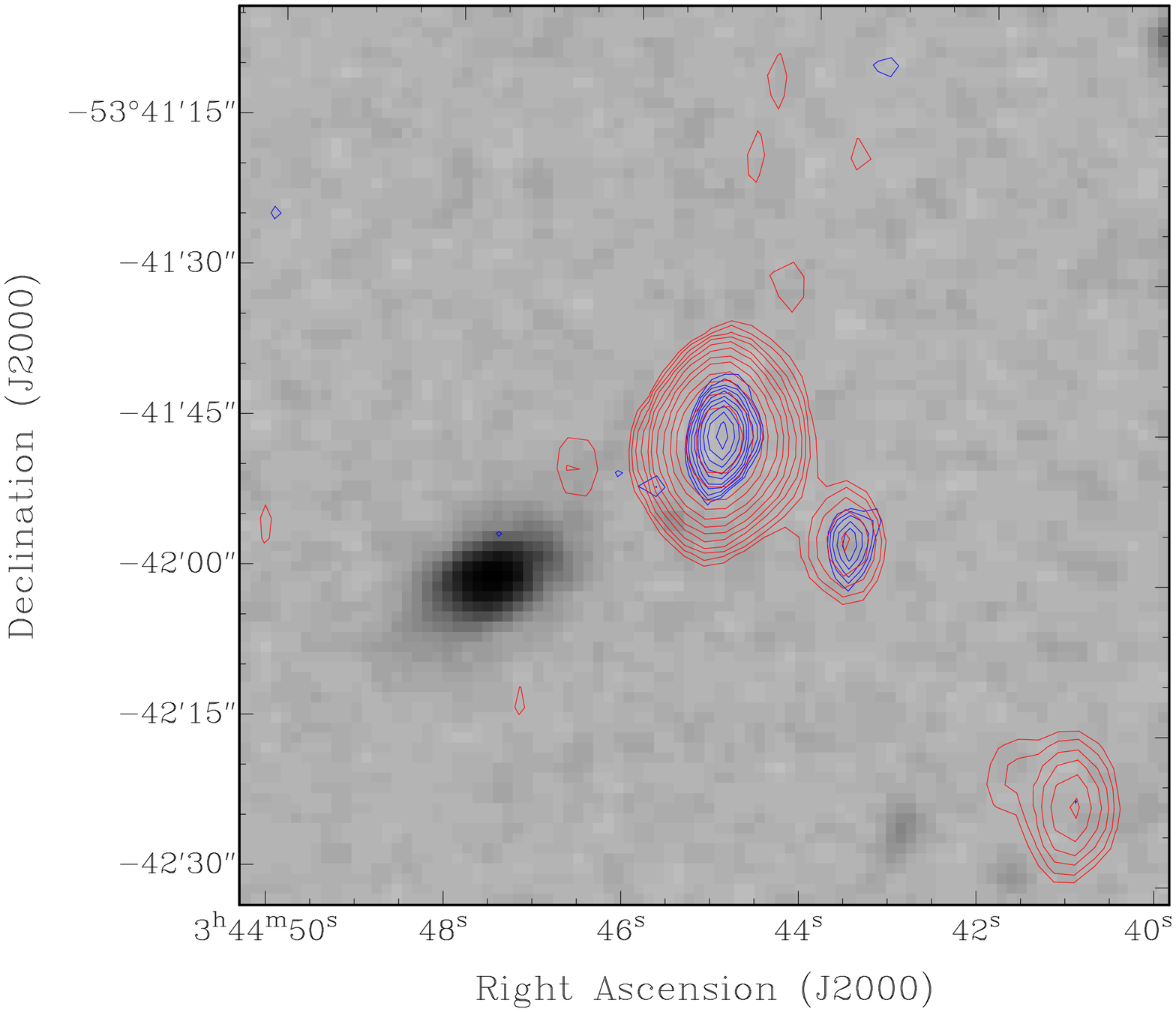}\includegraphics[angle=-90]{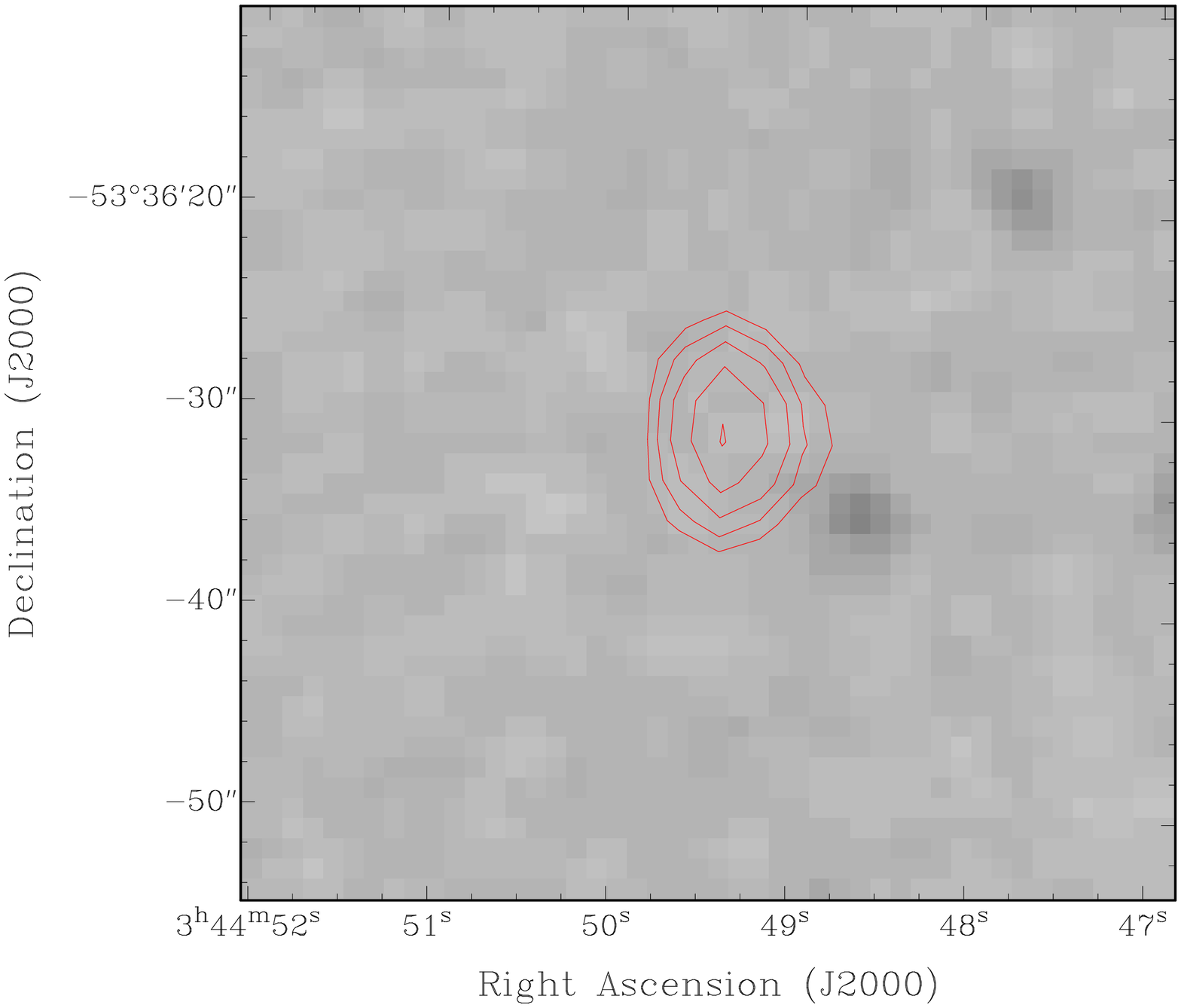}\includegraphics[angle=-90]{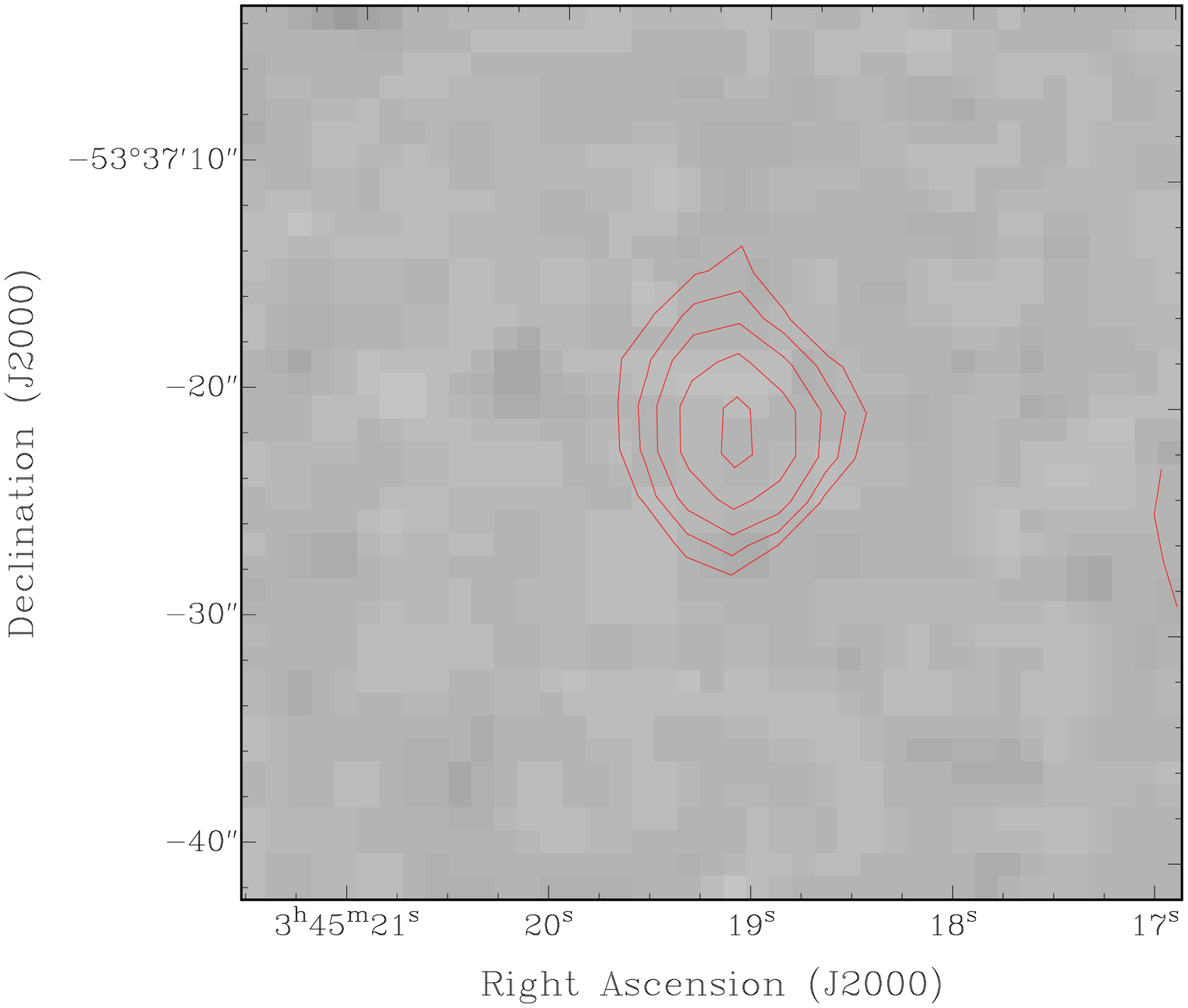}\includegraphics[angle=-90]{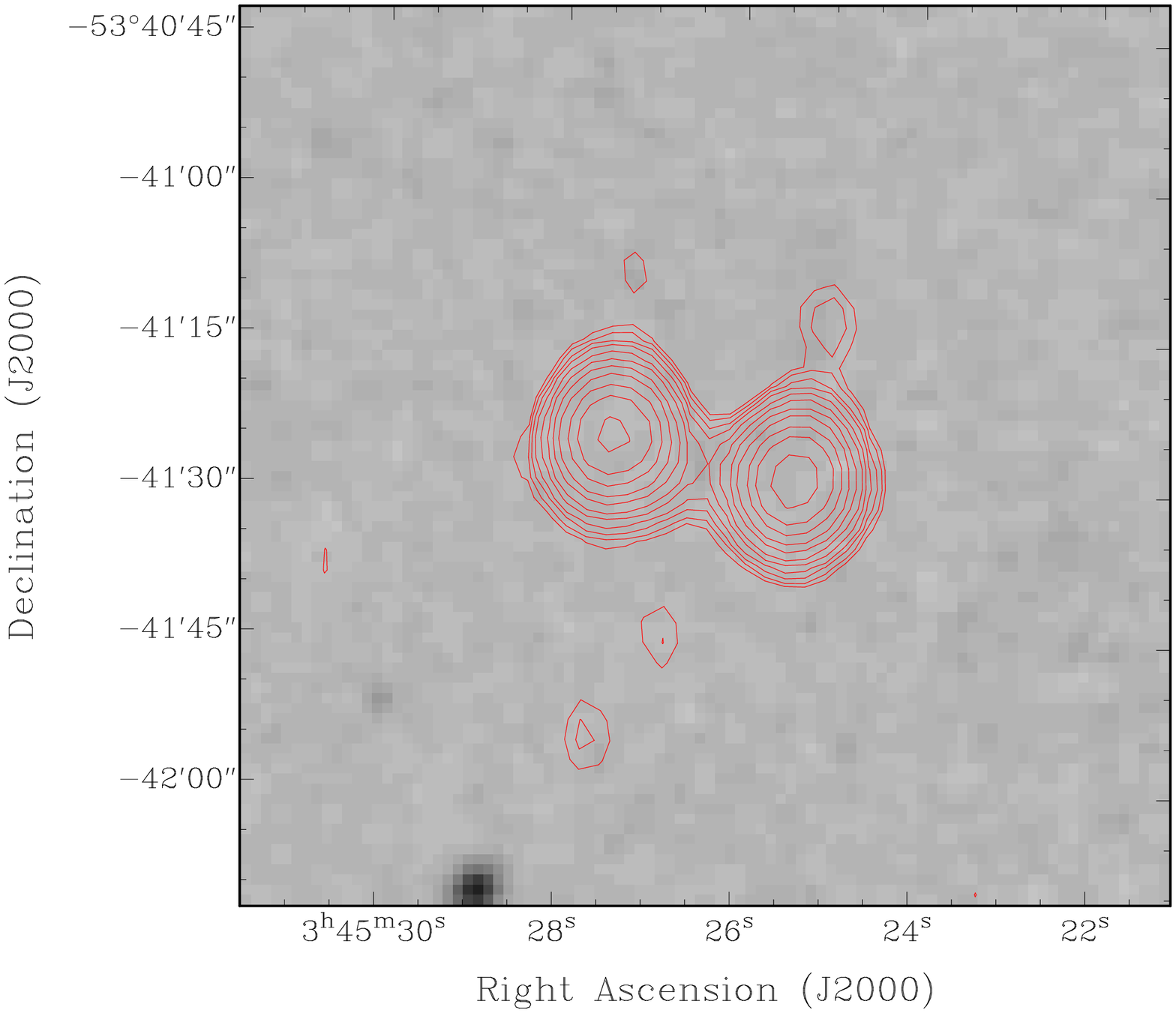}\includegraphics[angle=-90]{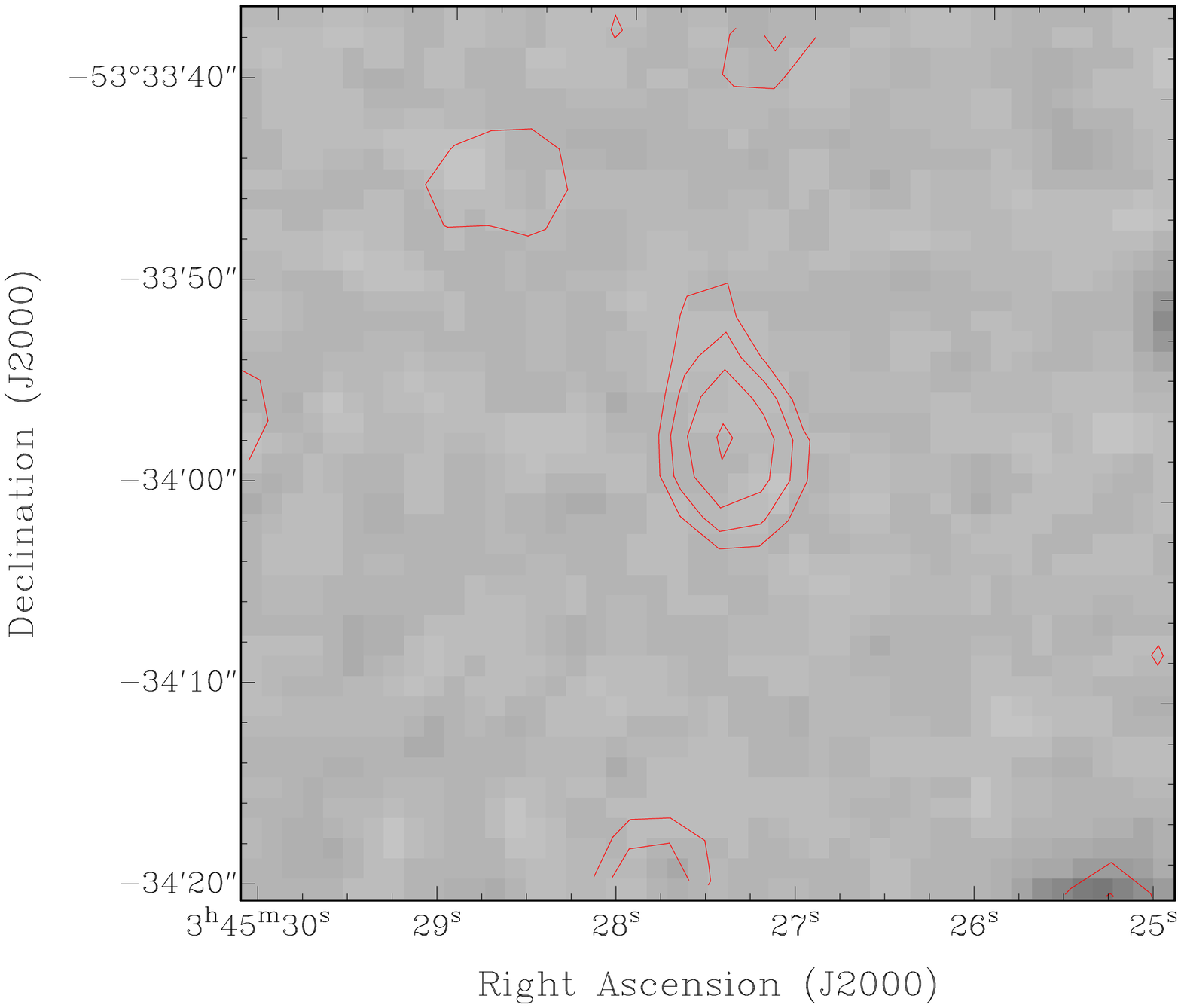}\includegraphics[angle=-90]{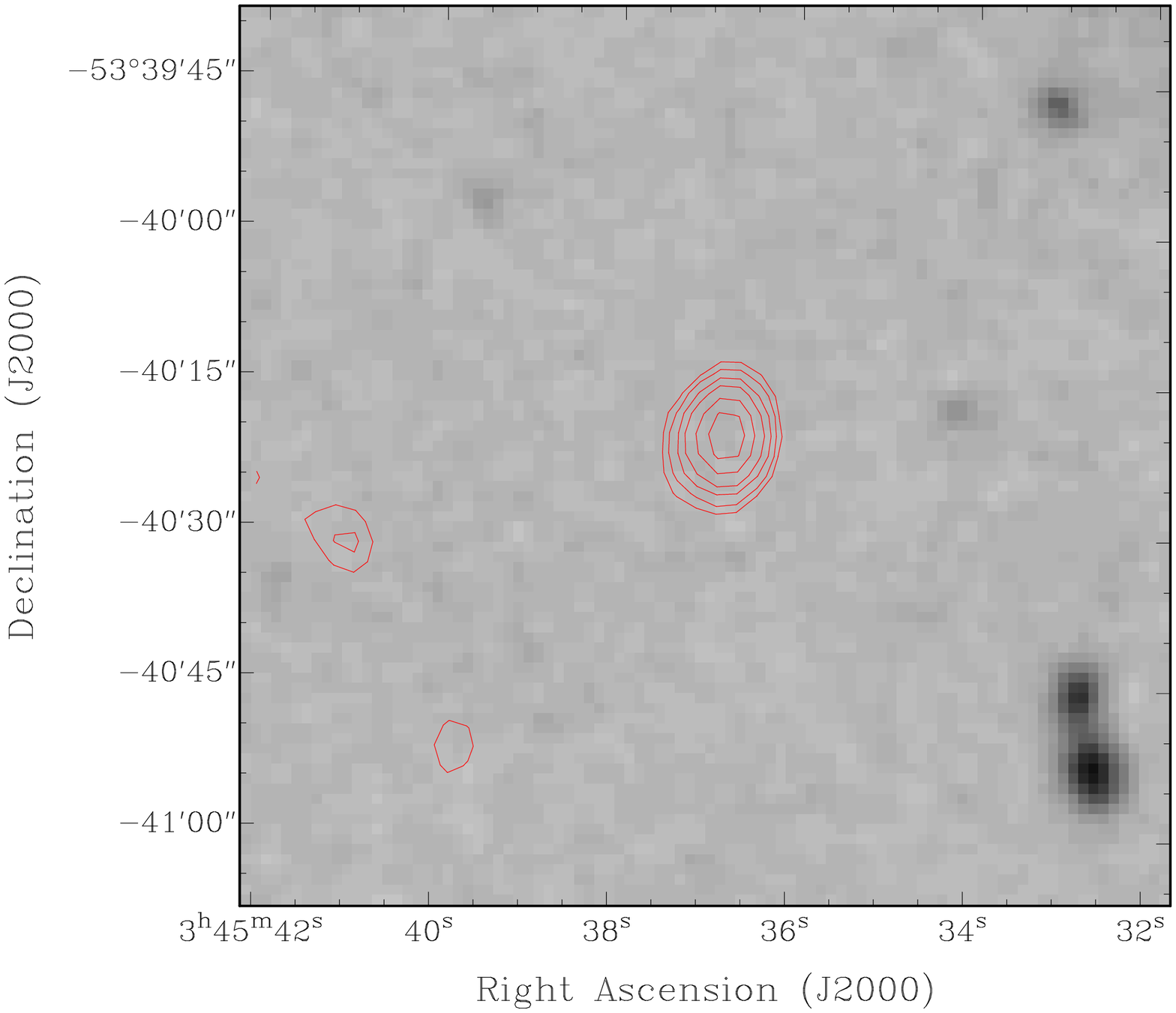}}
\caption{Postage stamps of the 36 radio sources with no optical counterparts detected within $23\farcm05$ of 03 42 53.00 -53 37 43.0. Red contours show the 1.4 GHz emission and blue the 2.5 GHz emission. Contours begin at
0.14 mJy beam$^{-1}$ and increase at intervals of root 2, except for the 5th image in the 2nd row \& the first image in the last row which begin at 0.2 mJy beam$^{-1}$ and 4th image in the 6th row \& the 5th image in the 5th row which begin at 0.4 mJy beam$^{-1}$ and 0.8 mJy beam$^{-1}$ respectively. Images are presented in the same order as given in Table \ref{tab:b}. 
}
\label{fig:rad_only}
\end{figure*}

Optical identifications were performed relative to both the SuperCOSMOS catalogue via a software routine, and the
Digitized Sky Survey by eye. For the automated routine searching the SuperCOSMOS catalogue we limited the blue magnitude of 
objects to be less than 23 which is a reasonable limit to ensure objects are real. Results of both processes were consistent in the case 
of separations between the radio and optical coordinates of less than 3 arcseconds. For extended sources which 
were more than 3 arcseconds from the optical host the automated routine broke down and these sources were all 
discovered by inspection of the DSS images overlaid with radio contours. 
 
Of the total there are 62/110 optical sources detected to the limits of the DSS, 29 of which have b$_{\rm{J}} \leq$ 19.5.
This compares well with the results for Shapley in which 26\% of sources are found to have b$_{\rm{J}} \leq$ 19.5 
(Venturi et al.~1997, 2000). All of the double radio galaxies have either no optical counterparts 
or have counterparts with b$_{\rm{J}} \geq$ 22.98 making them unlikely to be members of A3158. In contrast there
are five known Head-tailed galaxies associated with the nearby A3125/A3128 complex (Mao et al.~2008) and
five Head-tailed galaxies in the A3558 complex at the centre of the Shapley supercluster (Venturi et al.~2000, 
Miller 2005). The presence of tailed radio galaxies in clusters has been suggested as a signpost 
for ``cluster weather'' (Burns et al.~1998, Mao et al.~2008). Recently, it has been shown that head-tailed 
galaxies reside in only the most dense
regions of clusters and that the local environment is a strong factor 
in the generation of the radio tails (Mao et al.~2008). The lack of tailed radio galaxies in A3158 
might indicate that the cluster either lacks regions of high density or does not exhibit turbulent 
cluster weather either, because the cluster has not yet undergone such an interaction or because it is
seen in the late stages of a merger event, given the current multiwavelength data, the later is more probable.

\begin{figure}
   \begin{center}
    \includegraphics [width=8.5cm, height=7cm]{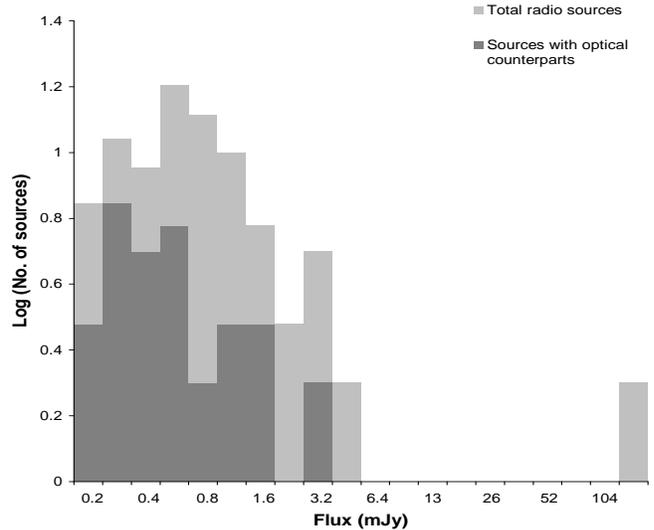}
   \caption{Distribution of radio sources with an optical counterpart versus flux density (dark
grey) compared to the total number of sources (light grey).}
   \label{fig:opthist}
 \end{center}
 \end{figure}

In Figure \ref{fig:opthist} we plot the histogram of the number of radio galaxies
found to have optical counterparts as a function of flux density (dark grey) compared to the 
total radio sample (light grey). The shape of each distribution is remarkably similar suggesting
that optical identification rate is uniform with respect to the flux density range.

Spectroscopic information from the region was obtained from the NASA Extragalactic Database (NED). 
Of the available spectroscopic data, 17 redshifts correspond to radio sources in our sample of which 14 
(82\%) correspond to a velocity range consistent with membership of the HRS. 

All of the radio galaxies identified with A3158 (15,000 km s$^{-1} \leq $ v $\leq$ 35,000 km s$^{-1}$) 
are faint with 1.4 GHz radio powers between 5.40 $\times 10^{21}$ WHz$^{-1}$ and 6.35 $\times 10^{22}$ WHz$^{-1}$
putting them all below the FRI category based on radio power which is a similar result to radio galaxies
associated with clusters in Shapley (Venturi et al.~2000). Radio powers in this regime may indicate starformation
as the driving mechanism. The brightest galaxy identified with the cluster
spectroscopically is the most massive cD galaxy which has a radio flux at 1.4 GHz of S$_{1.4}$=7.88 mJy.

\section{Radio Source Counts}
\begin{figure}
   \begin{center}
    \includegraphics [width=8.5cm, height=7cm]{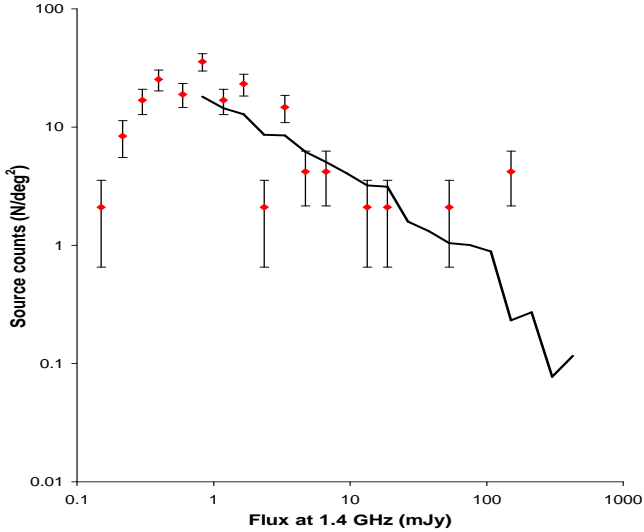}
   \caption{Log N-Log S plot for the galaxies in A3158 as compared with the background counts by
Prandoni et al.~(2000).}
   \label{fig:NS}
 \end{center}
 \end{figure}

In order to assess the dynamical state of the system, as determined via radio source counts, we
compared our findings to the background counts obtained by Prandoni et al.~(2001). Prandoni et al.~(2001) 
detected 2960 radio sources above 0.5mJy at 1.4 GHz over an area of 25.82deg$^2$ using the ATCA in a similar
observational set-up and thus provides the ideal background source comparison. The sample is
complete down to 0.7 mJy and provides the best determination of source counts in the range
0.7 $\leq$ S$_{1.4}$ $\leq$ 2 mJy using the ATCA to date. Our sample of 88 galaxies is
found over an area of 0.463 deg$^2$ and there are 52 sources above 0.7 mJy. Figure \ref{fig:NS} shows the resulting 
log N-log S plot over the flux density range from 0.15 to 512 mJy. The differential number of sources in 
each bin is computed as Ndeg$^{-2}$ and the errors are Poissonian. Each bin has a width of 
$\Delta$S = S $\times$ log(2.5).

The plot clearly shows that our source counts are incomplete below 0.5 mJy, however prior to this point there are several
bins in excess of the background found by Prandoni et al.~(2001). Specifically, the radio galaxies below 3.96 mJy are 
several deviations above the background, at the mean redshift of A3158 this corresponds to log P$_{1.4}$(WHz$^{-1}$)=22.50.
We performed a Kolmogoroff-Smirnov (KS) test against the two distributions above 0.7 mJy (the lower limit of the Prandoni
data), the D-statistic was 0.18 -- 0.19 depending on if we used only the average value of each data point per bin in the Prandoni
data or if we assumed uniform flux densities across each of their bins. This corresponds to a very low
chance that distributions are the same (3.3\% -- 8.2\%). In addition, we performed the same analysis for the distributions above 
3.96 mJy which returned a D-statistic of 0.22 -- 0.24 which corresponds to a probability of 61\% -- 70\% that the distributions are the same.
Despite the fact that we are in a very low-number regime for the sub-sample above 3.96 mJy, the improved chance of coincidence 
in sub-sample over the large sample is suggestive that the lower bins are more strongly different to the background distribution. 
Similar analysis on a subsample of 145 radio galaxies with S$_{1.4} \geq$ 2 mJy found over a 3.25 deg$^2$ area in Shapley 
(Venturi et al.~2000) did not show an excess over the background counts. 

The effect of sources with multiple components was also considered. Prandoni et al.~(2001) estimate that around 10\% of
sources in radio catalogues are due to different components of multi-component, extended radio galaxies being counted separately. In our data
this would correspond to only one source. Via inspection we found two examples where it might be the case that multiple components
of the same source had been counted separately. As a check we recomputed the D-statistic on our distribution as compared to Prandoni's data
assuming that these sources should be conflated. The D-statistic for the population above 0.7 mJy now gives 0.17 -- 0.18 corresponding 
to a probablity of 8.2\% -- 13\% that the distributions are drawn from the same population. For the population above 3.96 mJy the result
remains unchanged which demonstrates the robustness of this result.

Using the available spectroscopic data we examined the likelihood for radio galaxies in these low-flux bins to be
members of A3158. There are 13 sources with spectroscopic redshifts which correspond to sources in the excess 
low flux bins (0.7 $\leq$ S$_{1.4} \leq$ 3.96 mJy), of which 10 (77\%) are members of the HRS compared to 
only 1 in 3 (33\%) for sources with fluxes above this level. We compared the percentage of sources with measured
redshift which were in the HRS as a function of flux bin. The results are summarised in Table 
%\ref{tab:flux} 
6 and show
a very strong correlation with the likelihood of a source being in the HRS to increase inversely with radio flux. This strongly
suggests that the excess low-powered sources are in the cluster. Given that the change over between AGN and starforming 
galaxies as the dominate population occurs at powers below log P$_{1.4}$(WHz$^{-1}$)=22.8 for rich clusters 
(Miller \& Owen 2002) this result is strongly suggestive of an increased starforming population.

\section{Radio Luminosity Function}
The radio luminosity function (RLF) at 1.4 GHz was computed using the 88 radio sources detected
within $23\farcm05$ of the centre of the radio image. This corresponds to the region in which the primary
beam response of the ATCA is essentially uniform and the attenuation negligible.

\begin{table*}
   \begin{center}

 \begin{tabular}{l|l|l|l|l|c|c|c|c|c|c}
        \hline

  Radio            &Coordinates   &  S$_{1.4}$ & Optical ID & & z & Ref. & Mag. &Name\\
  RA$_{\rm{J}2000}$     & DEC$_{\rm{J}2000}$&  mJy           & RA$_{\rm{J}2000}$  & DEC$_{\rm{J}2000}$&&&&\\
         \hline

03 41 11.43&  -53 19 26.7&	1.14 $\pm$ 0.03   & &                &&&&\\
03 41 41.79&  -53 59 26.0&	205.57 $\pm$ 2.06 & &                &&&&\\
03 42 39.82&  -54 04 08.5&	2.07 $\pm$ 0.04   & &                &&&&\\
03 42 40.55&  -53 09 51.9&	3.24 $\pm$ 0.04   & &                &&&&\\
03 42 49.22&  -53 10 06.1&	15.65 $\pm$ 0.16  & &                &&&&\\
03 42 56.86&  -54 02 31.1&	8.73 $\pm$ 0.09   & &                &&&&\\
03 45 06.75&  -53 15 44.6&      4.76 $\pm$ 0.06   & &                &&&&\\
03 45 41.95&  -53 39 07.5&	0.48 $\pm$ 0.03   & &                &&&&\\	
03 45 47.18&  -53 36 40.8&	0.52 $\pm$ 0.03   & &                &&&&\\	
03 45 53.65&  -53 39 03.8&	0.54 $\pm$ 0.03   & &                &&&&\\ 
03 46 04.90&  -53 15 20.9&	120.44 $\pm$ 1.20 & &                &&&&\\	
03 47 02.60&  -53 44 39.7&	5.58 $\pm$ 0.06   & &                &&&&\\
03 40 13.27&  -53 32 00.5&      4.02 $\pm$ 0.05   & 03 40 13.19   &-53 32 00.9& 0.061442 $\pm$ 0.000334 & 5 & 16.10 &2MASX J03401314-5331596\\
03 40 48.04&  -53 18 18.3&      4.07 $\pm$ 0.05   & 03 40 48.09   &-53 18 19.0&&&22.42&\\
03 40 52.42&  -53 53 09.7&      6.19 $\pm$ 0.07   & 03 40 52.46   &-53 53 10.8&&&22.56&\\
03 41 28.69&  -53 17 21.5&      7.54 $\pm$ 0.08   & 03 41 29.32   &-53 17 23.6&&&16.90&\\
03 41 50.78&  -54 02 29.5&      61.74 $\pm$ 0.62  & 03 41 50.73   &-54 02 29.3&&&18.10& 2MASX J03415075-5402298\\
03 43 33.79&  -54 11 19.1&      44.89 $\pm$ 0.45  & 03 43 33.71   &-54 11 19.8&&&17.55&\\
03 44 16.85&  -54 04 23.7&      31.74 $\pm$ 0.32  & 03 44 16.88   &-54 04 22.5&&&15.81& ESO 156-IG 009 NED01\\
03 44 23.84&  -53 59 31.9&	25.55 $\pm$ 0.29  & 03 44 23.84   &-53 59 31.9&&&22.57&\\
03 45 41.98&  -53 32 51.4&      4.36 $\pm$  0.05  & 03 45 42.04   &-53 32 52.0&&&20.99&\\
03 45 47.33&  -53 22 25.2&	0.56 $\pm$ 0.03   & 03 45 47.42   &-53 22 25.4&&&18.00& 2MASX J03454747-5322261\\
        \hline	
 \end{tabular}
 \caption{Details of detected radio source outside the area considered for the RLF.Columns 1 and 2 give the J2000 coordinates 
of the radio source; column 3 is the total flux at 1.4 GHz, note that as these sources are detected at the edges of the beam the
flux has been corrected for attenuation using the standard ATCA values (Wieringa \& Kesteven 92); columns 4 and 5 give the position of
the corresponding optical source from the DSS (where detected); column 6 gives the redshift of the optical source; column 7 is the
reference for the redshift where 1 is Katgert et al.~(1998), 2 is Loveday et al.~(1996), 3 is Smith et al.~(2004), 4 is De Vaucouleurs
et al.~(1991) and 5 is Lucey et al.~(1983); finally column 8 gives the optical source name as returned by the NASA Extragalactic Database.}
 \end{center}
 \label{tab:c}
\end{table*}

The integral RLF for different optical magnitude cuts offs is shown in Figure 
%\ref{uni}
7.
The plot expresses the probability of each galaxy, satisfying the selection criteria, having a radio source
with radio power $\ge$ log P$_{1.4}$. It has been well established that brighter
optical galaxies are more likely to host radio emission (Ledlow \& Owen 1996; Venturi et al. 1997) and our results
are consistent with this property with the curves exhibiting the same overall shape. Of course this is not that
surprising for integral plots expressed on a log scale as they are not comprised of independent data (i.~e.~ all of 
the data in the brighter optical magnitude curves is also contained in the fainter ones) and on a log scale small 
changes will be washed out. Nevertheless given limitations of the plots these results, though coarse, do suggest
that brighter galaxies within clusters are more likely to host radio emission.

In order to compare the bivariate RLF for A3158 with the cluster sample of Ledlow \& Owen (1996), hereafter LO96, our 
sample had to be matched to their sensitivity limits after being corrected for different cosmologies. This 
meant that the sample had to be restricted to sources with fluxes over 2.00 mJy which corresponds to 
logP$_{1.4}$(WHz$^{-1}$)=21.78, the LO96 lower limit, and b$_{\rm{J}} \leq$17.78 assuming that 
b$_{\rm{J}}$=B-0.2(B-V) (Fukugita et al.~1995) which corresponds to M$_R \leq$ -20.5 in LO96. There are nine galaxies (eight
ellipticals and one spiral) in the sample of 52 with fluxes greater than 2.00 mJy, of which only five
have blue magnitudes less than 17.78. Our sample is complete to a lower flux density than LO96 and if we examine
the number of sources having the same magnitude limit but with S$_{1.4} \geq$ 0.82 mJy we have a total of 15 galaxies.

The total number of galaxies with b$_{\rm{J}} \leq$17.78 in the region under consideration was obtained from 
SuperCOSMOS. This gave 112 galaxies. The total number of galaxies with blue magnitude less than 17.78 from SuperCOSMOS included those
of all morphological types over all redshifts. In order to compute the fraction of early-type galaxies we 
assumed that the spectroscopic samples of Rose et al.~(2002) and Fleenor et al.~(2005) were 
representative and that all galaxies for which no emission lines were detected represent early types. We determined the
fraction of galaxies likely to be found in the cluster by examining the percentage of galaxies in the Rose 
and Fleenor samples between 15,000 km s$^{-1}$ and 30,000 km s$^{-1}$ which corresponds to velocity coverage of 
the Horologium-Reticulum Supercluster (Fleenor et al 2006). This gave 78\% for the neighbouring A3125/A3128 complex
and 70\% for the HRS field examined by Fleenor. We averaged these results and adopted 74\% as the percentage of
galaxies likely to be in the HRS. These results were similar to the values obtained for the 
Shapley supercluster (Bardelli et al.~1998). We then computed
the total fraction of non-emission line galaxies from both the Rose and Fleenor samples finding them to be 
44\% in both samples. As a check we also examined the fraction of non-emission line galaxies in the restricted 
velocity range from 15,000 km s$^{-1}$ to 30,000 km s$^{-1}$ . The velocity restricted samples
returned 44\% and 47\% emission line galaxies. These values seemed at odds with known properties of clusters
(ie that the majority of galaxies would be elliptical) and results for similar studies in Shapley which found over 
80\% ellipticals. The problem relates to the fact that it is not uncommon for early-types to have at least one emission
line, usually [OII]$\lambda$3727 and occasionally two with [NII]$\lambda$6583 being likely (Kennicutt 1992). Thus,
a better indicator is not the total lack of emission lines but rather the use of absorption features as the main
indicator of redshift. Unfortunately the spectroscopic samples only specify if there were emission or absorption lines
present, there is no indication of how many. We have assumed that all objects for which either absorption only or
both emission and absorption features where used to generate the redshift correspond to early-type galaxies. Doing this the fraction of
early-types were found to be 77\% and 71\% for the Rose et al (2002) and Fleenor et al (2005) samples respectively. In
the velocity limited subsample we obtained 78\% and 74\% respectively. This small difference in the velocity restricted 
examination is likely due to the differences in how the spectroscopic samples were obtained. Whereas the 
spectroscopic sample of Rose et al.~(2002) covered the neighbouring clusters A3125 and A3128, the Fleenor 
sample covered the inter-cluster regions of the supercluster. Given both samples and the velocity restricted
analysis of the Rose data, which is a better match to A3158, all returned $\sim$77\% we adopted this value. Correcting for 
this fraction gives 64 galaxies which are likely to be both early-type and within A3158.

Figure \ref{fig:RLF} shows the resultant bivariate RLF for A3158 along with data for a sample of clusters from LO96
and data for the A3558 complex located in the centre of the Shapley supercluster generated by Venturi et al.~(2000). Details of
the bivariate luminosity function data are found in Table 
7
%\ref{tab:rlf}
. If the LO96 cluster sample were representative of A3158
we would expect 18 radio sources with S$_{1.4} \leq$ 2.00 mJy compared to the five observed.  Like the data for A3558 we see
that the RLF is lower than for LO96 meaning the probability for a radio source to be found in A3158 with radio power $\geq$ 10$^{21.77}$
W Hz$^{-1}$ is lower than in other cluster environments and the field. However, in the bin just fainter than the LO96 cut off we
find a sharp increase in the likelihood of having radio emission in A3158 which is again consistent with the excess in the low end 
of the source counts. We note that our estimation of the fraction of early-type galaxies in A3158 is possibly an overestimate, but
that even if the number of early-types is half of what we have used the bivariate RLF will still be significantly below both that found
for LO96 and Shapley except in the bin just fainter than the LO96 cut-off. Thus, even if our estimate is vastly wrong the general 
result remains unchanged.

 \begin{figure}
   \begin{center}
    \includegraphics[width=8.5cm]{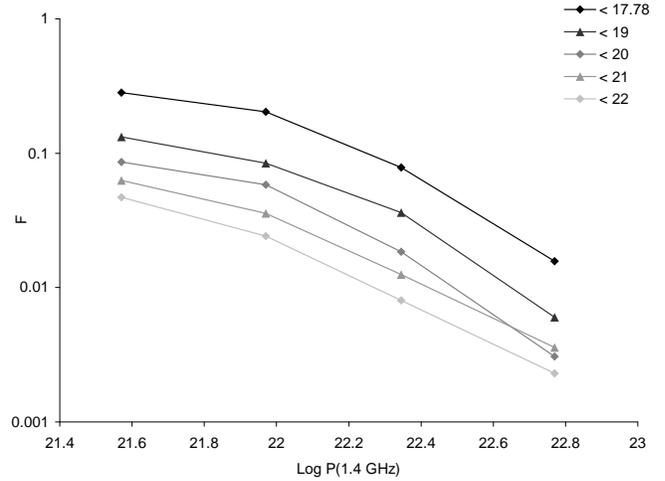}
   \caption{Radio luminosity curves for A3158 at different optical magnitude limits. The magnitude limits 
correspond to b$_{\rm{J}}$, the highest of which corresponds to M$_R \leq$ -20.5 in LO96 and is the same data 
as plotted in the bivariate RLF shown in Figure \ref{fig:RLF}.}
   \label{uni}
 \end{center}
 \end{figure}

\begin{table}
\label{tab:flux}
   \begin{center}

 \begin{tabular}{l|c|c|c|c|c}
        \hline

  Flux bin          &  0.7-0.99  &  0.99-1.4  &  1.4-1.98  &  1.98-2.80  &  2.80-3.96  \\
  (mJy)             &            &            &            &             &             \\
             \hline 
  A3158             &   2        &     3      &     3      &     -       &      2      \\
  background        &   -        &     1      &     1      &     -       &      1      \\
  \% in A3158       &  100\%     &    75\%    &    75\%    &     -       &      66\%   \\
         \hline

 \end{tabular}
 \caption{Table giving details of the 13 sources with measured redshifts and radio fluxes
between 0.7 and 3.96 mJy. The table shows an increasing trend for the sources to be found
in A3158 with decreasing flux. This suggests that the excess in the source counts found over
this flux range is likely to be associated with the cluster.}
 \end{center}
\end{table}

The cross over point between AGN and star-formation dominated radio emission depends on the
richness of the system. Typically it can be said that above P$_{1.4 GHz} = \times 10^{23} W Hz^{-1}$
AGNs dominate (Condon et al.~2002, Mauch \& Sadler 2007). However, for rich clusters, where 
the fraction of early-type galaxies is higher, this shifts slightly to  P$_{1.4 GHz} = \times 10^{22.8} W Hz^{-1}$
(Miller \& Owen 2002). Considering, that A3158 has a richness class of two, and
the fact that the break point between the two populations can only ever be determined statistically, it is
fair to say our results are broadly consistent with a star-forming population. This is further supported by the excess
in source counts below P$_{1.4 GHz} = \times 10^{22.5} W Hz^{-1}$ as discussed in Section 5 above.

We examined the colours of the galaxies which corresponded to the excess i.~e.~ those with 0.7 $\leq$S$_{1.4} \leq$3.9 mJy.
Finding there to be a substantially higher number of blue galaxies in these flux bins (48\%) as compared to either higher bins (26\%). Examination
of the colour magnitude diagram (B-R vs R) for the region for all galaxies in the SuperCOSMOS catalogue shows that 13 out of 27 galaxies 
corresponding to radio sources in the flux range  0.7 $\leq$S$_{1.4} \leq$3.9 mJy are bluer than the Red Cluster Sequence (RCS) as compared to only
7 galaxies which are redder (the rest are on the sequence) i.~e.~ 48\% of objects are bluer than the sequence compared to 26\% which are redder. 
In comparison, at higher flux values the percentage of objects above and below the RCS is the same (25\%). Without spectroscopic confirmation of
cluster membership, it is impossible to categorically argue that the larger number of objects bluer than the RCS in the flux range from 
0.7 $\leq$S$_{1.4} \leq$3.9 mJy definitely correspond to blue galaxies in the cluster. However, the bias in the scatter of objects in this 
flux bin to the blue compared to an even distribution about the RCS line at the higher flux bin is suggestive of a blue, low radio power population.

In addition, we examined the spatial distribution of these objects, finding them to be strongly elongated about the NW-SE axis of 
the cluster aligned toward the A2135/A3128 system.  In Figure \ref{fig:zoom} we show the DSS image of the entire cluster with 
radio objects with a detected optical host marked with either circles, ellipses or squares
based on the radio power bins used in the source count analysis. Pink circles show the positions of radio 
galaxies with radio fluxes at 1.4 GHz of less than 0.7 mJy, purple circles denote radio galaxies with fluxes in the range 0.7 - 0.98 mJy, 
blue ellipses are for galaxies with radio fluxes in the range 0.98 to 3.96 mJy and red squares are sources with fluxes greater than 3.96 mJy. 
Source with fluxes between 0.7 and 3.96 mJy represent those for which an excess over the background counts is detected. Sources which have been 
spectroscopically confirmed as members of A3158 are marked with crosses. Figure \ref{fig:zoom2} shows a magnified region around the cD galaxies 
where most of the faint, blue radio galaxies reside. The Figure shows radio emission from two of the three cD galaxies and, what appears to be, 
a chain of low-powered radio galaxies linking them.

  \begin{figure}
   \begin{center}
    \includegraphics [width=8.5cm, height=7cm]{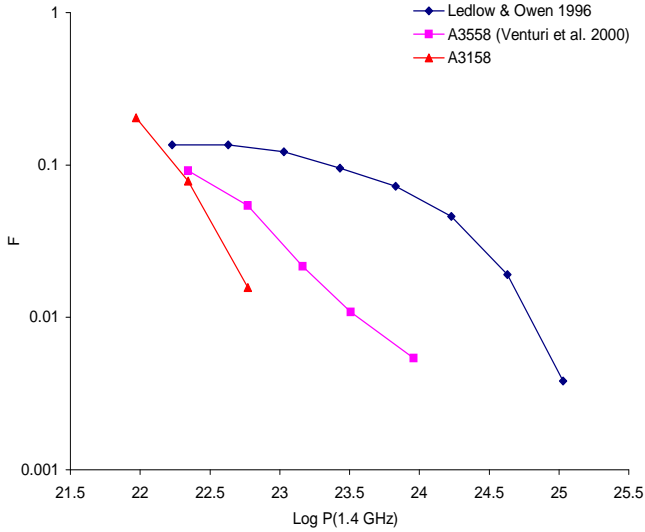}
   \caption{Integrated bivariate Radio Luminosity Function for A3158 (red triangles) for all galaxies with b$_{\rm{J}} \leq$ 
17.78 and S$_{1.4} \geq $ 0.88 mJy compared to data for the A3558 complex in Shapley (Venturi et al.~(2000) (pink squares) and
the RLF for clusters of galaxies from Ledlow \& Owen (1996) (blue diamonds).}
   \label{fig:RLF}
 \end{center}
 \end{figure}

\begin{table}
   \begin{center}

 \begin{tabular}{c|c|c|c}
        \hline

   $\Delta$ log P$_{1.4}$& $\Delta$ S$_{1.4}$ & Fractional BLF  & Integral BLF\\
         \hline 
  21.77 -- 22.17 &  0.82 -- 2.06   & 8/64 & 0.1250\\
  22.17 -- 22.52 &  2.06 -- 5.15   & 4/64 & 0.0625\\ 
  22.57 -- 22.97 &  5.15 -- 12.88  & 1/64 & 0.0156\\
  22.97 -- 23.36 &  12.88 -- 32.19 & 0/64 & 0\\
  23.36 -- 23.76 &  32.19 -- 80.47 & 0/64 & 0\\
  23.76 -- 24.16 &  80.47 -- 201.2 & 0/64 & 0\\
         \hline

 \end{tabular}
 \caption{Details of the bivariate RLF for A3158.}
 \end{center}
 \label{tab:rlf}
\end{table}

  \begin{figure*}
\vspace{-1.0cm}
   \begin{center}
    \includegraphics[width=15cm, angle=-90]{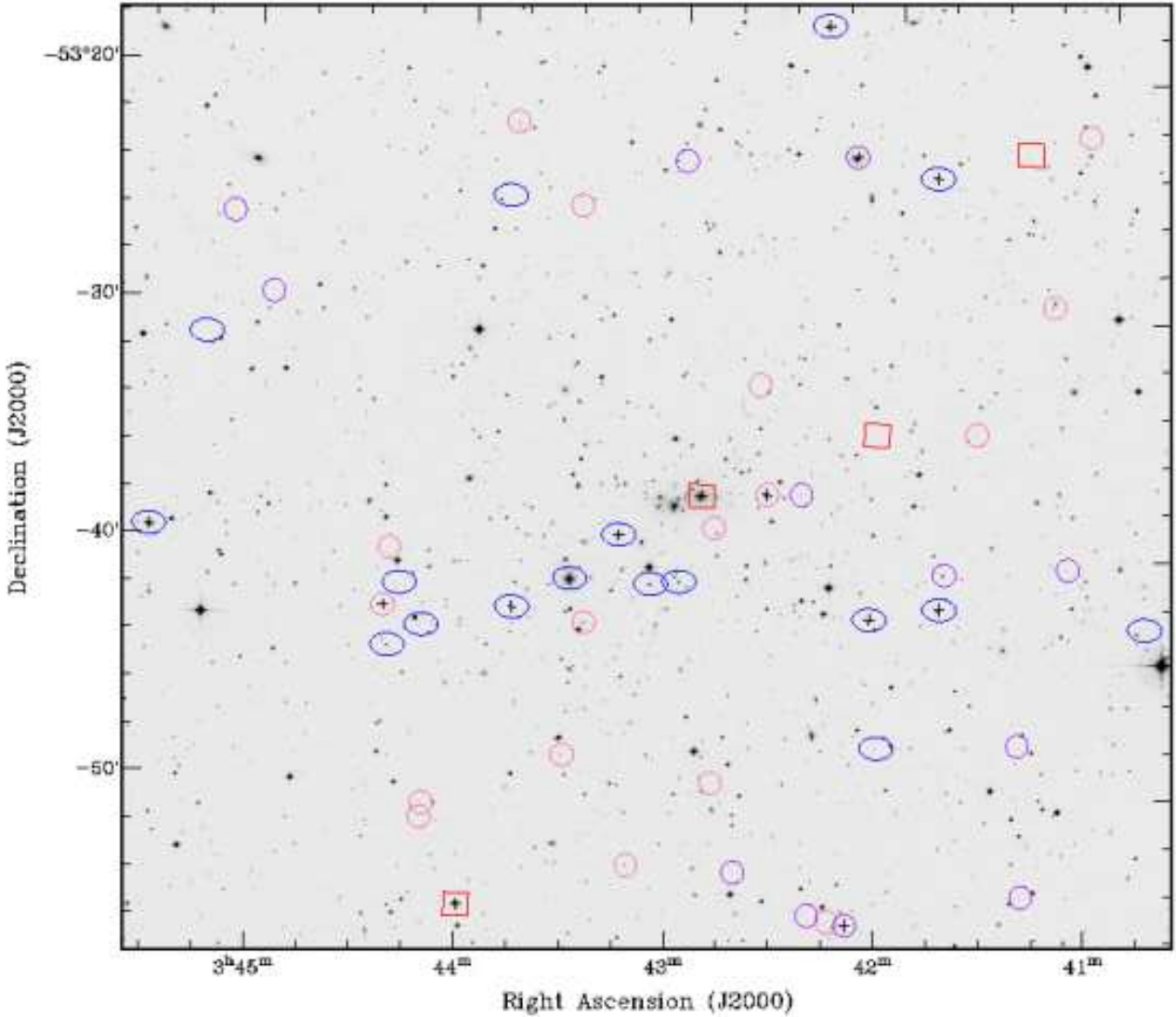}
\vspace{-0.4cm}
   \caption{Digitized Sky Survey image of A3158 annotated with the positions of the radio galaxies which have optical 
counterparts. The radio galaxy population has been divided based on the radio power bins used in the source count analysis
where pink circles show the positions of radio galaxies with radio fluxes at 1.4 GHz of less than 0.7 mJy, purple circles denote radio
galaxies with fluxes in the range 0.7 - 0.98 mJy, blue ellipses are for galaxies with radio fluxes in the range 0.98 to 3.96 mJy and red
squares are sources with fluxes greater than 3.96 mJy. Source with fluxes between 0.7 and 3.96 mJy represent those for which an excess 
over the background counts is detected. Sources which have been spectroscopically confirmed as members of A3158 are marked with crosses.
 Note the East-west chain of optical galaxies, including two of the three cD galaxies, 
which are associated with radio sources with fluxes between 0.98 to 3.96 mJy . This radio emission corresponds to the low-power 
excess detected in the cluster. The alignment is significant as it follows the overall cluster elongation and direction of the 
bridge toward the nearby A3125/A3128 complex.}
   \label{fig:zoom}
 \end{center}
 \end{figure*}

\section{Discussion}

Several studies have been conducted to examine the way in which the cluster environment affects the radio galaxy population either
via composite studies such as that undertaken in Ledlow and Owen (1996), ``LO96'', or detailed studies of individual clusters or regions.
Unfortunately, the results of the individual studies are often conflicting, possibly as a result of low number statistics, however some
clear trends are emerging with respect to the three main indicators of the status of the radio population: source counts, radio
luminosity functions and colour analysis of faint excesses. Of these, source counts provide an effective way to search for enhancements
over the expected background. To date there have been several source count studies undertaken for parts of the Shapley Supercluster
(Venturi et al.~1997, 2000, 2001, 2002) all of which have returned results which are consistent with the predicted background suggesting
that the increased optical density of the supercluster environment is not at all reflected in the radio galaxy population. However,
all of these studies have been relativity shallow, typically only examining sources with log P$_{1.4}$ $\geq$ 21.8 and are thus, barely probing the
regime of starforming galaxies. 

The situation with radio luminosity functions and in particular those for early-type galaxies seems to be complex.
Venturi et al.~(1997) examined the RLF for A3556 in Shapley and found it to be consistent with that found in other 
environments, hence concluding that, in this case, the dynamical state of the cluster played no role in the likelihood 
of radio galaxies forming. However, in a 
further study Venturi et al.~(2000) presented a detailed study of the A3558 complex in the center of the Shapley supercluster finding
that neither local nor the cluster environment 
played a significant role in the power distribution of
radio sources in the region.  Therefore, Venturi et al.~(2000) conclude that 
the probability of forming low-powered radio sources (log P$_{1.4} \geq$ 22.5 ) in Shapley is lower
than in both the field and (composite) cluster environments.  Furthermore,
they suggest that cluster merging switches off previously active radio 
galaxies, i.~e.~, mergers anti-correlate with radio emission in early-type galaxies.
Next they examined the A3528 cluster complex in Shapley
again finding that the RLF was consistent with that found in other environments (Venturi et al.~2001). Finally, they
considered the radio emission in A3571 (Venturi et al.~2002) again finding that it was suppressed with respect to the expected
results from LO96. Giacintucci et al.~(2004) reobserved the outskirts of the A3558 complex and found the RLFs for both early and 
late-type galaxies to be consistent with the ``universal" RLFs of Ledlow and Owen (1996) and Gavazzi \& Boselli (1999) respectively, 
suggesting that the deficit of radio galaxies in the previous observation was entirely due to the rich cluster A3558 itself. 

Venturi et al.~(2001, 2002) attribute the difference in the RLFs to different stages in the merger history of 
the regions playing a critical role in the determination of radio emission in cluster galaxies. In the case of early mergers suggested 
for A3556 and the A3528 complex they suggested that the radio galaxy population has not yet been affected, whereas for 
the late-stage merger in A3571 and A3558, in the very core of Shapley, the radio galaxy population has been switched off by the merging. 
Given that cluster mergers are particularly long lived ($\sim$ Gyr) compared to the lifetimes of radio sources which are typically 
only 10$^6$ -10$^8$ years it seems likely that the radio population could give vital clues as to the exact point in a cluster's 
dynamical history, thus providing to be an extremely important evolutionary indicator. This is particularly true if 
the radio emission is considered together with other signposts of cluster dynamics such as X-ray observations.

  \begin{figure*}
\vspace{-0.5cm}
   \begin{center}
    \includegraphics[width=10cm, angle=-90]{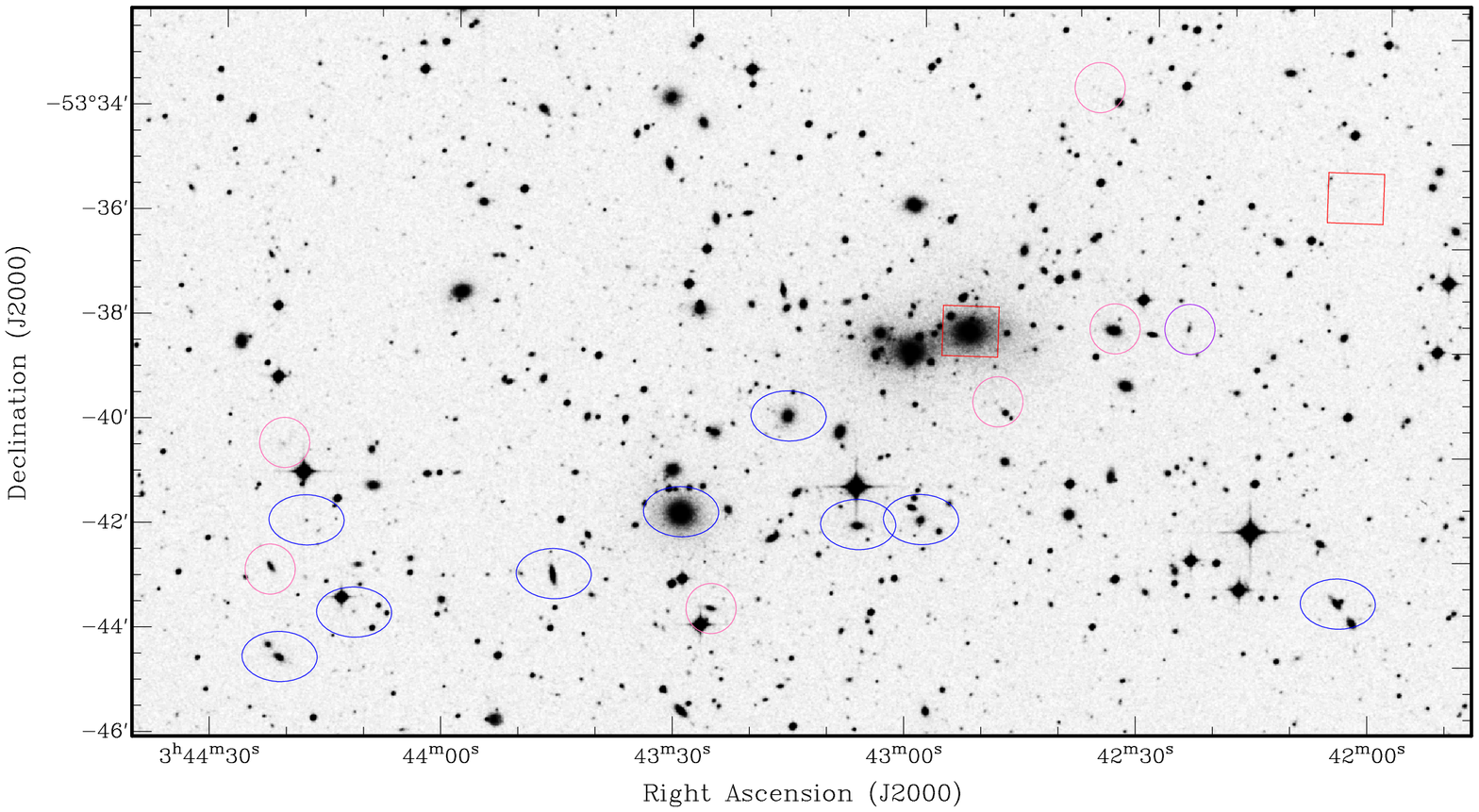}
%\vspace{-2.4cm}
   \caption{Digitized Sky Survey image of the centre of A3158 annotated with the positions of the radio galaxies which have optical 
counterparts as in Figure \ref{fig:zoom}. The image shows the chain of low-powered radio galaxies that correspond to the excess seen in
the source counts. It shows that this chain of galaxies is aligned in the direction of the three cD galaxies.}
   \label{fig:zoom2}
 \end{center}
 \end{figure*}

Despite not being evident via an excess in source counts, Giacintucci et al.~(2004) found a population of low powered (log P$_{1.4} 
\sim$ 22) galaxies on the outskirts of the A3558 complex, many of which showed a blue excess which they suggested are likely to be 
starbursts, although they noted that their observations provided only
weak evidence that the fraction of radio-emitting galaxies was higher than in non-merging environments. Miller (2005) conducted a deeper observations of the A3558 complex over 7 deg$^2$ and, on the basis of colour and radio 
luminosity, confirmed the presence of an increased starforming population associated with the recent merging of SC 1329-313 and 
A3562 with the rest of the supercluster. An excess of low-powered blue galaxies was also observed
in the merging clusters A2125 (Owen et al.~ 2005), A2255 (Miller et al.~2003) and in A2111 (Miller et al.~2006) which despite 
lacking statistical evidence of substructure at right angles to the plane of the sky, is likely to be dynamically active. In both 
cases the authors attribute the excess of dusty starburst galaxies to the merger activity. In all cases an increase in 
low-power, blue galaxies either confirmed via spectroscopic analysis as starburst (Owen et al.~2005) or assumed through colour 
analysis (Giacintucci et al.~2004) has been associated with recently merging clusters.

In the case of A3158 we see both an excess of low-powered radio galaxies (log P$_{1.4} \sim$ 22) in the source counts and a suppression
of the early-type RLF with respect to universal indicators. Thus, we seem to be seeing both a suppression of AGN as observed in the
late mergers in A3571 and A3558 (Venturi et al.~2000 and 2002) and an increase in the low-powered, blue galaxies, likely to be starbursts,
as seen in the mergers for A2125, A2255, A2111 (Owen et al.~2005, Miller et al.~2003 \& Miller et al.~2006 respectively) and 
the outskirts of A3558 (Giacintucci et al.~2004). In addition, the axis of the starforming galaxy candidates is elongated and aligned
in the direction which connects A3158 to the more massive A3125/A3128 complex which lies at the southern centre of the HRS. This is similar to the
distribution of star-forming galaxies seen in A2125.
Additionally the lack of HT galaxies is similar to both A3571 and A3558
and was used by Venturi et al.~(2001) to argue for these clusters being in the later stages of a dynamical encounter.

\section{Conclusions}
We present the the first of a series of papers investigating radio emission in the Horologium-Reticulum Supercluster. In this paper
we present radio imaging at 1.4 and 2.5 GHz of the cluster A3158, from which we detect 109 radio sources of which 64 are found to have
optical counterparts. By considering a reduced subsample of 87 sources within a $23\farcm05$ radius, over which the primary beam attenuation of the ATCA
is negligible, we are able to investigate the statistical properties of radio emission in the cluster. Our results can be summarised as
follows.

(1) The radio source counts in A3158 are unlikely to be consistent with the background source counts and in particular there appears
to be an excess of galaxies with S$_{1.4} \leq$ 3.96 mJy; a KS test of the entire sample against the background only returns a $\sim$3\% - 8\% chance
that the distributions are drawn from the same population, while a KS test over the sample with S$_{1.4} \leq$ 3.96 mJy gives
a much 
% only a slightly
higher chance ($\sim$60\%) that the distributions are drawn from the same population suggesting that the low-powered excess is the significant difference. 

(2) The percentage of radio galaxies associated with the cluster as determined by spectroscopic redshifts is highly correlated to radio flux/power
which is consistent with the excess source counts being the result of low-powered radio galaxies associated with A3158; in addition colours
of these galaxies suggest they are blue which is likely to indicate starformation as a driving mechanism.

(3) The alignment of the candidate starforming galaxies associated with the increased source counts follows the bridge of galaxies connecting
A3158 to the A3125/A3128 complex suggestive of merger-induced formation.

(4) The radio luminosity function for early-type galaxies is significantly lower than for other composite cluster environments and the
field; this result is similar to the results for the late merger in A3558 obtained by Venturi et al.~(2000).

Thus, if we can use the radio galaxy population as an indicator of merger history it is likely that A3158 is observed in a late merger state with
a significant excess of low-powered, blue galaxies aligned along the axis connecting this cluster to the A3125/A3128 complex. These galaxies are likely
the result of merger-induced starformation.

\section*{Acknowledgments}

We thank the anonymous referee for their careful reading of the manuscript and insightful
comments. MJ.-H. acknowledges the support of IRGS grant J0015125 \& J0016583 administered by the 
University of Tasmania. M.S. was supported in this work via a University of Tasmania Vacational 
Scholarship in Astrophysics. J.A.G acknowledges the support of a Dave Warren Scholarship 
administered by the University of Tasmania. A.-M.B. was supported by a University of Tasmania 
Honours Scholarship. This research was supported by the DFG cluster of excellence ``Origin 
and Structure of the Universe''. A portion of this work was supported  by NSF grants 
AST-9900720 and AST-0406443 to the University of North Carolina-Chapel Hill.
The Australia Telescope Compact Array is part of the 
Australia Telescope which is funded by the Commonwealth of Australia for 
operation as a National Facility managed by CSIRO. This research has made use of 
the NASA/IPAC Extragalactic Database (NED) which is operated by the Jet 
Propulsion Laboratory, California Institute of Technology, under contract with 
the National Aeronautics and Space Administration. The Digitized Sky Surveys were 
produced at the Space Telescope Science Institute under U.S. Government grant 
NAG W-2166. The images of these surveys are based on photographic data obtained 
using the Oschin Schmidt Telescope on Palomar Mountain and the UK Schmidt 
Telescope. The plates were processed into the present compressed digital form with 
the permission of these institutions.

%\bsp

\label{lastpage}

\end{document}